%% file: dpm-mp.tex
\documentclass[letterpaper]{llncs}

\usepackage[utf8]{inputenc}
\usepackage[T1]{fontenc}
\usepackage{textcomp}
\usepackage{mathptmx}
\usepackage{fancyvrb}
\usepackage{algorithm}
\usepackage{algpseudocode}
\usepackage{mathtools}
\usepackage{bbm}
\usepackage[flushleft]{threeparttable}
\usepackage{etoolbox}\AtBeginEnvironment{algorithmic}{\scriptsize} 
\usepackage[usenames,dvipsnames]{xcolor}
\usepackage{hyperref}
\usepackage{adjustbox}
\usepackage{amssymb}
\usepackage{pgfplots}
\usepackage{pgfplotstable}
\usepackage{filecontents}
\usepgfplotslibrary{units}
\usetikzlibrary{arrows, calc, spy, patterns, backgrounds}

\pgfplotsset{
	my ybar legend/.style={
		legend image code/.code={
			\draw [##1] (0cm,-0.6ex) rectangle +(1.75em,1.1ex);
		},
	},
}

\usepackage{booktabs}

\usepackage{graphicx, caption, subcaption}
\usepackage{acro}

\usepackage{relsize}
\usepackage[shrink=20]{microtype}
\usepackage{csquotes}
\usepackage{listings}
\usepackage{algpseudocode}
\usepackage[curves]{struktex}
\usepackage{mathrsfs}
\usepackage{multicol}

\input{macros}

\acuse{CPU}
\input{todo}
\usepackage{cleveref}

\SaveVerb{Emmy}.Emmy.
\SaveVerb{Meggie}.Meggie.
\SaveVerb{HazelHen}.Hazel Hen.
\SaveVerb{SuperMUCNG}.SuperMUC-NG.
\SaveVerb{MPIcommsize}.MPI_COMM_SIZE.
\SaveVerb{MPI}.MPI.
\SaveVerb{OpenMP}.OpenMP.
\SaveVerb{MPIOpenMP}.MPI+OpenMP.
\SaveVerb{NUMA}.NUMA.
\SaveVerb{STREAM}.STREAM.
\SaveVerb{Cosine}.Cosine.
\SaveVerb{DIV}.DIV.
\SaveVerb{SELL}.SELL.
\SaveVerb{SELL11}.SELL-1-1.
\SaveVerb{SELL321}.SELL-32-1.
\SaveVerb{SpMV}.SpMV.
\SaveVerb{static}.static.
\SaveVerb{H}.H.
\SaveVerb{Topi128}.Topi-128-128-64.
\SaveVerb{2Topi128}.Topi-128-64-64.

\newif\ifblind
\blindfalse

\makeatletter
\newcommand{\verbatimfont}[1]{\renewcommand{\verbatim@font}{\ttfamily#1}}
\makeatother

\begin{document}
\title{Desynchronization and Wave Pattern Formation in MPI-Parallel and Hybrid Memory-Bound Programs}
\titlerunning{Structure Formation in \verb.MPI. Programs}
\ifblind
\else
\author{Ayesha Afzal\inst{1}, Georg Hager\inst{1}, and Gerhard Wellein\inst{1,2}}
\authorrunning{A.\ Afzal et al.} 
\institute{Erlangen Regional Computing Center (RRZE), 91058 Erlangen, Germany,\\
\email{ayesha.afzal@fau.de, georg.hager@fau.de}
\and
Department of Computer Science, University of Erlangen-N\"urnberg, 91058 Erlangen, Germany,\\
\email{gerhard.wellein@fau.de}}
\fi
\maketitle
\pgfkeys{/pgf/number format/.cd,1000 sep={\,}}

\begin{abstract}
  Analytic, first-principles performance modeling of
  distributed-memory parallel codes is notoriously imprecise. Even for
  applications with extremely regular and homogeneous
  compute-communicate phases, simply adding communication time to
  computation time does often not yield a satisfactory prediction of
  parallel runtime due to deviations from the expected simple lockstep
  pattern caused by system noise, variations in communication time,
  and inherent load imbalance. In this paper, we highlight the
  specific cases of provoked and spontaneous desynchronization of
  memory-bound, bulk-synchronous pure MPI and hybrid MPI+OpenMP
  programs. Using simple microbenchmarks we observe that although
  desynchronization can introduce increased waiting time per process,
  it does not necessarily cause lower resource utilization but can lead
  to an increase in available bandwidth per core. In case of
  significant communication overhead, even natural noise can shove the
  system into a state of automatic overlap of communication and
  computation, improving the overall time to solution. The saturation
  point, i.e., the number of processes per memory domain required to
  achieve full memory bandwidth, is pivotal in the dynamics of this
  process and the emerging stable wave pattern. We also demonstrate
  how hybrid MPI-OpenMP programming can prevent desirable
  desynchronization by eliminating the bandwidth bottleneck among
  processes.  A Chebyshev filter diagonalization application is used
  to demonstrate some of the observed effects in a realistic setting.
\end{abstract}

\section{Introduction}\label{sec:intro}

In principle, a parallel computer should be a deterministic
system. Given some code and hardware specifications, it should be
possible to predict the runtime of the program and measure it
consistently in repeated experiments. Analytic, first-principles
performance models such as \Rl~\cite{roofline:2009} or
ECM~\cite{sthw15,hofmann2019bridging} approximate this goal on the
core and socket level. Although residual deviations and statistical
variations remain, these models can yield valuable
insights into the hardware bottlenecks of computation despite the
simplifications that go into the model assumptions.  One of these is
the notion that all cores or hardware threads execute the same code on
different data, which is often true for programs exploiting
thread-level loop parallelism. With message passing, however, the
dependencies among instruction streams (processes) are less tight, and
communication overhead complicates the picture. Ideally, one would like
to add communication models such as the Hockney
model~\cite{Hockney:1994} or its refinements on top of \Rl or ECM, but
this is too simplistic: System noise,
variations in network bandwidth and latency, load
imbalance, and strong one-off delays can cause global effects such
as desynchronization and traveling idle
waves~\cite{AfzalHW19,markidis2015idle}. Using threaded MPI processes
changes the phenomenology and dynamics of the system,
because socket-level bottlenecks
(i.e., memory bound vs.\ core bound) play a decisive role.
It is therefore necessary to shed light on the dynamic processes
that lead to desynchronization and global structure formation
in pure MPI and threaded MPI programs.

An \emph{idle wave} is a period of idleness caused by a strong
delay in computation or communication on one process of an MPI
program.  It travels across the MPI processes with a speed that is
governed by the particular communication characteristics (distance of
communication, eager vs.\ rendezvous mode, etc.) and interacts
with other idle waves, computational noise, and system noise in a nonlinear way. 
In this paper, we extend a previous study on the dynamics of
idle waves with core-bound pure-MPI programs~\cite{AfzalHW19}
towards the memory-bound case, i.e., codes with a low computational intensity.
These exhibit saturation characteristics when running on multiple cores
connected to a single memory interface (the \emph{contention
  domain}\footnote{This is usually identical to a ccNUMA domain and
  often, but not always, a full CPU socket.}).
The basic mechanisms are investigated using parallel microbenchmarks
that are amenable to straightforward node-level performance modeling
and can be easily altered to mimic different application
characteristics.

We start by comparing the dynamics of traveling idle waves generated
by injected one-off delays between core-bound and memory-bound MPI
programs with negligible communication overhead and perfect load balance. More
complex dynamics can be observed in the memory-bound case within the memory domain and
when crossing domain boundaries (sockets, nodes)\@. Even after the
idle wave is gone, a distinctive ``computational wave'' pattern
prevails that is governed by the topological properties of the MPI
program (inter-process communication dependencies, boundary
conditions) and the location of the memory bandwidth saturation point,
i.e., the number of processes required for full memory bandwidth
utilization.
In case of significant communication overhead, a massive one-off delay
is not required to provoke the wave pattern; the natural system noise
or a single, small disturbance of regularity in computation or
communication time is sufficient. Based on these observations, we study
the impact of using threaded MPI processes. Multithreading has an
influence on the bandwidth saturation point, and filling the
contention domain with a multi-threaded MPI process effectively
generates a bandwidth-scalable code.
This answers the
long-standing question why a nonreflective introduction of OpenMP
threading into an MPI-only code can cause a slowdown even if
OpenMP-specific overheads are negligible.  Finally, we employ an
application code implementing Chebyshev Filter Diagonalization
(ChebFD) for a topological insulator problem to show the relevance
of our findings in a real-world scenario.

The configuration parameter space of MPI and hybrid MPI-OpenMP
parallel programs is huge. Here we restrict ourselves
to simple, bidirectional point-to-point communication using
eager or rendezvous protocols (depending on the message size)\@.

This paper is organized as follows: 
\Cref{sec:environment} details our experimental environment and methodology.
In \Cref{sec:idlewave} we study the propagation of an injected, one-off delays,
contrasting memory-bound with core-bound scenarios. Computational wavefronts
in memory-bound programs emerging from idle waves are covered in \Cref{sec:nocomm}\@.
\Cref{sec:sigcomm} deals with the spontaneous formation of wavefronts
and its consequences on program performance, and in \Cref{sec:chebFD}
we showcase some of the observed effects using an application code.
\Cref{sec:relatedwork} covers related work and \Cref{sec:conclusion}
concludes the paper and gives an outlook to future work.

\paragraph{Contributions}\label{sec:contributions}

This work makes the following novel contributions:
\begin{itemize}
\item We show the characteristics of idle waves traveling through
  memory-bound MPI applications on multicore clusters, and how they
  differ from the core-bound case studied in prior work~\cite{AfzalHW19}.
\item We show that the forced emergence of computational wave patterns
  via desynchronization by one-off delays only occurs with 
  memory-bound code.
\item We show that the average available memory bandwidth per core in an
  established computational wave (desynchronized state) is larger
  than in the synchronous state while the core is executing application
  code. 
  The wave settles in a state where the number of active processes
  per contention domain is near the memory bandwidth saturation point.
\item We show how natural system noise leads to
  spontaneous desynchronization and computational wave formation
  if there is significant communication overhead.
\item We show that desynchronization can lead to automatic overlap
  of communication and computation, reducing overall time to solution.
  Significant intra-node communication overhead can reduce this gain.
\item We show that the introduction of threaded MPI processes
  can prevent the formation of computational waves and automatic
  communication overlap if one process is used per contention domain,
  effectively recovering the characteristics of a scalable pure MPI
  code.
\end{itemize}

\section{Experimental environment and methodology}\label{sec:environment}

\subsection{Cluster test bed and external tools}\label{sec:hwsw}

In order to ensure that our observed phenomenology is not specific to a
singular hardware or software setup, four different clusters were
used to conduct various  experiments:
\begin{itemize}
\item \UseVerb{Emmy}\footnote{\ifblind{URL redacted for double-blind review}\else\url{https://anleitungen.rrze.fau.de/hpc/emmy-cluster}\fi},
  a QDR-InfiniBand cluster with dual-socket nodes comprising ten-core Intel Xeon
  ``Ivy Bridge'' CPUs ans Hyper-Threading (SMT) enabled,
\item \UseVerb{Meggie}\footnote{\ifblind{URL redacted for double-blind review}\else\url{https://anleitungen.rrze.fau.de/hpc/meggie-cluster}\fi},
  an Omni-Path cluster with dual-socket nodes comprising ten-core Intel Xeon
  ``Broadwell'' CPUs and Hyper-Threading (SMT) disabled,
\item \UseVerb{HazelHen}\footnote{\url{https://hlrs.de/systems/cray-xc40-hazel-hen}},
  a Cray XC40 with Aries interconnect and 12-core Intel Xeon ``Haswell'' CPUs,
\item \UseVerb{SuperMUCNG}\footnote{\url{https://doku.lrz.de/display/PUBLIC/SuperMUC-NG}},
  an Omni-Path cluster with dual-socket nodes comprising 24-core Intel Xeon
  ``Skylake SP'' CPUs.
\end{itemize}
Details of the hardware and software environments on these systems
can be found in Table~\ref{tab:system}.
\begin{table}[tb]
	\centering
	\begin{adjustbox}{width=1\textwidth}
	\begin{threeparttable}
	\caption{Key hardware and software specifications of systems} 
	\label{tab:system}
	\begin{tabular}[fragile]{lcccc}
		\toprule
		Systems  & \UseVerb{Emmy}   & \UseVerb{Meggie} & \UseVerb{HazelHen} (CRAY XC40)  &  \UseVerb{SuperMUCNG}  \\
		\midrule
		Intel Xeon Processor  & Ivy Bridge EP  & Broadwell EP       &   Haswell EP		& Skylake SP \\    
		Processor Model      & E5-2660 v2   & E5-2630 v4            &   E5-2680 v3 			& Platinum 8174 \\
		Base clock speed &\SI{2.2}{\giga \Hz}       & \SI{2.2}{\giga \Hz}  & \SI{2.5}{\giga \Hz}  &  \SI{3.10}{\giga \Hz}(\SI{2.3}{\giga \Hz} used\mbox{$\ast$})  \\
		Physical cores per    & 20    &20             & 24			& 48 \\
		dual socket node &     &             & 			&   \\
		LLC size & \SI{25}{\mega \byte}    &\SI{25}{\mega \byte}             & \SI{30}{\mega \byte}			& \SI{33}{\mega \byte}  \\
		Memory per node (type)& \SI{64}{\giga \byte} (DDR3)    & \SI{64}{\giga \byte} (DDR4)            & \SI{128}{\giga \byte} (DDR4) 			& \SI{96}{\giga \byte} (DDR4)  \\
		Theor. memory bandwidth & \SI{51.2}{\giga \byte / \second} & \SI{68.3}{\giga \byte / \second}& \SI{68.3}{\giga \byte / \second} &  \SI{128}{\giga \byte / \second}\\
		Node interconnect    & QDR InfiniBand      & Omni-Path           & Cray Aries			& Omni-Path\\
		Interconnect topology & Fat-tree & Fat-tree & Dragonfly & Fat-tree \\
		Raw bandwidth per &\SI{40}{\giga \bit \per \second}       &    \SI{100}{\giga \bit \per \second}  &  \SI{126}{\giga \bit \per \second} & \SI{100}{\giga \bit \per \second}  \\
		link and direction &     &             & 			&   \\
		\midrule
		Software  \\
		\midrule
		Compiler    & Intel \CPP{} v2019.4.243         & Intel \CPP{} v2019.4.243       & Cray \CPP{} v8.7.10			& Intel \CPP{} v2019.4.243\\
		Message passing library & Intel \verb.MPI. v2019u4         & Intel \verb.MPI. v2019u4       & Cray MPICH v7.7.6			& Intel \verb.MPI. v2019u4\\
		Operating system    & CentOS Linux v7.7.1908      &  CentOS Linux v7.7.1908           & SESU Linux ENT. Server 12 SP3 			&  SESU Linux ENT. Server 12 SP3\\ 
		\midrule
		Tools  \\
		\midrule
		\verb.ITAC.    & v2019u4         & v2019u4      & \mbox{$\S$}			& v2019\\
		\verb.LIKWID.   & 5.0.0         & 5.0.0      & 
		\mbox{$\S$}			& 4.3.3\\
		\bottomrule
	\end{tabular}
\begin{tablenotes}
	\small
	\item \mbox{$\ast$} A power cap is applied on
          \UseVerb{SuperMUCNG}, i.e., the CPUs run by
          default on a lower than maximum clock speed (\SI{2.3}{\giga
            \Hz} instead of \SI{3.10}{\giga \Hz}).
	\item \mbox{$\S$} \CPP{} high-resolution \verb.Chrono. clock for timing measurement.
\end{tablenotes}
  \end{threeparttable}
\end{adjustbox}
\end{table}

We used
\ac{ITAC}\footnote{\url{https://software.intel.com/en-us/trace-analyzer}}
for timeline visualization (except on \UseVerb{HazelHen}, where
traces were recorded by explicit timing measurements), the \CPP{}
high-resolution \verb.Chrono. clock for timing, and
\verb.likwid-perfctr. from the \verb.LIKWID. tool
suite\footnote{\url{http://tiny.cc/LIKWID}} for memory bandwidth
measurements.

\subsection{Experimental parameters and methodology}

We took a number of measures to create a reproducible experimental
environment and minimize any noise from system sources. On the
\UseVerb{Emmy} and \UseVerb{Meggie} systems, we ran all multi-node experiments on
nodes connected to a single leaf switch. Core-thread affinity
was enforced.  The computational workload for the core-bound case was
a number of back-to-back divide instructions (\verb.vdivpd.), which
have a low but constant throughput on Intel architectures if
``simple'' denominators are avoided. Except for the application case
study, the memory-bound workload comprised simple kernels like STREAM triad.
One-off idle periods
were generated by massively extending one computational phase.

Most microbenchmark experiments were performed on two nodes only, since
the basic phenomenology is visible even on this scale. Bidirectional
point-to-point communication between MPI processes employed a
standard
\verb.MPI_Isend./\verb.MPI_IRecv./\verb.MPI_Waitall. sequence. Before
actual measurements were taken, at least two warm-up time steps with barrier
synchronization were performed to allow the MPI and OpenMP runtimes
to settle and eliminate first-call overhead. We only report statistical
variation in measurements where the relative spread was larger than
5\%.
Unless otherwise stated, the clock speed of processors was fixed.
On \UseVerb{SuperMUCNG}
the active power capping feature leads to an effective clock speed of
2.3\,\GHZ, which was validated by the \verb.likwid-perfctr. tool.

\input{figures/IdleWave}
\section{Idle wave mechanisms for memory-bound code}\label{sec:idlewave}

In~\cite{AfzalHW19},
idle waves were
shown to have \emph{nonlinear} characteristics, i.e., colliding waves
interact and partially cancel each other. Noise, i.e., short
delays from different sources such as load imbalance, varying
communication characteristics, or system noise, causes the decay
of traveling idle waves. In this section, we compare the known dynamics
of idle waves between core-bound code and memory-bound code.
For brevity and to avoid confusion, we will call the two phenomena
\emph{core-bound} and \emph{memory-bound idle wave}, respectively.
We also restrict ourselves to the case of negligible communication
overhead, i.e., a small communi\-cation-to-compu\-tation ratio.

\subsection{Idle wave propagation speed}

\Cref{fig:IdleWave}a shows a traveling idle wave on the 
\UseVerb{SuperMUCNG} system with core-bound code.
The leading and the trailing edges of the wave
are parallel, and due to the communication characteristics
(bidirectional next-neighbor, eager mode, closed ring) the waves
emanating from the idle injection cancel each other after one half
round trip. The memory-bound code in \Cref{fig:IdleWave}b shows
a very different pattern: Since the available memory bandwidth per core
declines after the saturation point, the length of any particular execution
phase on any particular MPI process depends on how many other processes
are executing user code at the same time on the same contention domain.
If $b(N)$ is the STREAM memory bandwidth
with $N$ processes, transferring a data volume of $V$ bytes with a single process
takes a time of 
\bq\label{eq:texec}
T_\mathrm{exec}=\frac{NV}{b(N)}\eos
\eq
In the saturation phase, where $N>N_\mathrm{sc}$, $b(N)\sim \text{const.}$
and thus $T_\mathrm{exec}\sim NV$, i.e., as the front (back) of the
idle wave progresses through the cores of a socket and more (fewer)
cores participate in code execution, the time per iteration
goes up (down)\@. Hence, the forward and backward edges of the idle wave ripple
through the system at \emph{variable} propagation speeds.

The expression
for the silent-system idle wave propagation speed from~\cite{AfzalHW19}
still holds, but with modifications. Instead of
the whole idle wave velocity, we can only draw conclusions for either
of its two edges at a single moment in time since the execution time
obeys the relation (\ref{eq:texec})\@. The \emph{local} velocity is
\bq\label{eq:PropSpeed}
v_\mathrm{silent}(N) = \frac{\sigma\cdot d}{NV/b(N)+T_\mathrm{comm}}~
\left[\frac{\mbox{ranks}}{\mbox{s}}\right]\cma
\eq
where $N$ is the number of processes executing code. This means that
$\nu_\mathrm{silent}$ can be different for processes on the same contention
domain, which will be investigated further in the next section. 
$T_\mathrm{comm}$ is the communication time, $d$ parameterizes the distance of communicating
processes, and $\sigma\in\{1,2\}$ is a correction factor that depends on communication
characteristics, e.g., communication patterns (uni- vs.\ bidirectional),
flavors (multiple split-waits vs.\ one wait-for-all),
and protocols (eager vs.\ rendezvous)~\cite{AfzalHW19}.

Note that (\ref{eq:PropSpeed}) even holds for hybrid MPI/OpenMP
programs that communicate only outside parallel regions.  In this
case, $N$ is the number of active multi-threaded MPI processes on a
socket. If the process spans the full socket, $N=1$ and the propagation
speed does not vary. This setting will be analyzed in Sect.~\ref{sec:Hybrid}\@.

\input{figures/ComputationalWave}

\subsection{Idle wave decay}

In~\cite{AfzalHW19} it was shown that noise, i.e., small statistical disturbances
of the pure lock-step pattern, cause the
decay of traveling idle waves, possibly to the point where a one-off
injection does not even impact the time to solution of the program.
In a noise-free system, a core-bound idle wave does not decay, but eventually
interacts with itself or with the boundaries of an open process topology.

The propagation and decay mechanisms of memory-bound idle waves are much
different since
the propagation speed of the trailing and leading edges is strongly influenced by
topological domain boundaries, specifically those between adjacent
contention domains. Together with the contention effect,
decay occurs even on a silent system. \Cref{fig:IdleWave}(b,c) shows
the basic phenomenology: As the idle wave
progresses through the contention domain (from core 5 to 23
as shown in the upper section of \Cref{fig:IdleWave}b and in the upper
half of \Cref{fig:IdleWave}c),
the trailing edge is gradually getting steeper as fewer cores
participate in the computation (cores 6--16 \circled1) because more bandwidth
becomes available per core. On the other hand, idle phases are
emanating from the end of the domain (core 23 \circled2) because the next
contention domain (core 24 and up, \circled3) is still executing with all cores
and is thus slower per core. These small idle waves propagate up and
interact with the main idle wave on cores 17--23 \circled4, effectively causing
its partial decay. The same occurs on the second contention domain at cores
39--47 \circled5 and, in reverse direction due to the wrapping around of the wave,
on the fourth domain on cores 72--80 \circled6\@.

Domain boundaries and the memory bottleneck are just as important for
the leading edge dynamics.  Within the domain where the one-off delay
was injected (cores 6--23 \circled7), the leading edge of the idle wave is not
straight but shows a slowdown as time progresses. This is because the
number of active cores on the contention domain increases as the wave
propagates, and the available memory bandwidth per core goes down as
soon as contention sets in. Eventually, the leading edge
hits the boundary to the next contention domain.  Right after this
point \circled8 the first domain is free of any delay and the bulk-synchronous
execution is restored there. The idle wave is now progressing entirely through
the second domain. Since the first domain is in synchronous state
and there is idleness on the second domain, more bandwidth is available
per core on the latter, so the computation phases are shorter
and there is waiting time (small red boxes in the timeline graph \circled9).
The key observation here is that the second domain does not go back
to a synchronized state; computation alternates with waiting times
on every process, but enough
cores are active concurrently to saturate the memory bandwidth.
Hence, the overall throughput of the second domain is the same
as on the first but the processes are out of sync. Finally,
after the preset number of time steps has passed, the computation
terminates. Processes that have collected less idle time because of the
decay of the injected idle wave (on the second to fourth domain)
finish early, as shown by the dashed blue line in \Cref{fig:IdleWave}b \circled{10}.
A distinctive wave-like pattern emerges across all
contention domains but the one in which the idle wave was injected.
We call this pattern a ``computational wavefront.''

\section{Induced computational wavefronts}\label{sec:nocomm}

In this section, we will further analyze the generating mechanisms of
computational wavefronts with memory-bound MPI code that emerge from singular one-off delays. We
restrict ourselves to the case of negligible communication
overhead. \emph{Spontaneous} wavefront formation and significant
communication overhead are linked and will be covered in \Cref{sec:sigcomm}\@.

\input{figures/SaturationCurves}

\input{figures/CompWaveActiveProcesses}

\input{figures/CompWaveShapeAndSlope}

\input{figures/CompWaveStability}

\subsection{Wavefront amplitude vs.\ processes per contention domain}\label{sec:bwsat}

A computational wavefront is a stable structure that can be visualized
by marking the wallclock time of a specific time step on each MPI
process in a bulk-synchronous iterative application. In a fully
synchronized state, the pattern is a straight line perpendicular to
the time axis.  Desynchronization causes wave-like patterns like the
one shown in \Cref{fig:IdleWave}b.  We have shown above that the
memory-bound nature of the code is crucial for desynchronization, so
we start with a series of experiments with progressively more severe
memory bottlenecks. \Cref{fig:ComputationalWave} shows six timelines
of memory-bound MPI programs on the \UseVerb{Emmy} system (parameters
as in \Cref{fig:IdleWave}) after injecting a one-off delay.  The
difference among the six cases is the number of MPI processes per
contention domain (socket). In the scalable regime (up to $N=3$ cores
per socket) the idle wave causes no visible computational wave. As
soon as the bandwidth bottleneck becomes relevant, i.e., when using
more cores leads to less bandwidth available per core, (here at
$N\gtrsim 4$), the damping effect on the idle wave sets in although it
is weak at first (\Cref{fig:ComputationalWave}c,d).  Our experiments
show, however, that even in this regime a stable computational wave
persists, albeit with a low amplitude \circled1. At strong saturation ($N\gtrsim 7$)
the fully developed wave is clearly visible. 
In all cases, the desynchronization prevails
even after the idle wave has died out, and even on contention domains
that were never traversed by it (cores 20--29 in
\Cref{fig:ComputationalWave}f \circled2)\@. Note also that
the socket on which the idle wave was originally injected is
still synchronized. 

This shows that
strong computational wave patterns require
a strong memory bandwidth saturation. Note that wave 
patterns will also form without initial one-off idle injection,
but this is a very slow process so we provoked it by ``kicking''
the system. This ``kick'' will not be required  when there is
significant communication overhead. See \Cref{sec:sigcomm} for
details.

\subsection{Saturation point and wavefront amplitude}\label{sec:saturation}

There is still the question  whether the
saturation point, i.e., how many processes are needed to attain maximum
memory bandwidth, plays any role.  Our benchmark platforms exhibit
different characteristics in this respect, as shown in
\Cref{fig:SaturationCurves}a: The Broadwell CPUs on
\UseVerb{Meggie} have the convenient property that the saturated
memory bandwidth depends only weakly on the clock speed, so we set the
core frequency to a constant 1.2\,\GHZ\ or activated ``Turbo Mode.''
The latter led to clock frequency varying from 3.0\,\GHZ\ (1 core) to
2.4\,\GHZ\ (full socket) along the scaling curve.  On
\UseVerb{SuperMUCNG} with its 24 cores per contention domain and fixed
2.3\,\GHZ\ clock speed, we employed a modified variant of the
Sch\"onauer vector triad that has a higher computational cost
(\verb.A(:)=B(:)+cos(C(:)/D(:)).)  in order to increase
$N_\mathrm{sc}$ from about 14 to 20 cores. As a side effect,
the saturation point becomes more sharply defined.
On \UseVerb{Emmy}, using nontemporal (NT) stores for
the STREAM triad the single-core bandwidth is about a factor
of two lower than with standard stores, shifting the saturation
point further out.

In \Cref{fig:SaturationCurves}b--f these variants are tested
for their reaction to injected idle waves when using all cores on
the contention domain.
The data shows
that the more data hungry the serial code (i.e., the earlier the
saturation point), the stronger the damping. This
was expected from the analysis in \Cref{fig:IdleWave}\@. In addition,
an early saturation point causes a large amplitude of the generated
computational wavefront (compare \Cref{fig:SaturationCurves}c and d,
and \Cref{fig:SaturationCurves}e and f)\@. Thus,
the saturation point impacts
the amplitude of the computational wavefront. Since the wavefront is
defined by a constant time step ID across processes, a large wave
amplitude indicates a larger inter-process skew, i.e., stronger
desynchronization, which causes longer waiting times within MPI
calls despite negligible communication volume. Since the computational wave
survives even long after the idle wave has died out, it is impossible
for these waiting times to cause reduced memory bandwidth utilization
(else the still-synchronized contention domain would eventually
catch up)\@. It thus seems that there are is always a sufficient
number of computing processes within the computational wave to still
reach bandwidth saturation.  \Cref{fig:CompWaveActiveProcesses}a
shows the average number of computing processes within the fully developed wave
for the three cases in Figures~\ref{fig:ComputationalWave}(f),
\ref{fig:SaturationCurves}(e), and \ref{fig:SaturationCurves}(f)\@.
Comparing with \Cref{fig:SaturationCurves}a it is evident that
this number is very close to the bandwidth saturation point (at
7, 13, and 20 cores, respectively)\@. Hence, the computational
wave settles at an amplitude that allows for just enough active
processes to saturate the memory bandwidth, but not more. The inevitable
waiting times caused by desynchronization are perfectly overlapped
with user code execution.

\subsection{Influence of communication patterns and injection length}

In \Cref{fig:CompWaveShapeAndSlope} we 
investigate how the shape and slope of an induced computational wave
depends on the communication pattern (distance of point-to-point
communication) and topology (open vs.\ periodic boundary
conditions)\@. In \Cref{fig:CompWaveShapeAndSlope}a we injected
a short idle period into a code with open boundary conditions
and next-neighbor bidirectional communication. The corresponding
idle wave in negative rank direction dies at rank 0, as expected~\cite{AfzalHW19}\@.
The idle wave in the positive rank direction hardly travels beyond
the next contention domain (node 0, socket 1) before dying out,
but a computational wave prevails on that domain
in the form of a single ramp with a slope of $-40\,\text{rank/s}$\@.
Doubling the duration of the injection (\Cref{fig:CompWaveShapeAndSlope})
leads to a longer idle wave that extends across three sockets in positive
rank direction, and so does the generated computational wave.
Its slope, however, is the same as in the previous case. The strength of
the initial idle wave thus has no influence on the local slope of the
computational wave.

The experiment in \Cref{fig:CompWaveShapeAndSlope}c shows the
influence of communication patterns. Each MPI process communicates
with its next neighbor in positive rank direction and with its next-
and next-to-next neighbors in negative rank direction; moreover, the
topology was changed to periodic boundary conditions. The idle wave
can now roll over the system boundary and eventually annihilates
itself. Its leading edges are governed by the known mechanisms
investigated in~\cite{AfzalHW19}: The idle wave in negative rank
direction is three times faster than the one in positive rank
direction. The resulting computational wave is continuous (because of
the boundary condition) and shows two distinct slopes, which are
different from the slopes of the idle wave but have the same 3:1
ratio. Hence, the slopes involved in the computational wave are influenced
by the same communication parameters that govern the slopes of the
idle wave, but the absolute slopes are different, which translates into different
wave amplitudes. As shown in the previous sections,
they depend on the saturation characteristics of the memory-bound code.

\section{Spontaneous computational wavefronts}\label{sec:sigcomm}

With negligible communication overhead, the desynchronization phenomena
described above can be observed when provoked by a rather strong
one-off delay injection. They only occur spontaneously, i.e.,
via the normal system noise, over very long time scales.
Moreover, although the available memory bandwidth
per process is larger in the desynchronized state, the runtime of the whole program,
i.e., the wall-clock time required for the slowest process to
reach the last time step, cannot be reduced in this scenario since
no significant overhead is overlapped with code execution.

In this section we show how computational wavefronts and desynchronization
can occur \emph{spontaneously} via natural system noise if there
is significant communication overhead, which paves the way towards
automatic communication-computation overlap.

\subsection{Pure MPI}

In \Cref{fig:PureMPI} we show four phases of a timeline of a
memory-bound STREAM triad code on four sockets of
\UseVerb{Emmy} and an initial communication overhead of $\approx
25\%$\@. One MPI process was run per core with bidirectional
next-neighbor communication, open boundary conditions, and
a message size of 5\,\MB. The synchronized
state from the beginning soon dissolves. After 100 time steps
(second phase), local wavefronts have emerged, but no global state
is reached yet. Within 500
time steps (third phase), a global wave has formed, and it persists
till the end of the program (50\,000 time steps)\@. Interestingly,
although the wavelength and amplitude of the computational wave
are rather constant, the pattern can shift across the MPI ranks over time:
After 26\,s of walltime the slowest process
is on socket 1, while after 2000\,s it is on socket 0\@. The cause
for such shifts are small perturbations (natural noise), whose
close investigation is left for future work. 

The overall MPI time per process goes up when entering the wave state
as expected because waiting time is added on top of actual
communication time. However, since communication can be overlapped
with execution, performance increases. In our particular case,
the total average (computation plus communication/waiting) time per
iteration goes down from $30\,\MS+10\,\MS=40\,\MS$ to $20\MS+17.5\MS=37.5\,\MS$,
i.e., by about 6\%.

\input{figures/PureMPI.tex}

\subsection{Latency- vs.\ bandwidth-dominated overhead}

There are two potential benefits from desynchronization: Better memory
bandwidth utilization by the application code and better network
interface utilization (not discussed here)\@.  These advantages are
partially offset by the memory bandwidth drawn by MPI communication of
large messages. For example, in the experiment in \Cref{fig:PureMPI},
each message had a size of 5\,\MB. In particular the intra-node
point-to-point communication can aggregate to a significant data
volume (at least 20\,\MB\ per process and time step in this case, and
probably more depending on the implementation of intra-node MPI),
reducing the bandwidth available to the application code. This is why
the theoretical speedup of 25\% could not be obtained.

\subsection{Threaded MPI processes}\label{sec:Hybrid}

All phenomenology discussed so far can also be observed with hybrid
MPI+OpenMP codes that communicate only outside OpenMP-parallel
regions. However, spanning an MPI process across several cores on a
contention domain is equivalent to reducing the number of cores, which
makes for weaker saturation characteristics as discussed in
\Crefrange{sec:bwsat}{sec:saturation}\@.  If the number of threads per
process is large enough to show linear bandwidth scaling across
processes, spontaneous wave formation and automatic overlap will
not occur.

\Cref{fig:Hybrid}a shows an injected idle wave on \UseVerb{Emmy} with
40 MPI processes by ten threads each, running the STREAM triad with
one process per contention domain, bidirectional next-neighbor
communication (negligible overhead), and periodic
boundary conditions. Since there is no bandwidth contention among
processes, the situation is very similar to \Cref{fig:IdleWave} and
\Cref{fig:ComputationalWave}a:
The idle wave is hardly damped and eventually cancels itself,
with no discernible desynchronization prevailing and no computational wave
following up. The memory-bound nature of the code is of no significance.

The property of scalable code to automatically eliminate idle waves by
the interaction of the trailing edge with system noise (which was
thoroughly studied in~\cite{AfzalHW19}) leads to the important
and general conclusion that spontaneous desynchronization does not occur
in this case.
\Cref{fig:Hybrid}b shows
a timeline of four MPI processes with ten threads each, running on
four contention domains of \UseVerb{Emmy}\@. System noise
causes a delay with subsequent desynchronization, which is quickly
dissolved and the system returns to the synchronized state.
One can argue that there is more to hybrid MPI+OpenMP
programming than optimizing communication overhead; ``full hybrid''
codes, in which one MPI process spans a full contention
domain (or more), do not profit from desynchronization and automatic
overlap since they enforce a lock-step across threads.

We have to add that we have deliberately chosen a simplified scenario
where the number and size of point-to-point messages sent between
processes does not depend on the number of threads per process.  In
real-world codes, many effects complicate matters, especially when
comparing pure MPI with MPI+OpenMP code for the same problem since the
number of messages and (probably) the communication volume
changes~\cite{Rabenseifner:2009}. A
thorough study of this problem area is left for future work.

\input{figures/Hybrid.tex}

\section{Chebychev Filter Diagonalization}\label{sec:chebFD}

\ac{ChebFD}~\cite{pieper2016high}
is a popular technique for calculating inner or extremal
eigenvalues of large sparse matrices. It is based on subspace
projection via polynomial filters constructed from Chebyshev
polynomials. ChebFD is applied in many problems in quantum physics and
chemistry, such as the study of topological materials (e.g., graphene)
or electronic structure calculations based on density functional
theory. Although the basic algorithm is just a sequence of simple
vector operations and sparse matrix-vector multiplications (SpMV), it
is amenable to loop fusion and blocking
optimizations~\cite{kreutzer2018chebyshev}\@.
\begin{algorithm}[tb]
\caption{Application of the ChebFD polynomial filter to block vectors.}
\begin{algorithmic}[1]  
    \State $\vec U$\ :=\ $\vec u_1,\dots,\vec u_{n_s}$\Comment{define block vector}
    \State $\vec W$\ :=\ $\vec w_1,\dots,\vec w_{n_s}$\Comment{define block vector}
    \State $\vec X$\ :=\ $\vec x_1,\dots,\vec x_{n_s}$\Comment{define block vector}
    \State $\mathrlap{\vec U}\phantom{\vec W} \gets (\alpha\vec H + \beta \mathbbm
        1) \vec X$ \Comment{\texttt{spmmv()}}
    \State $\vec W \gets 2 (\alpha\vec H + \beta \mathbbm 1) \vec U - \vec X$
    \Comment{\texttt{spmmv()}}
    \State $\mathrlap{\vec X}\phantom{\vec W} \gets g_0 c_0 \vec X + g_1 c_1
    \vec U + g_2 c_2 \vec W$ \Comment{\texttt{baxpy()+bscal()}}
    \For{ $p=3$\ to\ $n_p$ }
    \State $\mbox{swap} ( \vec W ,  \vec U )$
    \State $\vec W \gets 2 (\alpha\vec H + \beta \mathbbm 1) \vec U -  \vec W$ 
    \State $\mathrlap{\vec \eta_p}\phantom{\vec X} \gets \langle \vec W, \vec U \rangle$ 
    \State $\mathrlap{\vec \mu_p}\phantom{\vec X} \gets \langle \vec U, \vec U \rangle$ 
    \hfill{\smash{\raisebox{\dimexpr.5\normalbaselineskip+.5\jot}{$\left.\begin{array}{@{}c@{}}\\{}\\{}\\{}\\{}\end{array}\right\}\triangleright\text{\textsc{chebfd\_op($H,\vec U, \vec W,
                                    \vec X$)}}$}}}
    \State $\mathrlap{\vec X}\phantom{\vec W} \gets \vec X + g_p c_p \vec W$ 
    \EndFor
\end{algorithmic}
\label{alg:chebfd_vanilla}
\end{algorithm}

We use the scalable ChebFD implementation,
specifically the application of the polynomial filter to a block of vectors.
The compute kernels and implementation alternatives are available with the open-source
GHOST\footnote{\url{https://bitbucket.org/essex/ghost}} library for download.
This is the dominant part of the full ChebFD algorithm, which still requires
an orthogonalization procedure that is omitted here without loss of generality.
The code supports MPI+OpenMP parallelism.

\Cref{alg:chebfd_vanilla} shows the basic algorithm. $H$ is the Hamiltonian
matrix describing the physical system, while $U$, $W$, and $X$ are blocks of
$n_s$ vectors, with $n_s$ being the dimension of the search space.
The loop from line 7 to 13 iterates up to the polynomial degree $n_p$,
which determines how selective the polynomial filter will be. The goal
of the algorithm is the computation of the polynomial filter coefficients
$\{\eta_p\}$ and $\{\mu_p\}$, which requires global scalar products
(lines 10 and 11)\@. However, since these coefficients are not needed
until after the end of the calculation, the global reduction can be
postponed and leads to an algorithm without synchronization points
or global operations. The body of the $p$ loop can then be fused completely
into a single kernel \textsc{CHEBFD\_OP} for better cache reuse.
Our implementation uses a blocking optimization
that processes blocks of $n_b$ vectors at a time for improved cache
efficiency. Details can be found in~\cite{kreutzer2018chebyshev}\@.

Our specific application case is a topological insulator of size
$128\times 64\times 64$ with periodic boundary conditions. This leads
to a Hamiltonian of dimension $2^{21}$ and $2.71\times 10^6$ nonzeros.
The full working set is about 6.7\,\GB\ (double precision matrix,
4-byte indices, plus all block vectors) when using $n_s=128$ search
vectors and a polynomial filter degree $n_p=500$, which are realistic
values. The optimistic code balance assuming perfect
cache reuse on the block vectors is~\cite{kreutzer2018chebyshev}
\bq B_c =
\frac{260/n_b+80}{146}\frac{\byte}{\flop}\cma
\eq
which is well beyond the machine balance of all current CPUs even for
large $n_b$, rendering the code memory bound according to a naive Roof{}line
model. In reality, the $n_b=32$ case is already close to core bound
since intra-cache data transfers begin to limit the performance of the
code on some platforms, such as \UseVerb{Emmy}~\cite{Kreutzer:2015}:
\Cref{fig:ChebFD}a shows performance vs.\ cores per socket for $n_b=2$
and $n_b=32$, and indeed the latter cannot fully saturate the bandwidth
and achieves only 41\,\GFS\ out of the bandwidth-bound Roof{}line limit of
66\,\GFS. \Cref{fig:ChebFD}b shows strong scaling from 2--10 nodes
for both cases with 1 to 10 threads per MPI process. At $n_b=2$, fewer
threads have a clear advantage while the situation is reversed at $n_b=32$\@.
The more saturating code ($n_b=2$) has ample opportunity for desynchronization
without threading (which is shown in the timeline comparison
in~\Cref{fig:ChebFD}c)\@. The more scalable code ($n_b=32$) shows no spontaneous
desynchronization without threading, and the fully hybrid code
can benefit from the reduced number of MPI messages.
\input{figures/ChebFD}

\section{Related Work}\label{sec:relatedwork}

There is very little research on idle wave propagation and pattern
formation in parallel code, especially in the context of memory-bound
programs. Hence, none of the existing prior work addressed spontaneous
pattern formation and desynchronization.
Markidis et al.~\cite{markidis2015idle} used a simulator to study idle
waves in MPI programs and their propagation for the first time. They
did not consider the socket-level character of the code, though, and
assumed a linear wave equation to govern the propagation of the waves.
Afzal et al.~\cite{AfzalHW19,AfzalEuroMPI19Poster} have investigated
the dynamics of idle waves in pure MPI programs with core-bound
code. Our work builds on theirs and significantly extends it towards
memory-bound code and spontaneous pattern formation.
Gamell et al.~\cite{Gamell:2015} noted the emergence of idle waves in
the context of failure recovery and failure masking of stencil codes,
but the speed of propagation, the memory-bound characteristics of the
application, and the corresponding damping mechanisms were not
studied.  B\"ohme et al.~\cite{Boehme:2016} presented a tool-based
approach to attribute propagating wait states in MPI programs to their
original sources, helping to identify and correct the root
issues. Global properties of such waves like damping and velocity, or
the interaction with memory-bound code, were ignored, however.

\section{Conclusion and outlook}\label{sec:conclusion}

We have shown how the memory-bound nature of load-balanced MPI
programs without explicit synchronization or global operations and
homogeneous communication characteristics is directly linked to the
damping of idle waves and to desynchronization effects. The key
concept is the \emph{computational wave}, a stable pattern marked by
different processes reaching a given step in a simulation at different
times. Such patterns can be provoked by injected idle waves or emerge
spontaneously; rapid, spontaneous pattern formation caused by natural
system noise is only possible with significant communication
overhead. In a desynchronized state, the time spent in MPI routines is
larger but the available memory bandwidth per process is higher. There
is evidence that a computational wave settles in a state where the
number of processes concurrently running user code within a contention
domain is very close to the bandwidth saturation
point. Desynchronization also enables automatic hiding of
communication overhead, which can in some cases improve the
performance of a program. This overlap may not be perfect due to the
MPI communication requiring part of the memory bandwidth. Using a
single, multi-threaded MPI process per contention domain effectively
recovers a scalable code (from the viewpoint of memory
bandwidth)\@. In this case, automatic overlap does not occur and
(induced or spontaneous) delays die out automatically.  While above
results were obtained using simple microbenchmark codes on four
different cluster systems, we have demonstrated the emergence of
computational waves and the detrimental effect of full hybrid mode
using a Chebyshev Filter application from quantum physics.

Although we could uncover some of the mechanisms behind
the computational wave formation in a qualitative way, a detailed
quantitative understanding of these effects is still out of reach.
For example, there is no actual mathematical \emph{proof} of stability
for computational waves, or a proof of instability for the
bulk-synchronous state.  We have also just scratched the surface of
how threaded MPI processes, natural system noise, and network contention
change the underlying mechanisms. For example, even with core-bound code
there may be a strong bottleneck on the network interface if parallel program
is strongly communication bound, and desynchronization does occur in this
case as well. It
will be helpful to have a controlled, noise-free experimental
environment in which all relevant aspects, from code characteristics
to communication parameters and contention effects, can be influenced
at will. To this end, we are working on a high-performance simulation
tool that goes far beyond existing simulators such as
LogGOPSim~\cite{hoefler-loggopsim}\@.

\ifblind
\else
\section*{Acknowledgments}
This work was supported by KONWIHR, the Bavarian Competence Network
for Scientific High Performance Computing in Bavaria, under project name ``OMI4papps.''
We are indebted to LRZ Garching for granting CPU hours on SuperMUC-NG.
\fi


\bibliographystyle{splncs03}
\bibliography{references}

\end{document}

%% file: macros.tex
\usepackage{acro}
\newcommand{\acrodef}[2]{\DeclareAcronym{#1}{short={#1},long={#2}}}
\acrodef{AABB}{axis-aligned bounding box}
\acrodef{API}{application programming interface}
\acrodef{ASIC}{application specific integrated circuit}
\acrodef{AST}{abstract syntax tree}
\acrodef{BRAM}{block RAM}
\acrodef{CB}{compute-bound}
\acrodef{CER}{communication-to-execution ratio}
\acrodef{CG}{conjugate gradient}
\acrodef{ChebFD}{Chebyshev filter diagonalization}
\acrodef{CPI}{cycles per instruction}
\acrodef{CPU}{central processing unit}
\acrodef{CUDA}{compute unified device architecture}
\acrodef{CST}{concrete syntax tree}
\acrodef{DPM}{delay propagation mechanism}
\acrodef{DOF}{degree of freedoms}
\acrodef{DOM}{delay overlapping mechanism}
\acrodef{DPOM}{delay propagation and overlapping mechanisms}
\acrodef{DSL}{domain-specific language}
\acrodef{DVFS}{dynmic voltage frequency scaling}
\acrodef{FD}{finite difference}
\acrodef{FEM}{finite element method}
\acrodef{FFC}{FEniCS Form Compiler}
\acrodef{FFT}{Fast Fourier transform}
\acrodef{FIFO}{first in first out}
\acrodef{FLOPS}{floating point operations per second}
\acrodef{FPGA}{field-programmable gate array}
\acrodef{FV}{finite volume}
\acrodef{GMRES}{generalized minimal residual}
\acrodef{GPU}{graphics processor unit}
\acrodef{GS}{Gauss-Seidel}
\acrodef{GUI}{graphical user interface}
\acrodef{HDL}{hardware description language}
\acrodef{HHG}{hierarchical hybrid grid}
\acrodef{HLS}{high-level synthesis}
\acrodef{HPC}{high-performance computing}
\acrodef{IP}{intellectual property}
\acrodef{ITAC}{Intel trace analyzer and collector}
\acrodef{IR}{intermediate representation}
\acrodef{JIT}{just-in-time}
\acrodef{KPM}{Kernel Polynomial Method}
\acrodef{LFA}{local Fourier analysis}
\acrodef{LBM}{Lattice Boltzmann}
\acrodef{LoC}{lines of code}
\acrodef{LZR}{Leibniz Supercomputing Centre}
\acrodef{MB}{memory-bound}
\acrodef{MPI}{Message Passing Interface}
\acrodef{NDG}{nodal discontinuous Galerkin}
\acrodef{NDGTD}{nodal discontinuous Galerkin time domain}
\acrodef{NIC}{network interface controller}
\acrodef{OMP}{OpenMP}
\acrodef{OS}{operating system}
\acrodef{P2P}{peer-to-peer}
\acrodef{PDE}{partial differential equation}
\acrodef{PGAS}{partitioned global address space}
\acrodef{PPnR}{post place and route}
\acrodef{PPS}{processes per socket}
\acrodef{QDR}{quad data rate}
\acrodef{RAM}{random access memory}\acuse{RAM}
\acrodef{RBGS}{red-black Gauss-Seidel}
\acrodef{RDMA}{remote direct memory access}
\acrodef{RHS}{right-hand side}
\acrodef{RRZE}{Regional Computer Center Erlangen} 
\acrodef{RTL}{register transfer level}
\acrodef{SHM}{shared memory}
\acrodef{SPIR}{standard portable intermediate representation}
\acrodef{SPL}{software product lines}
\acrodef{SIMD}{single instruction, multiple data}
\acrodef{SMP}{symmetric multiprocessing}
\acrodef{SMT}{simultaneous multithreading}
\acrodef{STL}{Standard Template Library}
\acrodef{TLB}{translation lookaside buffer}
\acrodef{TPDL}{target platform description language}
\acrodef{WF}{wavefront}
\acrodef{XML}{eXtensible Markup Language}

\usepackage{amsmath}
\usepackage{amsfonts}

\usepackage{relsize}
\usepackage{xspace}
\newcommand{\CPP}{C\nolinebreak[4]\hspace{-.05em}\raisebox{.23ex}{\relsize{-1}{++}}}

\newif\iftitle
\titletrue

\newcommand{\bq}{\begin{equation}}
\newcommand{\eq}{\end{equation}}
\newcommand{\bytes}{\mbox{B}}
\newcommand{\byte}{\mbox{byte}}
\newcommand{\second}{\mbox{s}}
\newcommand{\MS}{\mbox{ms}}

\newcommand{\seconds}{\mbox{s}}
\newcommand{\flop}{\mbox{flop}}

\newcommand{\bit}{\mbox{bit}}

\newcommand{\GFS}{\mbox{G\flop/\second}}

\newcommand{\GHZ}{\mbox{GHz}}

\newcommand{\GB}{\mbox{GB}}

\newcommand{\MB}{\mbox{MB}}

\newcommand{\eos}{~.}
\newcommand{\cma}{~,}

\newcommand{\Rl}{Roof{}line\xspace}

\newcommand*\circled[1]{\tikz[baseline=(char.base)]{
		\node[shape=circle,draw,inner sep=0.3pt] (char) {#1};}}
\newcommand*\circledfilled[1]{\tikz[baseline=(char.base)]{
		\node[shape=circle,fill=white,draw,inner sep=0.3pt] (char) {#1};}}

\definecolor{myblue}{RGB}{37,165,203}
\definecolor{myred}{RGB}{175,32,67}

%% file: todo.tex
\usepackage{todonotes}
\usepackage{setspace}
\makeatletter
\renewcommand{\todo}[2][]{\@todo[caption={#2},#1]{\begin{spacing}{0.5}\fontfamily{phv}\fontseries{mc}\selectfont{#2\vspace{-1em}}\end{spacing}}}
\makeatother

\colorlet{ghcolor}{ProcessBlue}
\colorlet{aycolor}{RubineRed}

\usepackage[binary-units=true]{siunitx}

%% file: figures/IdleWave.tex
\begin{figure}[tb]
	\centering
	\begin{adjustbox}{width=0.9\textwidth}
	\begin{subfigure}[t]{.265\textwidth}
		\begin{tikzpicture}
		\put(0,-2.2) {\includegraphics[width=0.92\textwidth,height=0.18 \textheight]{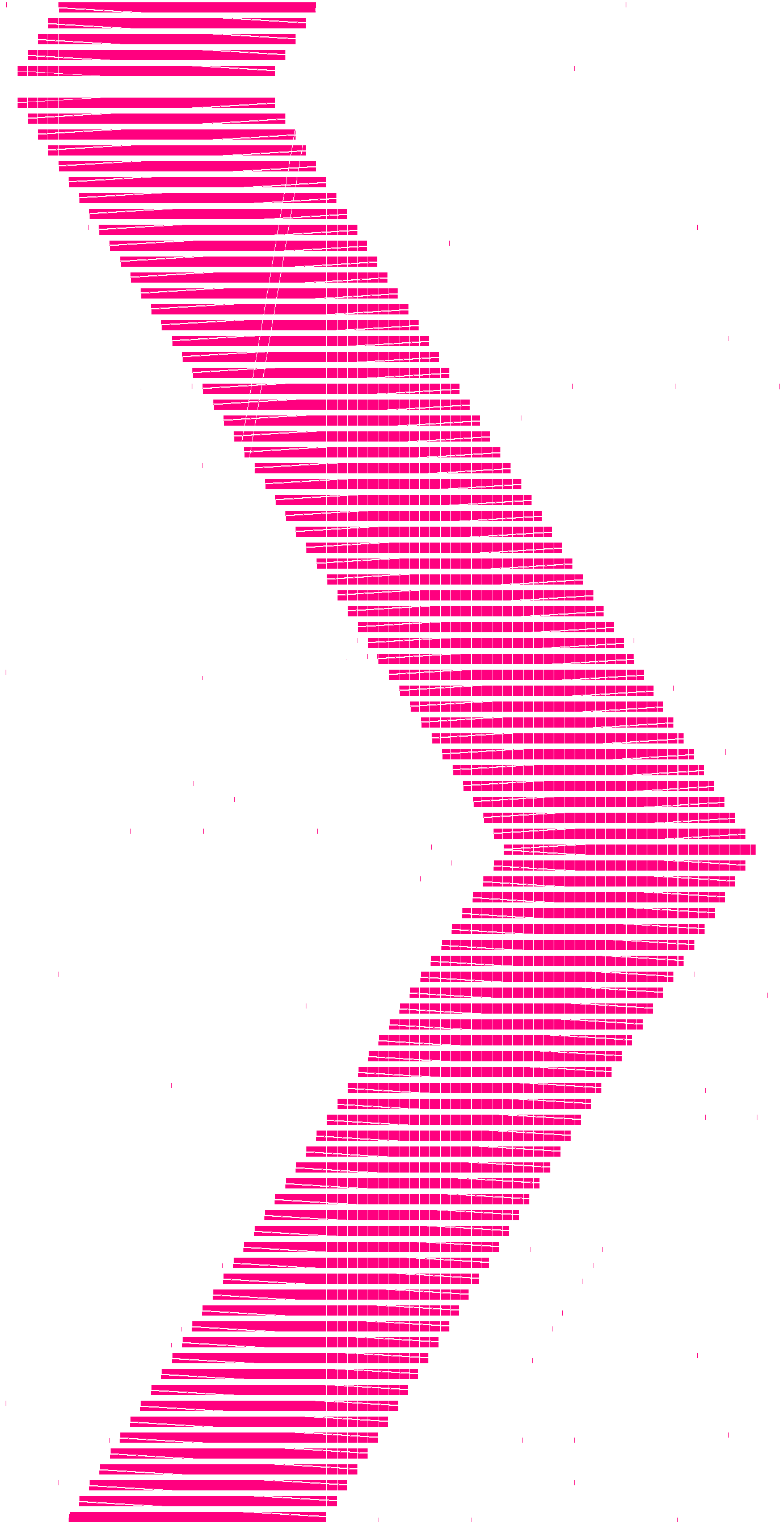}}
		\put(0,-3) {\begin{axis}[
		width=1.4\textwidth,height=0.264\textheight,
		ylabel = {Rank},
		y label style={at={(-0.1,0.5)}},
		xlabel = {Walltime},
		x label style={at={(0.5,0.08)}},
		x tick label style={font=\scriptsize},
		y tick label style={font=\scriptsize}, 
		xmin=1, xmax=50,
		xmajorticks=false,
		ymin=0, ymax=95,
		ytick={12,24,36,48,60,72,84,96}, 
		yticklabels={\textbf{83},,\textbf{59},,\textbf{34},,\textbf{11}},
		]
		\end{axis}}
		\node [font=\small] at (1.2,-1.1){(a) Scalable workload}; 
		\draw [semithick, dotted] (-0.5,2.53 ) -- (2.95,2.53 );
		\draw [semithick, dotted] (-0.5,1.65) -- (2.95,1.65);
		\draw [semithick, dotted] (-0.5,0.7825) -- (2.95,0.7825); 
		\node [font=\small] at (0.54,3.205){\tikz \fill [blue] (0,0) rectangle (0.96,0.02);};
		\node at (2.14,2.6) {\rotatebox{-41}{\tiny \textbf{leading edge}}};
		\node at (0.8,2.08) {\rotatebox{-41}{\tiny \textbf{trailing edge}}};
    		\draw [thick,densely dashed,blue] (2.9,-0.1) -- (2.9,3.4);
		\node at (-0.85,3.1) {\tiny \textbf{Socket 0}}; 
		\node at (-0.85,2.9) {\tiny \textbf{Node 0}}; 
		\node at (-0.85,2.2) {\tiny \textbf{Socket 1}}; 
		\node at (-0.85,2) {\tiny \textbf{Node 0}}; 		
		\node at (-0.85,1.4) {\tiny \textbf{Socket 0}}; 
		\node at (-0.85,1.2) {\tiny \textbf{Node 1}}; 		
		\node at (-0.85,0.5) {\tiny \textbf{Socket 1}}; 
		\node at (-0.85,0.3) {\tiny \textbf{Node 1}};  		
		\end{tikzpicture}
	\end{subfigure}
	\hspace{4.4em}	
	\begin{subfigure}[t]{.265\textwidth}
		\begin{tikzpicture}
		\put(-0.5,-2.2) {\includegraphics[width=0.92\textwidth,height=0.18 \textheight]{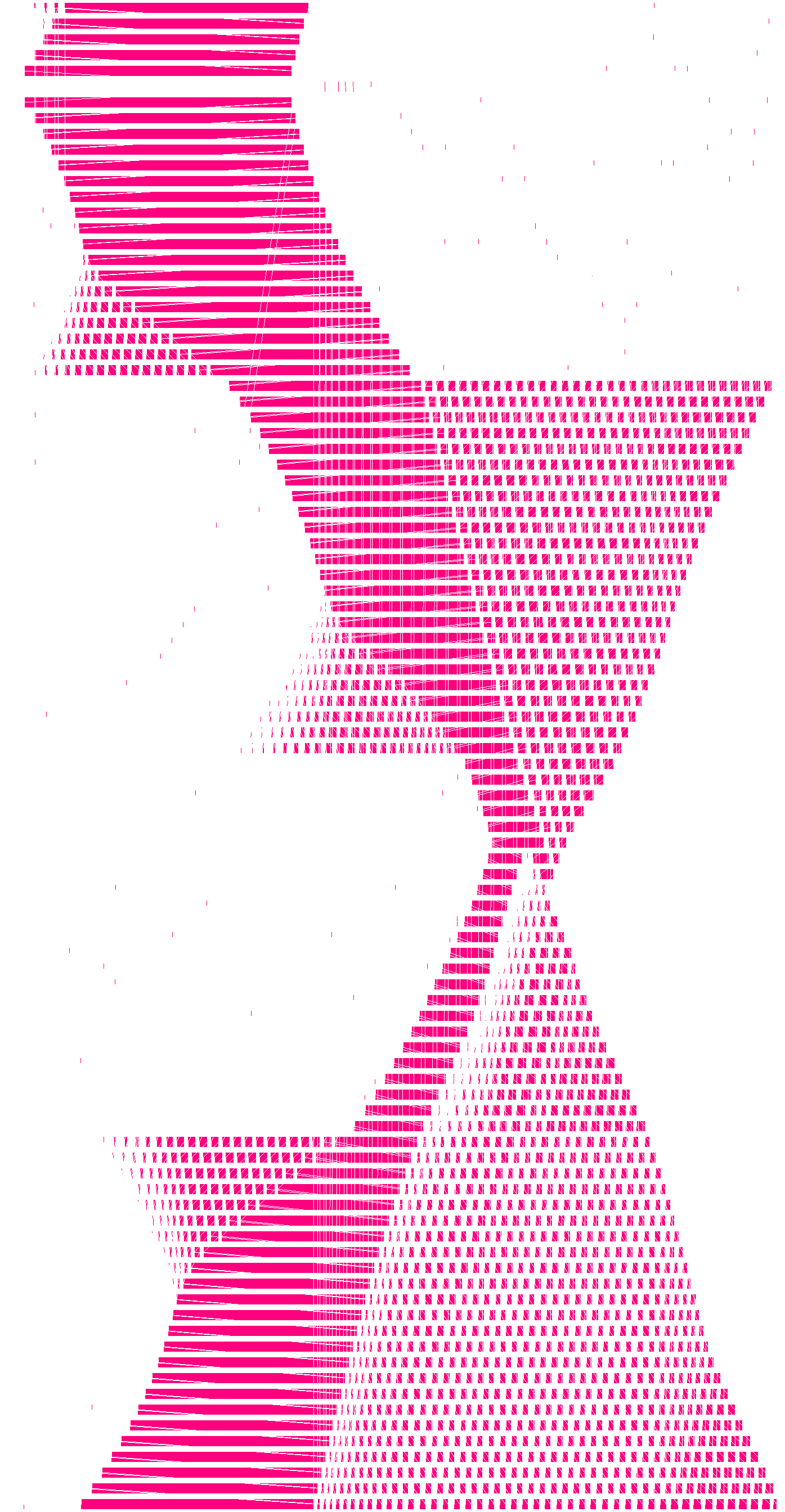}}
		\put(0,-3) {\begin{axis}[
		width=1.4\textwidth,height=0.264\textheight,
		y label style={at={(0.17,0.5)}},
		xlabel = {Walltime},
		x label style={at={(0.5,0.08)}},
		x tick label style={font=\scriptsize},
		y tick label style={font=\scriptsize}, 
		xmin=1, xmax=50,
		xmajorticks=false,
		ymin=0, ymax=95,
		ytick={12,24,36,48,60,72,84,96}, 
		yticklabels={\textbf{83},,\textbf{59},,\textbf{34},,\textbf{11}},
		]
		\coordinate (pt) at (125,860);
		\coordinate (c2) at (6,588);
		\draw (pt) rectangle (c2);
		\end{axis}}
		\node [font=\small] at (1.2,-1.1){(b) Saturating workload}; 
		\draw [semithick, dotted] (-0.5,2.53 ) -- (2.95,2.53 );
		\draw [semithick, dotted] (-0.5,1.65) -- (2.95,1.65);
		\draw [semithick, dotted] (-0.5,0.7825) -- (2.95,0.7825); 
		\node [font=\small] at (0.58,3.2){\tikz \fill [blue] (0,0) rectangle (0.99,0.02);};	
		\draw [thick,densely dashed,blue] (2.93,2.5) -- (2.93,3.4);
		\draw [thick,densely dashed,blue] (2.1,1.34) -- (2.93,2.5);
		\draw [thick,densely dashed,blue] (2.1,1.34) -- (2.93,-0.1);
		\node [font=\small] at (0.9,1.9){\circledfilled5};
		\node [font=\small] at (0.3,0.5){\circledfilled6};
		\node [font=\small] at (1.45,3){\circledfilled7};
		\node [font=\small] at (2,2.9){\circledfilled8};
		\draw [->,line width=0.2mm] (1.8,2.84) -- (1.6,2.5) ;
		\node [font=\small] at (2.2,2){\circledfilled9};   
		\node [font=\small] at (2.6,1.3){\circledfilled{10}};                     
		\end{tikzpicture}
	\end{subfigure}
	\hspace{1.1em}	
	\begin{subfigure}[t]{.265\textwidth}
		\begin{tikzpicture}
			\put(2,-2.5) {\includegraphics[width=0.86\textwidth,height=0.18 \textheight]{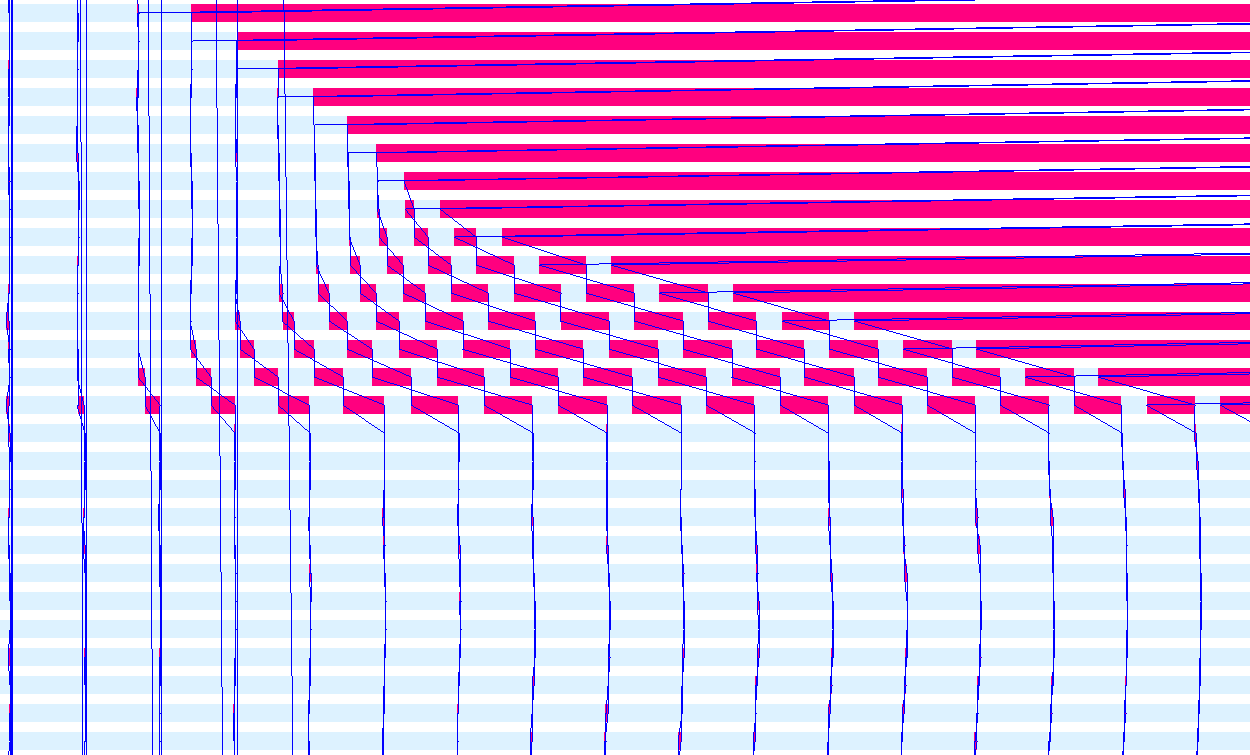}}
			\put(0,-3) {\begin{axis}[
			width=1.4\textwidth,height=0.264\textheight,
			y label style={at={(0.17,0.5)}},
			xlabel = {Walltime},
			x label style={at={(0.5,0.08)}},
			x tick label style={font=\scriptsize},
			y tick label style={font=\scriptsize}, 
			xmin=1, xmax=15,
			ymin=8, ymax=36, 
			y dir=reverse,
			ytick={9,12,15,18,21,24,27,30,33,35}, 
			xmajorticks=false,
			]
		\end{axis}}
		\node [font=\small] at (1.2,-1.1){(c) Zoom in of (b)}; 
		\draw [semithick, dotted] (-0.5,1.45) -- (2.95,1.45);
		\node [font=\small] at (0.3,2.9){\circledfilled1};
		\node [font=\small] at (1.8,1.8){\circledfilled4};   
		\node [font=\small] at (-0.4,1.6){\circledfilled2}; 
		\draw [->,line width=0.2mm] (-0.2,1.6) -- (0.3,1.5);
		\node [font=\small] at (1.5,0.7){\circledfilled3}; 
		\end{tikzpicture}
	\end{subfigure}	
	\end{adjustbox}
	\caption{Timelines of idle waves through MPI code (one process per core)
          with different
          workload characteristics, negligible communication overhead, and
          bidirectional next-neighbor communication in a closed ring topology
          on \protect\UseVerb{SuperMUCNG}. The $y$ axis is the MPI rank
          and the $x$ axis is wall-clock time. Red indicates waiting time
          (within the MPI library) while white or light blue
          denote user code ($50$ iterations). The injected delay of about $25$
          execution phases is shown in dark blue.
          (a) Core-bound code with execution phase of 10\,\MS,
          (b) memory-bound STREAM triad code (overall data transfer volume of
          4.8\,\GB, evenly distributed across all cores for a computation phase of
          11.5\,\MS), (c) zoom-in of marked area in (b).}
	\label{fig:IdleWave}
\end{figure}
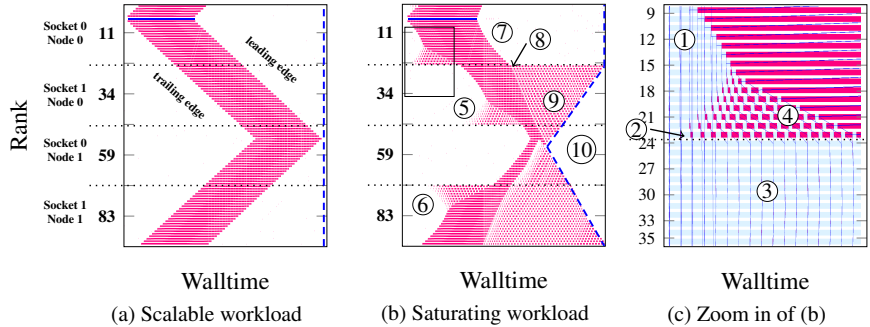

%% file: figures/ComputationalWave.tex
\begin{figure}[tb]
	\centering
	\begin{adjustbox}{width=0.835\textwidth}
	\begin{minipage}{\textwidth}
			\begin{subfigure}[t]{0.24\textwidth} 
			\begin{tikzpicture}
			\put(-27.9,-2.2) {\includegraphics[width=0.92\textwidth,height=0.178 \textheight]{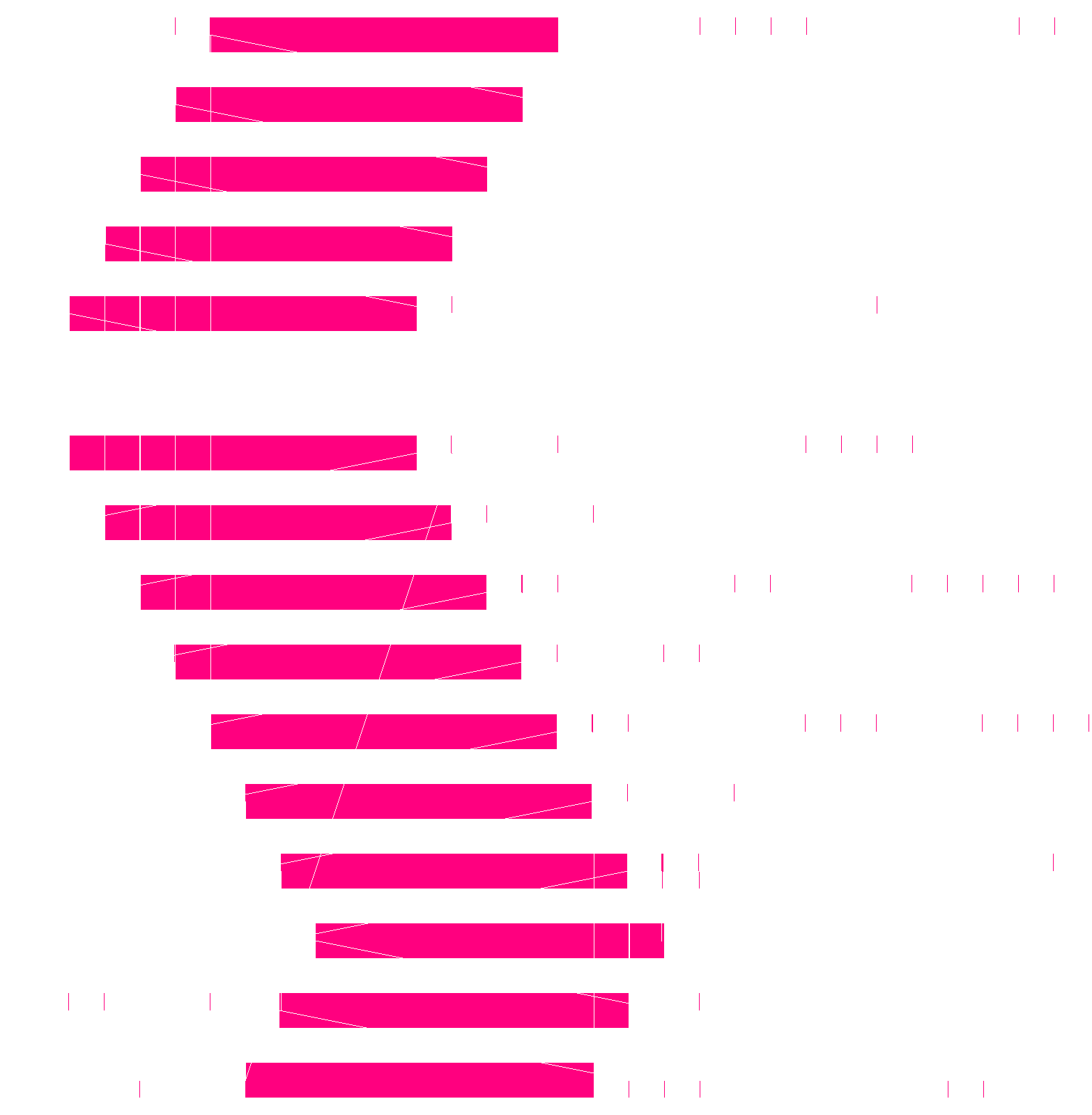}}
			\put(-25,-3) {\begin{axis}[
				width=1.4\textwidth,height=0.26\textheight,
				ylabel = {Rank},
				y label style={at={(0.17,0.5)}},
				xlabel = {Time step},
				x label style={at={(0.5,0.08)}},
				x tick label style={font=\scriptsize},
				y tick label style={font=\scriptsize}, 
				xmin=1, xmax=35,
				ymin=0, ymax=15,
				xtick={35},
				ytick={0,3,7,11,15}, 
				xticklabels={\textbf{20}},
				yticklabels={\textbf{15},\textbf{12},\textbf{8},\textbf{4},\textbf{0}},
				]
				\end{axis}}
			\node [font=\footnotesize] at (0.3,-1.1){(b) Cores $N  = 2$}; 
			\draw [semithick, dotted] (-0.85,2.974 ) -- (1.7,2.974 ); 
			\draw [semithick, dotted] (-0.85,2.544 ) -- (1.7,2.544 ); 
			\draw [semithick, dotted] (-0.85,2.114 ) -- (1.7,2.114 ); 
			\draw [semithick, dotted] (-0.85,1.664 ) -- (1.7,1.664 ); 
			\draw [semithick, dotted] (-0.85,1.224 ) -- (1.7,1.224 ); 
			\draw [semithick, dotted] (-0.85,0.784 ) -- (1.7,0.784 ); 
			\draw [semithick, dotted] (-0.85,0.344 ) -- (1.7,0.344 ); 
			\draw [thick,densely dashed,blue] (1.58,-0.1) -- (1.58,3.3);
			\node [font=\small] at (-0.4,2.2){\tikz \fill [blue] (0,0.10) rectangle (0.832,0.22);};
			\end{tikzpicture}
		\end{subfigure}
		\hspace{-2.8em}
		\begin{subfigure}[t]{0.24\textwidth} 
			\begin{tikzpicture}
			\put(-3,-2.2) {\includegraphics[width=0.88\textwidth,height=0.178 \textheight]{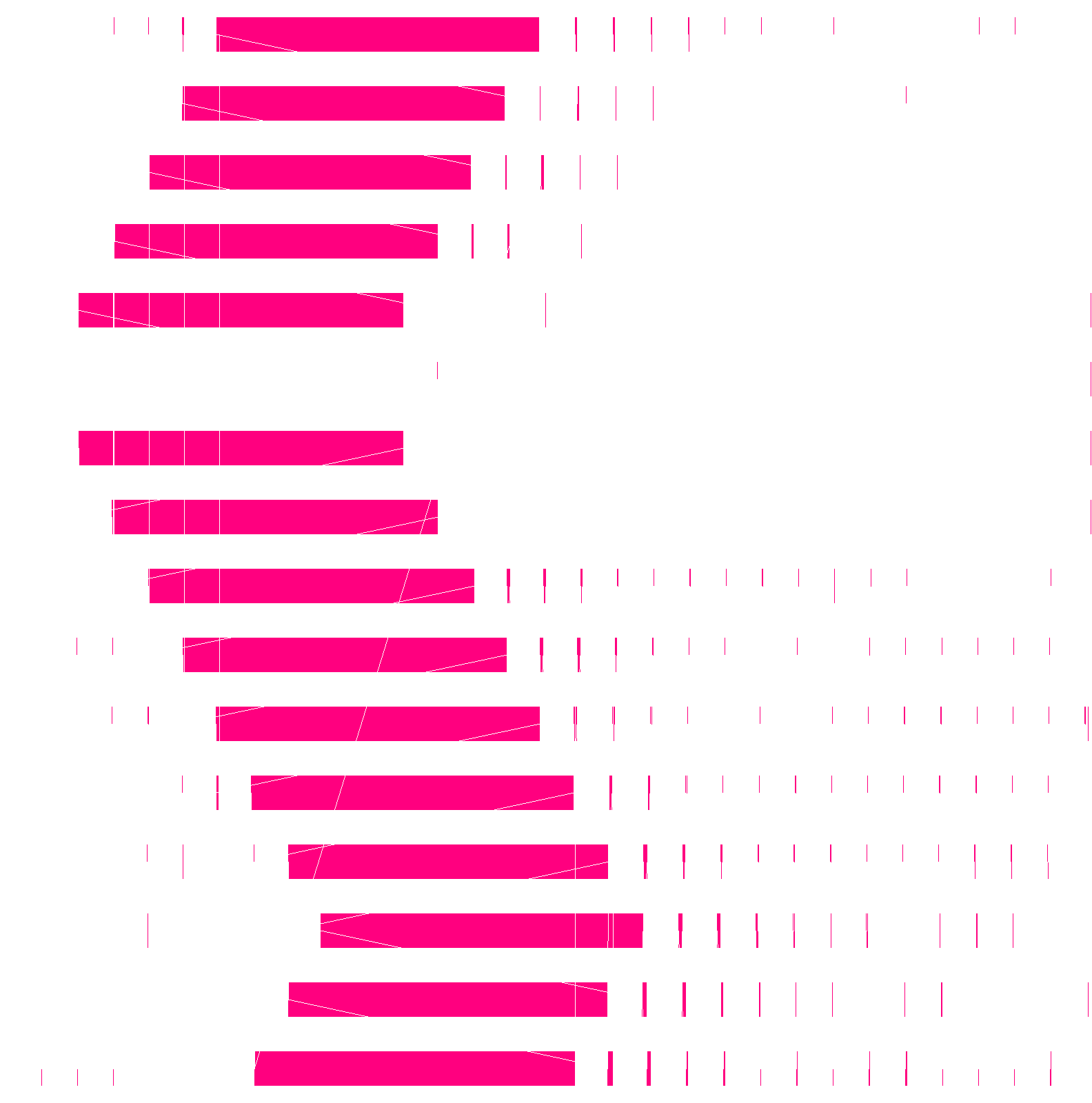}}
			\put(0,-3) {\begin{axis}[
				width=1.4\textwidth,height=0.26\textheight,
				y label style={at={(-0.1,0.5)}},
				xlabel = {Time step},
				x label style={at={(0.5,0.08)}},
				x tick label style={font=\scriptsize},
				y tick label style={font=\scriptsize}, 
				xmin=1, xmax=35,
				ymin=0, ymax=15,
				xtick={35},
				ytick={0,3,7,11,15}, 
				xticklabels={\textbf{20}},
				yticklabels={\textbf{15},\textbf{12},\textbf{8},\textbf{4},\textbf{0}},
				]
				\end{axis}}
			\node [font=\small] at (1.2,-1.1){(c) Cores $N = 4$}; 
			\draw [semithick, dotted] (0,2.53 ) -- (2.5,2.53 );
			\draw [semithick, dotted] (0,1.65) -- (2.5,1.65);
			\draw [semithick, dotted] (0,0.7825) -- (2.5,0.7825); 
			\node [font=\small] at (0.46,2.19){\tikz \fill [blue] (0,0.11) rectangle (0.76,0.215);};
                        \node [font=\small] at (2.2,1.0){\circledfilled1};
			\draw [thick,densely dashed,blue] (2.48,1.6) -- (2.48,3.3);
			\draw [thick,densely dashed,blue] (2.46,0.8) -- (2.48,1.6);
			\draw [thick,densely dashed,blue] (2.46,0.8) -- (2.48,-0.1);
			\end{tikzpicture}
		\end{subfigure}
		\hspace{1em}
		\begin{subfigure}[t]{0.24\textwidth} 
			\begin{tikzpicture}
			\put(-3,-2.2) {\includegraphics[width=0.88\textwidth,height=0.178 \textheight]{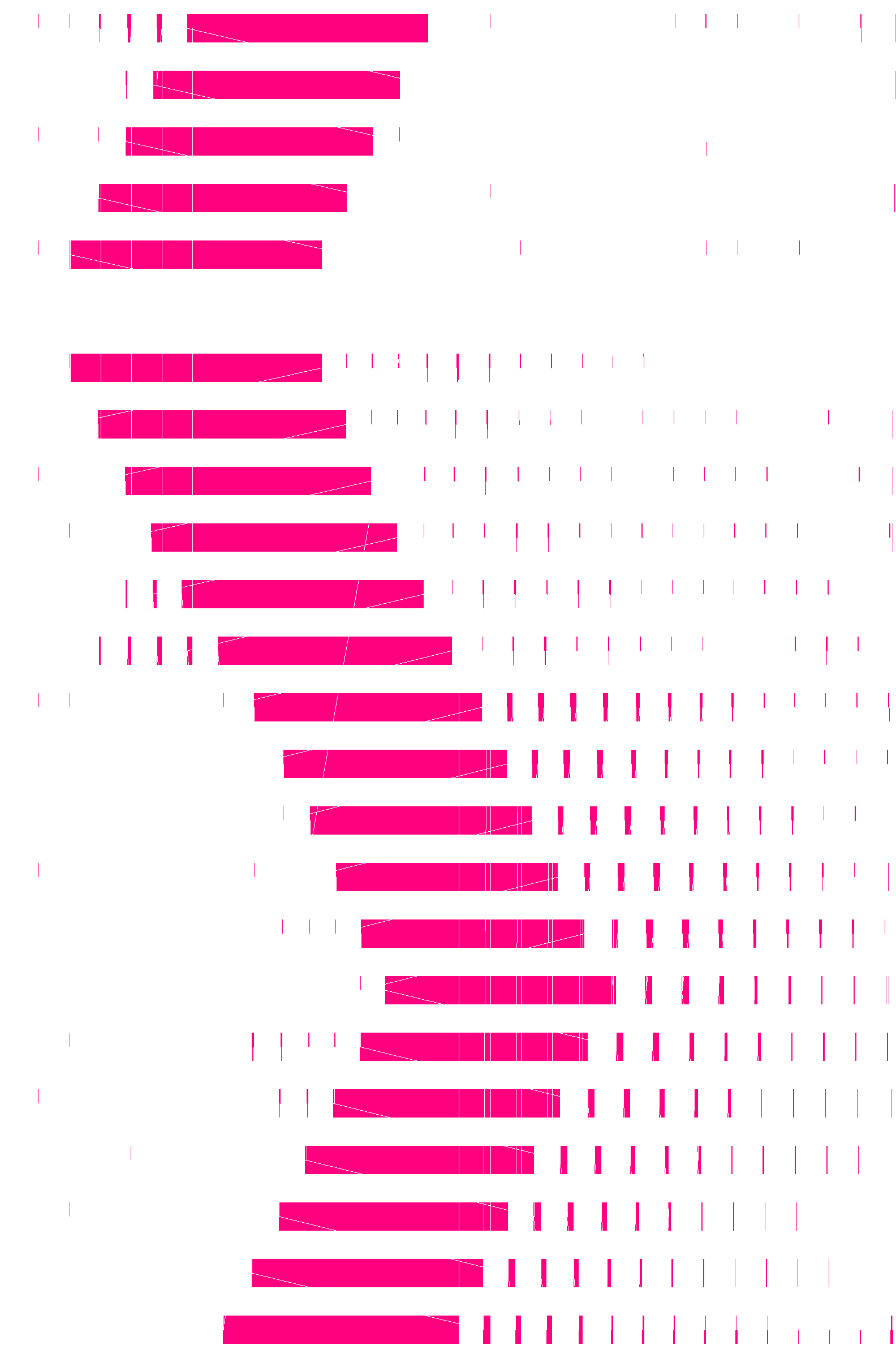}}
			\put(0,-3) {\begin{axis}[
				width=1.4\textwidth,height=0.26\textheight,
				y label style={at={(-0.1,0.5)}},
				xlabel = {Time step},
				x label style={at={(0.5,0.08)}},
				x tick label style={font=\scriptsize},
				y tick label style={font=\scriptsize}, 
				xmin=1, xmax=35,
				ymin=0, ymax=23,
				xtick={35},
				ytick={0,3,7,11,15,19,23}, 
				xticklabels={\textbf{20}},
				yticklabels={\textbf{23},\textbf{20},\textbf{16},\textbf{12},\textbf{8},\textbf{4},\textbf{0}},
				]
				\end{axis}}
			\node [font=\small] at (1.2,-1.1){(d) Cores $N = 6$}; 
			\draw [semithick, dotted] (0,2.53 ) -- (2.5,2.53 );
			\draw [semithick, dotted] (0,1.65) -- (2.5,1.65);
			\draw [semithick, dotted] (0,0.7825) -- (2.5,0.7825); 
			\node [font=\small] at (0.46,2.56){\tikz \fill [blue] (0,0.11) rectangle (0.73,0.19);};
            \node [font=\small] at (2.2,1.0){\circledfilled1};
            \draw [thick,densely dashed,blue] (2.48,1.6) -- (2.48,3.3);
            \draw [thick,densely dashed,blue] (2.45,0.8) -- (2.48,1.6);
            \draw [thick,densely dashed,blue] (2.45,0.8) -- (2.48,-0.1);
			\end{tikzpicture}
		\end{subfigure}
		\hspace{1em}
		\begin{subfigure}[t]{0.24\textwidth} 
			\begin{tikzpicture}
			\put(-3,-2.2) {\includegraphics[width=0.88\textwidth,height=0.178 \textheight]{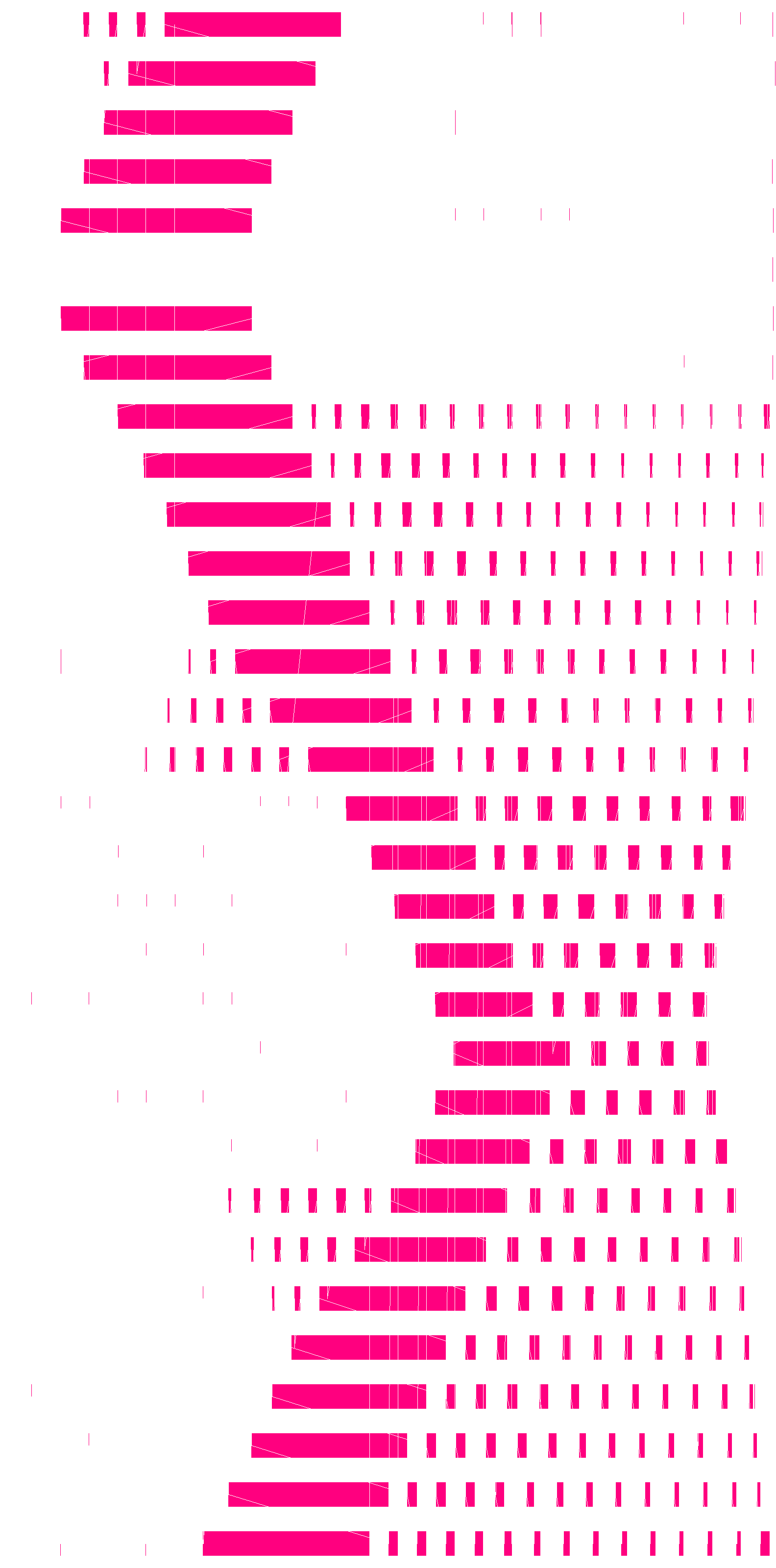}}
			\put(0,-3) {\begin{axis}[
				width=1.4\textwidth,height=0.26\textheight,
				y label style={at={(-0.1,0.5)}},
				xlabel = {Time step},
				x label style={at={(0.5,0.08)}},
				x tick label style={font=\scriptsize},
				y tick label style={font=\scriptsize}, 
				xmin=1, xmax=35,
				ymin=0, ymax=31,
				xtick={35},
				ytick={0,7,15,23,31}, 
				xticklabels={\textbf{20}},
				yticklabels={\textbf{31},\textbf{24},\textbf{16},\textbf{8},\textbf{0}},
				]
				\end{axis}}
			\node [font=\small] at (1.2,-1.1){(e) Cores $N = 8$}; 
			\draw [semithick, dotted] (0,2.53 ) -- (2.5,2.53 );
			\draw [semithick, dotted] (0,1.65) -- (2.5,1.65);
			\draw [semithick, dotted] (0,0.7825) -- (2.5,0.7825); 
			\node [font=\small] at (0.41,2.77){\tikz \fill [blue] (0,0.105) rectangle (0.635,0.17);};
    		\draw [thick,densely dashed,blue] (2.48,2.5) -- (2.48,3.3);
			\draw [thick,densely dashed,blue] (2.38,1.6) -- (2.48,2.5);
			\draw [thick,densely dashed,blue] (2.24,1.2) -- (2.38,1.6);
			\draw [thick,densely dashed,blue] (2.24,1.2) -- (2.34,0.8);
			\draw [thick,densely dashed,blue] (2.34,0.8) -- (2.48,-0.1);
			\end{tikzpicture}
		\end{subfigure}
	\end{minipage}
	\end{adjustbox}
	\begin{adjustbox}{width=0.835\textwidth}
	\begin{minipage}{\textwidth}
			\begin{subfigure}[t]{0.24\textwidth} 
			\begin{tikzpicture}
			\put(-117.9,-2.2) {\includegraphics[width=0.88\textwidth,height=0.178 \textheight]{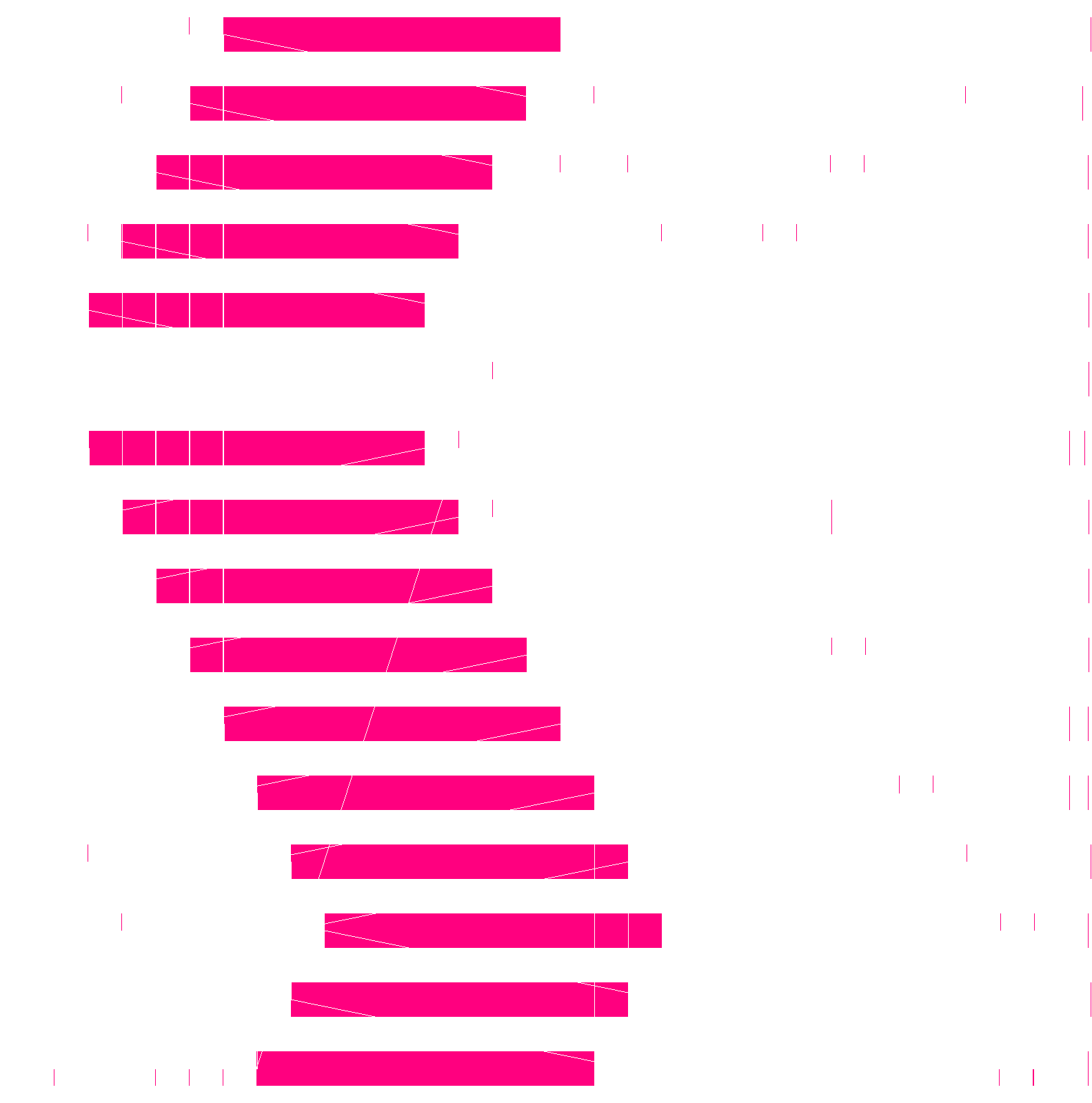}}
			\put(-115,-3) {\begin{axis}[
				width=1.4\textwidth,height=0.26\textheight,
				ylabel = {Rank},
				y label style={at={(-0.1,0.5)}},
				xlabel = {Time step},
				x label style={at={(0.5,0.08)}},
				x tick label style={font=\scriptsize},
				y tick label style={font=\scriptsize}, 
				xmin=1, xmax=35,
				ymin=0, ymax=15,
				xtick={35},
				ytick={0,4,8,12,15}, 
				xticklabels={\textbf{20}},
				yticklabels={\textbf{15},\textbf{12},\textbf{8},\textbf{4},\textbf{0}},
				]
				\end{axis}}
			\node [font=\small] at (-3.2,-1.1){(a) Cores $N = 1$}; 
			\draw [semithick, dotted] (-5.2,3.184 ) -- (-1.5,3.184 ); 
			\draw [semithick, dotted] (-5.2,2.974 ) -- (-1.5,2.974 ); 
			\draw [semithick, dotted] (-5.2,2.754 ) -- (-1.85,2.754 );
			\draw [semithick, dotted] (-5.2,2.544 ) -- (-1.85,2.544 ); 
			\draw [semithick, dotted] (-5.2,2.334 ) -- (-1.85,2.334 );
			\draw [semithick, dotted] (-5.2,2.114 ) -- (-1.85,2.114 ); 
			\draw [semithick, dotted] (-5.2,1.884 ) -- (-1.85,1.884 );
			\draw [semithick, dotted] (-5.2,1.664 ) -- (-1.85,1.664 ); 
			\draw [semithick, dotted] (-5.2,1.444 ) -- (-1.85,1.444 );
			\draw [semithick, dotted] (-5.2,1.224 ) -- (-1.85,1.224 ); 
			\draw [semithick, dotted] (-5.2,1.004 ) -- (-1.85,1.004 );		
			\draw [semithick, dotted] (-5.2,0.784 ) -- (-1.5,0.784 ); 
			\draw [semithick, dotted] (-5.2,0.564 ) -- (-1.5,0.564 );
			\draw [semithick, dotted] (-5.2,0.344 ) -- (-1.5,0.344 ); 
			\draw [semithick, dotted] (-5.2,0.124 ) -- (-1.5,0.124 );
    		\draw [thick,densely dashed,blue] (-1.58,-0.1) -- (-1.58,3.3);
			\node [font=\small] at (-3.55,2.2){\tikz \fill [blue] (0,0.11) rectangle (0.8,0.21);};
			\node at (-3.8,1.1) {\rotatebox{-71}{\tiny \textbf{trailing edge}}};
			\node at (-2.8,1.4) {\rotatebox{-71}{\tiny \textbf{leading edge}}};
			\node at (-1.75,1.8) {\rotatebox{90}{\tiny \textbf{\textcolor{blue}{Computational wavefront}}}};
			\node at (-4.85,3.32) {\tiny \textbf{Node 0}}; 
			\node at (-4.85,3.10) {\tiny \textbf{Node 1}}; 
			\node at (-4.85,2.87) {\tiny \textbf{Node 2}}; 
			\node at (-4.85,2.66) {\tiny \textbf{Node 3}}; 		
			\node at (-4.85,2.44) {\tiny \textbf{Node 4}}; 
			\node at (-4.85,2.22) {\tiny \textbf{Node 5}}; 		
			\node at (-4.85,2.00) {\tiny \textbf{Node 6}}; 
			\node at (-4.85,1.77) {\tiny \textbf{Node 7}}; 
			\node at (-4.85,1.56) {\tiny \textbf{Node 8}}; 
			\node at (-4.85,1.34) {\tiny \textbf{Node 9}}; 
			\node at (-4.85,1.12) {\tiny \textbf{Node 10}}; 
			\node at (-4.85,0.90) {\tiny \textbf{Node 11}}; 		
			\node at (-4.85,0.67) {\tiny \textbf{Node 12}}; 
			\node at (-4.85,0.46) {\tiny \textbf{Node 13}}; 		
			\node at (-4.85,0.24) {\tiny \textbf{Node 14}}; 
			\node at (-4.85,0.02) {\tiny \textbf{Node 15}};	
			\end{tikzpicture}
		\end{subfigure}
		\hspace{2.8em}
		\begin{subfigure}[t]{0.48\textwidth} 
			\begin{tikzpicture}
			\pgfplotstableread{figures/Fig2_Compwave_Emmy_/withNT1.txt}\emmydata;
			\pgfplotstableread{figures/Fig2_Compwave_Emmy_/withNT2.txt}\emmydataA;
		\put(0,7.5) {	\begin{axis}[trim axis left, trim axis right, scale only axis,
			width=0.73\textwidth,height=0.178\textheight,
			xlabel = {Cores per socket, $N$},
			ylabel = {Memory bandwidth [\si{\giga \byte / \second}]}, 
			ymin=0,
			xtick={0,1,2,3,4,5,6,7,8,9,10}, 
			xticklabels={\textbf{0},\textbf{1},\textbf{2},\textbf{3},\textbf{4},\textbf{5},\textbf{6},\textbf{7},\textbf{8},\textbf{9},\textbf{10}}, 
			y label style={at={(0.12,0.5)},font=\footnotesize},
			x label style={at={(0.5,0.02)},font=\footnotesize},
			x tick label style={font=\scriptsize},
			y tick label style={font=\scriptsize, font=\bfseries}, 
			legend style = {nodes={inner sep=0.04em}, font=\tiny, cells={align=left}, anchor=west, at={(0.45,0.6)}},
			legend columns = 1,
			]
			\addplot[only marks, mark=square*,mark size =2 pt, PineGreen, error bars/.cd, y dir=both, y explicit,]
			table
			[
			x expr=\thisrow{Cores}, 
			y error minus expr=\thisrow{Median}-\thisrow{Min},
			y error plus expr=\thisrow{Max}-\thisrow{Median},
			]{\emmydata};
			
			\addplot[only marks, mark=square*,mark size =2 pt, Bittersweet, error bars/.cd, y dir=both, y explicit,]
			table
			[
			x expr=\thisrow{Cores}, 
			y error minus expr=\thisrow{Median}-\thisrow{Min},
			y error plus expr=\thisrow{Max}-\thisrow{Median},
			]{\emmydataA};
			\end{axis}}
			\node at (0.7,0.55) {\tiny \textbf{(a)}};
			\node at (1.05,1.15) {\tiny \textbf{(b)}};
			\node at (1.8,2.1) {\tiny \textbf{(c)}};
			\node at (2.4,2.85) {\tiny \textbf{(d)}};
			\node at (3.1,3.1) {\tiny \textbf{(e)}};
			\node at (3.95,3.1) {\tiny \textbf{(f)}};
			\end{tikzpicture}
		\end{subfigure}
			\hspace{-3.1em}
			\begin{subfigure}[t]{0.24\textwidth} 
		\begin{tikzpicture}
		\put(-3,-2.2) {\includegraphics[width=0.88\textwidth,height=0.178 \textheight]{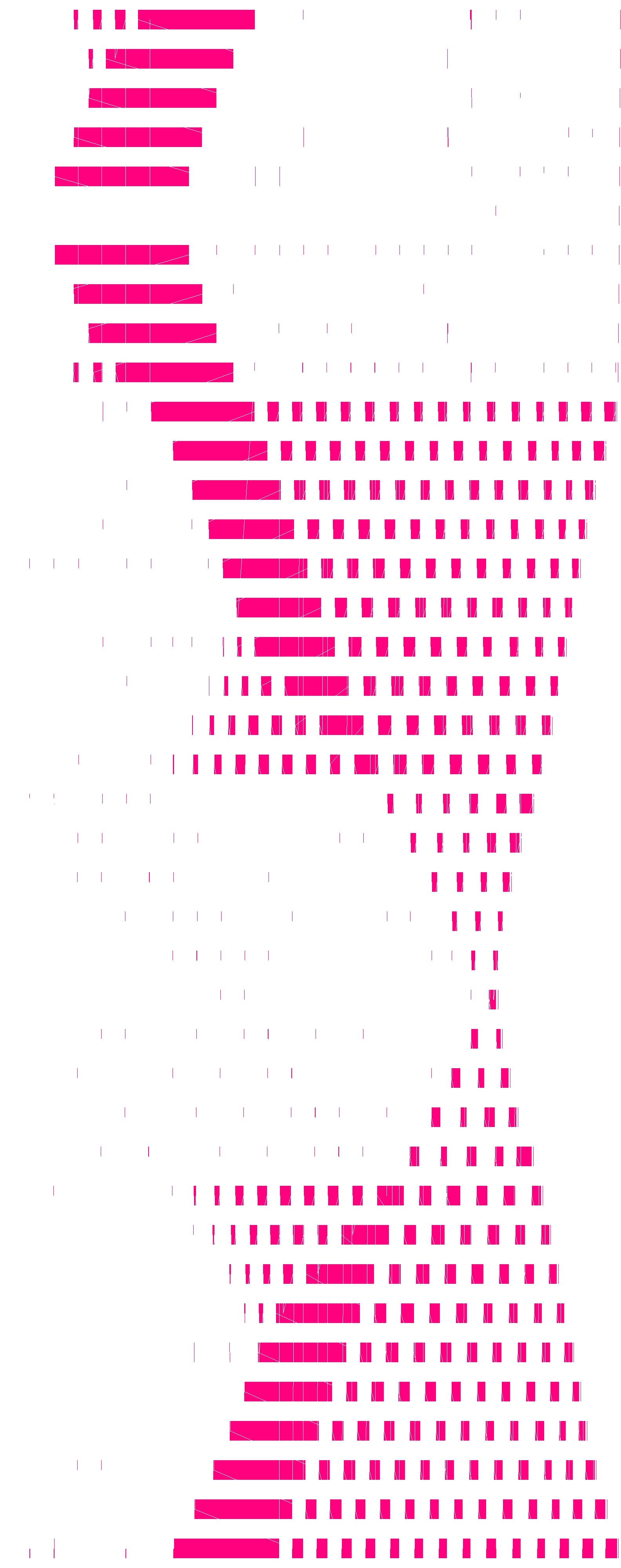}}
		\put(0,-3) {\begin{axis}[
			width=1.4\textwidth,height=0.26\textheight,
			y label style={at={(-0.1,0.5)}},
			xlabel = {Time step},
			x label style={at={(0.5,0.08)}},
			x tick label style={font=\scriptsize},
			y tick label style={font=\scriptsize}, 
			xmin=1, xmax=35,
			ymin=0, ymax=39,
			xtick={35},
			ytick={0,7,15,23,31,39}, 
			xticklabels={\textbf{20}},
			yticklabels={\textbf{39},\textbf{31},\textbf{24},\textbf{16},\textbf{8},\textbf{0}},
			]
			\end{axis}}
		\node [font=\small] at (1.2,-1.1){(f) Cores $N = 10$}; 
		\draw [semithick, dotted] (-1.1,2.53 ) -- (2.35,2.53 );
		\draw [semithick, dotted] (-1.1,1.65) -- (2.2,1.65);
		\draw [semithick, dotted] (-1.1,0.7825) -- (2.2,0.7825); 
    	\draw [thick,densely dashed,blue] (2.48,2.5) -- (2.48,3.3);
      	\draw [thick,densely dashed,blue] (1.98,1.2) -- (2.48,2.5);
      	\draw [thick,densely dashed,blue] (1.98,1.2) -- (2.48,-0.1);
		\node [font=\small] at (0.39,2.9){\tikz \fill [blue] (0,0.11) rectangle (0.555,0.165);};
		\node at (2.4,1.2) {\rotatebox{90}{\tiny \textbf{\textcolor{blue}{Computational wavefront}}}};
		\node at (-0.85,3.1) {\tiny \textbf{Socket 0}}; 
		\node at (-0.85,2.9) {\tiny \textbf{Node 0}}; 
		\node at (-0.85,2.2) {\tiny \textbf{Socket 1}}; 
		\node at (-0.85,2) {\tiny \textbf{Node 0}}; 		
		\node at (-0.85,1.4) {\tiny \textbf{Socket 0}}; 
		\node at (-0.85,1.2) {\tiny \textbf{Node 1}}; 		
		\node at (-0.85,0.5) {\tiny \textbf{Socket 1}}; 
		\node at (-0.85,0.3) {\tiny \textbf{Node 1}}; 	
                \node [font=\small] at (1.5,1.1){\circledfilled2};
		\end{tikzpicture}
	\end{subfigure}
		\end{minipage}
	\end{adjustbox}
	\caption{Idle wave-induced computational wavefront pattern
          formation with memory-bound (STREAM triad with nontemporal stores) workload on the
          \protect\UseVerb{Emmy} system with varying number of MPI
          processes per contention domain (socket). Overall triad data volume
          was 9.6\,\GB, other parameters as in
          \Cref{fig:IdleWave}\@. Middle panel:
          memory bandwidth versus number of cores for the STREAM triad
          benchmark on one socket.
          (a)--(f) Idle wave
          propagation with 1,\ldots,10 cores per contention domain over 20 time steps.
          Computational wavefronts are shown with blue dashed lines.}
	\label{fig:ComputationalWave}
\end{figure}
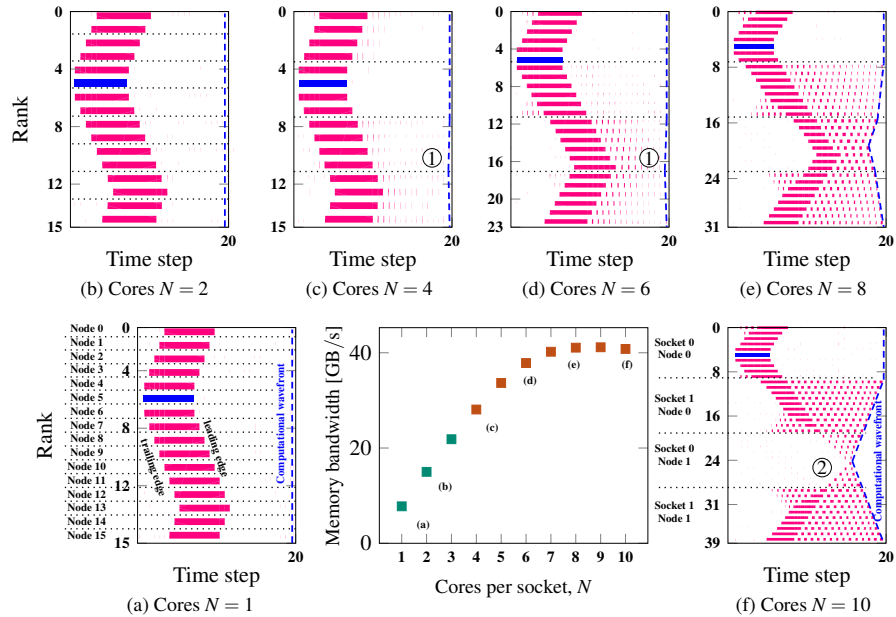

%% file: figures/SaturationCurves.tex
\begin{figure}[tb]
	\centering
	\begin{minipage}{\textwidth}
	\begin{adjustbox}{width=0.89\textwidth}
	\begin{subfigure}[t]{0.33\textwidth}
		\begin{tikzpicture}
		\pgfplotstableread{figures/Fig4_SaturationCurves_AllSystems_/Socket_BW/BW_HazelHen_STREAM_triad_2point5GHz.txt}\HLRSdata;
		\pgfplotstableread{figures/Fig4_SaturationCurves_AllSystems_/Socket_BW/BW_Meggie_STREAM_triad_turbo.txt}\meggiedata;
		\pgfplotstableread{figures/Fig4_SaturationCurves_AllSystems_/Socket_BW/BW_Meggie_STREAM_triad_1point2GHz.txt}\meggieLfreqdata;
		\pgfplotstableread{figures/Fig4_SaturationCurves_AllSystems_/Socket_BW/BW_SuperMUC-NG_DIV_triad_2point3GHz.txt}\supermucDTdata;
		\pgfplotstableread{figures/Fig4_SaturationCurves_AllSystems_/Socket_BW/BW_SuperMUC-NG_STREAM_triad_2point3GHz.txt}\supermucSTdata;
		\pgfplotstableread{figures/Fig2_Compwave_Emmy_/withNT.txt}\emmydataNTA;
		\pgfplotstableread{figures/Fig2_Compwave_Emmy_/BW_Emmy_STREAM_triad_2point2GHz.txt}\emmy;
		\begin{axis}[trim axis left, trim axis right, scale only axis,
		width=1.1\textwidth,height=0.178\textheight,
		xlabel = {Cores per socket, $N$},
		ylabel = {Memory bandwidth [\si{\giga \byte / \second}]}, 
		xmin=0,
		ymin=0,
		y label style={at={(0.05,0.5)},font=\footnotesize},
		x label style={font=\footnotesize},
		x tick label style={font=\scriptsize, font=\bfseries},
		y tick label style={font=\scriptsize, font=\bfseries},
		xmajorgrids,
		ymajorgrids,
		legend columns = 1, 
		legend style = {
			nodes={inner sep=0.04em}, font=\scriptsize, cells={align=left}, anchor=east, at={(1.01,1.225)},
			/tikz/column 1/.style={column sep=5pt,},
		},
		]
		
		\addlegendimage{only marks, mark=square*, PineGreen}
		\addlegendimage{only marks, mark=x, Sepia}
		\addlegendimage{only marks, mark=diamond*, Bittersweet}
		\addlegendimage{only marks, mark=*,blue}
		\addplot[ mark=square*,mark size =1.5 pt, PineGreen, error bars/.cd, y dir=both, y explicit,]
		table
		[
		x expr=\thisrow{Cores}, 
		y error minus expr=\thisrow{Median}-\thisrow{Min},
		y error plus expr=\thisrow{Max}-\thisrow{Median},
		]{\emmydataNTA};
		\addlegendentry{~STREAM triad, \protect\UseVerb{Emmy} @\SI{2.2}{\giga \Hz}, NT}
		
		\addplot[ mark=x,mark size =2.3 pt, Sepia, error bars/.cd, y dir=both, y explicit,]
		table
		[
		x expr=\thisrow{Cores}, 
		y error minus expr=\thisrow{Median}-\thisrow{Min},
		y error plus expr=\thisrow{Max}-\thisrow{Median},
		]{\emmy};
		\addlegendentry{~STREAM triad, \protect\UseVerb{Emmy} @\SI{2.2}{\giga \Hz}}
		
		\addplot[mark=diamond*, mark size =2 pt, Bittersweet, error bars/.cd, y dir=both, y explicit,]
		table
		[
		x expr=\thisrow{Cores}, 
		y error minus expr=\thisrow{Median}-\thisrow{Min},
		y error plus expr=\thisrow{Max}-\thisrow{Median},
		]{\meggiedata};
		\addlegendentry{~STREAM triad, \protect\UseVerb{Meggie} @turbo, NT}
		
		\addplot[mark=*,mark size =1.5 pt, blue, error bars/.cd, y dir=both, y explicit,]
		table
		[
		x expr=\thisrow{Cores}, 
		y error minus expr=\thisrow{Median}-\thisrow{Min},
		y error plus expr=\thisrow{Max}-\thisrow{Median},
		]{\meggieLfreqdata};
		\addlegendentry{~STREAM triad, \protect\UseVerb{Meggie} @\SI{1.2}{\giga \Hz}, NT}
		
 		\end{axis}
		\node [font=\small] at (6,-1.3){(a) Socket-level saturation characteristics};
		\end{tikzpicture}
	\end{subfigure}
	\hspace{5.5em}
	\begin{subfigure}[t]{0.33\textwidth}
		\begin{tikzpicture}
		\pgfplotstableread{figures/Fig4_SaturationCurves_AllSystems_/Socket_BW/BW_HazelHen_STREAM_triad_2point5GHz.txt}\HLRSdata;
		\pgfplotstableread{figures/Fig4_SaturationCurves_AllSystems_/Socket_BW/BW_Meggie_STREAM_triad_turbo.txt}\meggiedata;
		\pgfplotstableread{figures/Fig4_SaturationCurves_AllSystems_/Socket_BW/BW_Meggie_STREAM_triad_1point2GHz.txt}\meggieLfreqdata;
		\pgfplotstableread{figures/Fig4_SaturationCurves_AllSystems_/Socket_BW/BW_SuperMUC-NG_DIV_triad_2point3GHz.txt}\supermucDTdata;
		\pgfplotstableread{figures/Fig4_SaturationCurves_AllSystems_/Socket_BW/BW_SuperMUC-NG_STREAM_triad_2point3GHz.txt}\supermucSTdata;
		\put(0,14){\begin{axis}[trim axis left, trim axis right, scale only axis,
		width=1.1\textwidth,height=0.178\textheight,
		xlabel = {Cores per socket, $N$},
		xmin=0,
		ymin=0,
		y label style={at={(0.05,0.5)},font=\footnotesize},
		x label style={font=\footnotesize},
		x tick label style={font=\scriptsize, font=\bfseries},
		y tick label style={font=\scriptsize, font=\bfseries},
		xmajorgrids,
		ymajorgrids,
		legend columns = 1, 
		legend style = {
			nodes={inner sep=0.04em}, font=\scriptsize, cells={align=left}, anchor=east, at={(1.1,1.225)},
			/tikz/column 1/.style={column sep=5pt,},
		},
		]

		\addlegendimage{only marks, mark=diamond, Bittersweet}
		\addlegendimage{only marks, mark=star, blue}
		\addlegendimage{only marks, mark=triangle*,Sepia}
		\addplot[mark=diamond,mark size =1.5 pt, Bittersweet, error bars/.cd, y dir=both, y explicit,]
		table
		[
		x expr=\thisrow{Cores}, 
		y error minus expr=\thisrow{Median}-\thisrow{Min},
		y error plus expr=\thisrow{Max}-\thisrow{Median},
		]{\HLRSdata};
		\addlegendentry{~STREAM triad, Hazel Hen @\SI{2.5}{\giga \Hz}, NT}
		
		\addplot[mark=star, mark size =2.3 pt, blue, error bars/.cd, y dir=both, y explicit,]
		table
		[
		x expr=\thisrow{Cores}, 
		y error minus expr=\thisrow{Median}-\thisrow{Min},
		y error plus expr=\thisrow{Max}-\thisrow{Median},
		]{\supermucSTdata};
		\addlegendentry{~STREAM triad, SuperMUC-NG @\SI{2.3}{\giga \Hz}, NT}
		
		\addplot[ mark=triangle*, mark size =2 pt, Sepia, error bars/.cd, y dir=both, y explicit,]
		table
		[
		x expr=\thisrow{Cores}, 
		y error minus expr=\thisrow{Median}-\thisrow{Min},
		y error plus expr=\thisrow{Max}-\thisrow{Median},
		]{\supermucDTdata};
		\addlegendentry{~Slow Sch\"onauer triad, SuperMUC-NG @\SI{2.3}{\giga \Hz}, NT} 
		\end{axis}}
		\end{tikzpicture}
	\end{subfigure}
	\hspace{5em} 
	\begin{subfigure}[t]{.265\textwidth}
		\begin{tikzpicture}
		\pgfplotstableread{figures/Fig4_SaturationCurves_AllSystems_/timeRankViz_HazelHen_streamtriad.txt}\HLRSdata;
		\put(0,-3) {\begin{axis}[
			width=1.4\textwidth,height=0.26\textheight,
			ylabel = {Rank},
			y label style={at={(-0.1,0.5)}},
			xlabel = {Time step},
			x label style={at={(0.5,0.08)}},
			x tick label style={font=\scriptsize},
			y tick label style={font=\scriptsize}, 
			xmin=0, xmax=2.3e9,
			ymin=0, ymax=47,
			y dir=reverse,
			change x base,
			x SI prefix=giga,
			xtick={1e8,1.95e9,2.3e9},
			xticklabels={\textbf{1},\textbf{20},\textbf{24}},
			ytick={6,12,18,24,30,36,42,48}, 
			yticklabels={\textbf{5},,\textbf{17},,\textbf{29},,\textbf{41}},
			]
			\addplot[blue,only marks,mark=+,restrict x to domain=1e8:2.3e9] table[x expr=\thisrowno{2},y expr=\thisrowno{1}] {\HLRSdata};
			\end{axis}}
		\node [font=\small] at (1.2,-1.3){(b) \protect\UseVerb{HazelHen}@\SI{2.5}{\giga \Hz}};
		\draw [semithick, dotted] (-0.5,2.53 ) -- (2.95,2.53 );
		\draw [semithick, dotted] (-0.5,1.65) -- (2.95,1.65);
		\draw [semithick, dotted] (-0.5,0.7825) -- (2.95,0.7825);
		\node [font=\small] at (0.68,2.98){\tikz \fill [blue] (0,0) rectangle (1.12,0.03);};
		\node [font=\small] at (0.68,3.06){\tikz \fill [Magenta] (0,0) rectangle (1.12,0.03);};
		\node [font=\small] at (0.68,2.9){\tikz \fill [Magenta] (0,0) rectangle (1.12,0.03);};
		\node at (-0.85,3.1) {\tiny \textbf{Socket 0}}; 
		\node at (-0.85,2.9) {\tiny \textbf{Node 0}}; 
		\node at (-0.85,2.2) {\tiny \textbf{Socket 1}}; 
		\node at (-0.85,2) {\tiny \textbf{Node 0}}; 		
		\node at (-0.85,1.4) {\tiny \textbf{Socket 0}}; 
		\node at (-0.85,1.2) {\tiny \textbf{Node 1}}; 		
		\node at (-0.85,0.5) {\tiny \textbf{Socket 1}}; 
		\node at (-0.85,0.3) {\tiny \textbf{Node 1}}; 
		\end{tikzpicture}
	\end{subfigure}
	\end{adjustbox}
	\end{minipage}
	\begin{adjustbox}{width=0.8\textwidth}
		\begin{minipage}{\textwidth}
			\hspace{-1.5em}
			\begin{subfigure}[t]{0.24\textwidth}
				\begin{tikzpicture}	
					\put(-30,-2.2) {\includegraphics[width=0.84\textwidth,height=0.178 \textheight]{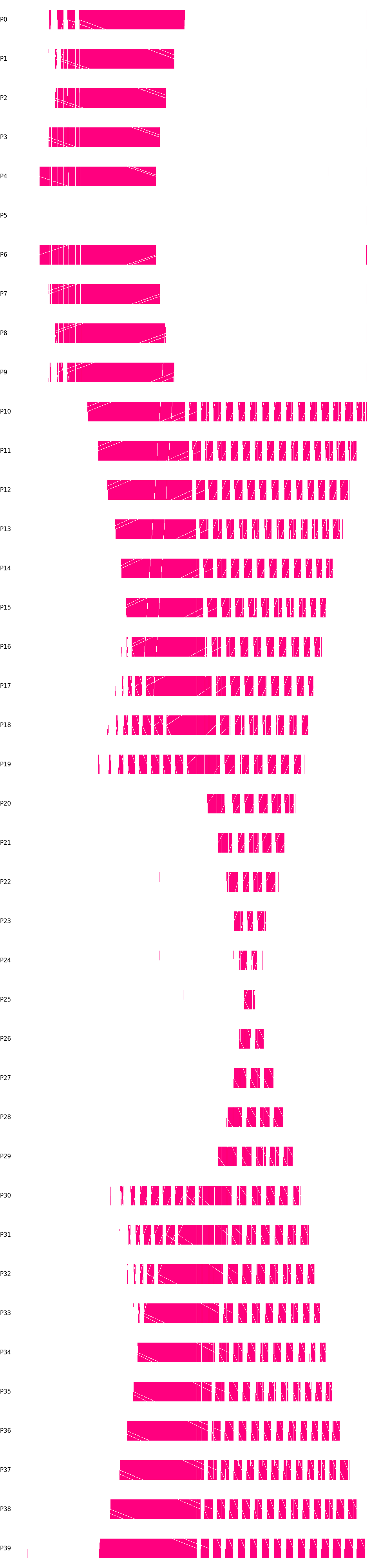}}			
					\put(-30,-3) {\begin{axis}[
						width=1.4\textwidth,height=0.26\textheight,
						ylabel = {Rank},
						y label style={at={(0.17,0.5)}},
						xlabel = {Time step},
						x label style={at={(0.5,0.08)}},
						x tick label style={font=\scriptsize},
						y tick label style={font=\scriptsize}, 
						xmin=1, xmax=35,
						ymin=0, ymax=39,
						xtick={35},
						ytick={5,10,15,20,25,30,35,40}, 
						xticklabels={\textbf{20}},
						yticklabels={\textbf{34},,\textbf{24},,\textbf{14},,\textbf{4}},
						]
				\end{axis}}
				\node [font=\footnotesize] at (0.3,-1.1){(c) \protect\UseVerb{Meggie}@turbo}; 
				\draw [semithick, dotted] (-1,2.53 ) -- (1.5,2.53 );
				\draw [semithick, dotted] (-1,1.65) -- (1.5,1.65);
				\draw [semithick, dotted] (-1,0.7825) -- (1.5,0.7825);
				\node [font=\small] at (-0.42,2.89){\tikz \fill [blue] (0,0.105) rectangle (0.76,0.155);};
				\draw [thick,densely dashed,blue] (1.41,2.5) -- (1.41,3.3);
				\draw [thick,densely dashed,blue] (0.67,1.2) -- (1.41,2.5);
				\draw [thick,densely dashed,blue] (0.67,1.2) -- (1.41,-0.1);
				\end{tikzpicture}
			\end{subfigure}
			\hspace{-2.5em}
			\begin{subfigure}[t]{0.24\textwidth}
				\begin{tikzpicture}
				\put(0,-2.2) {\includegraphics[width=0.84\textwidth,height=0.178 \textheight]{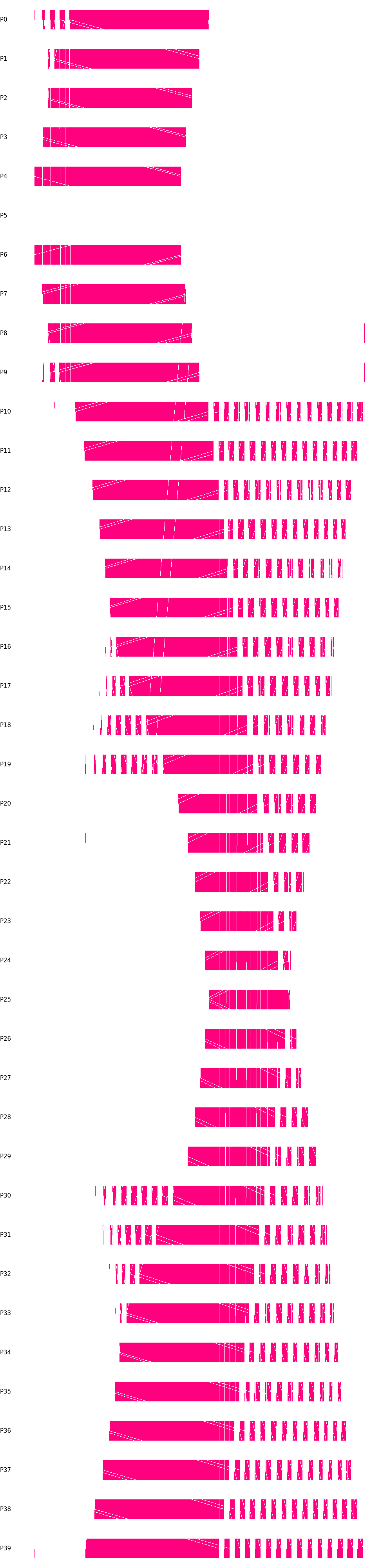}}
				\put(0,-3) {\begin{axis}[
						width=1.4\textwidth,height=0.26\textheight,
						y label style={at={(-0.1,0.5)}},
						xlabel = {Time step},
						x label style={at={(0.5,0.08)}},
						x tick label style={font=\scriptsize},
						y tick label style={font=\scriptsize}, 
						xmin=1, xmax=35,
						ymin=0, ymax=39,
						xtick={35},
						ytick={5,10,15,20,25,30,35,40}, 
						xticklabels={\textbf{20}},
						yticklabels={\textbf{34},,\textbf{24},,\textbf{14},,\textbf{4}},
						]
				\end{axis}}
				\node [font=\small] at (1.2,-1.1){(d) \protect\UseVerb{Meggie}@\SI{1.2}{\giga \Hz}}; 
				\draw [semithick, dotted] (0,2.53 ) -- (2.5,2.53 );
				\draw [semithick, dotted] (0,1.65) -- (2.5,1.65);
				\draw [semithick, dotted] (0,0.7825) -- (2.5,0.7825); 
				\node [font=\small] at (0.7,2.89){\tikz \fill [blue] (0,0.105) rectangle (0.95,0.155);};
				\draw [thick,densely dashed,blue] (2.45,2.5) -- (2.45,3.3);
				\draw [thick,densely dashed,blue] (1.95,1.2) -- (2.45,2.5);
				\draw [thick,densely dashed,blue] (1.95,1.2) -- (2.45,-0.1);
				\end{tikzpicture}
			\end{subfigure}	
			\hspace{1.7em}
			\begin{subfigure}[t]{0.24\textwidth}
				 \begin{tikzpicture}
				\put(0,-2.2) {\includegraphics[width=0.84\textwidth,height=0.178 \textheight]{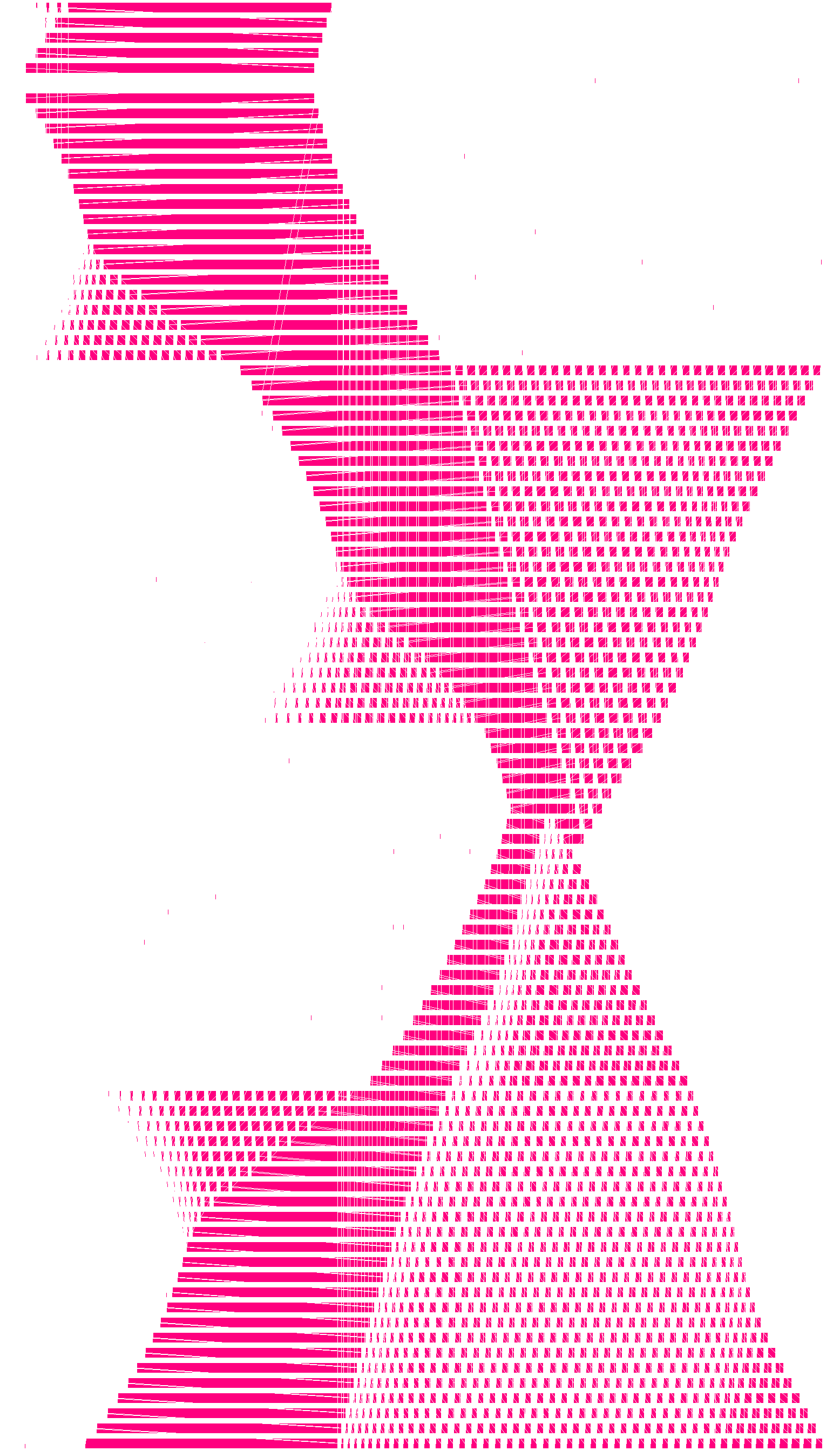}}
				\put(0,-3) {\begin{axis}[
						width=1.4\textwidth,height=0.26\textheight,
						y label style={at={(-0.1,0.5)}},
						xlabel = {Time step},
						x label style={at={(0.5,0.08)}},
						x tick label style={font=\scriptsize},
						y tick label style={font=\scriptsize}, 
						xmin=1, xmax=35,
						ymin=0, ymax=95,
						xtick={35},
						ytick={12,24,36,48,60,72,84,96}, 
						xticklabels={\textbf{50}},
						yticklabels={\textbf{83},,\textbf{59},,\textbf{34},,\textbf{11}},
						]
				\end{axis}}
				\node [font=\small] at (1.2,-1.1){(e) \protect\UseVerb{SuperMUCNG}@triad}; 
				\draw [semithick, dotted] (0,2.53 ) -- (2.5,2.53 );
				\draw [semithick, dotted] (0,1.65) -- (2.5,1.65);
				\draw [semithick, dotted] (0,0.7825) -- (2.5,0.7825); 
				\node [font=\small] at (0.49,3.16){\tikz \fill [blue] (0,0) rectangle (0.845,0.02);};
				\draw [thick,densely dashed,blue] (2.43,2.5) -- (2.43,3.3);
				\draw [thick,densely dashed,blue] (1.72,1.34) -- (2.43,2.5);
				\draw [thick,densely dashed,blue] (1.72,1.34) -- (2.43,-0.1);
				\end{tikzpicture}
			\end{subfigure}
			\hspace{1.7em}
			\begin{subfigure}[t]{0.24\textwidth}
				\begin{tikzpicture}
				\put(0,-2.2) {\includegraphics[width=0.83\textwidth,height=0.178 \textheight]{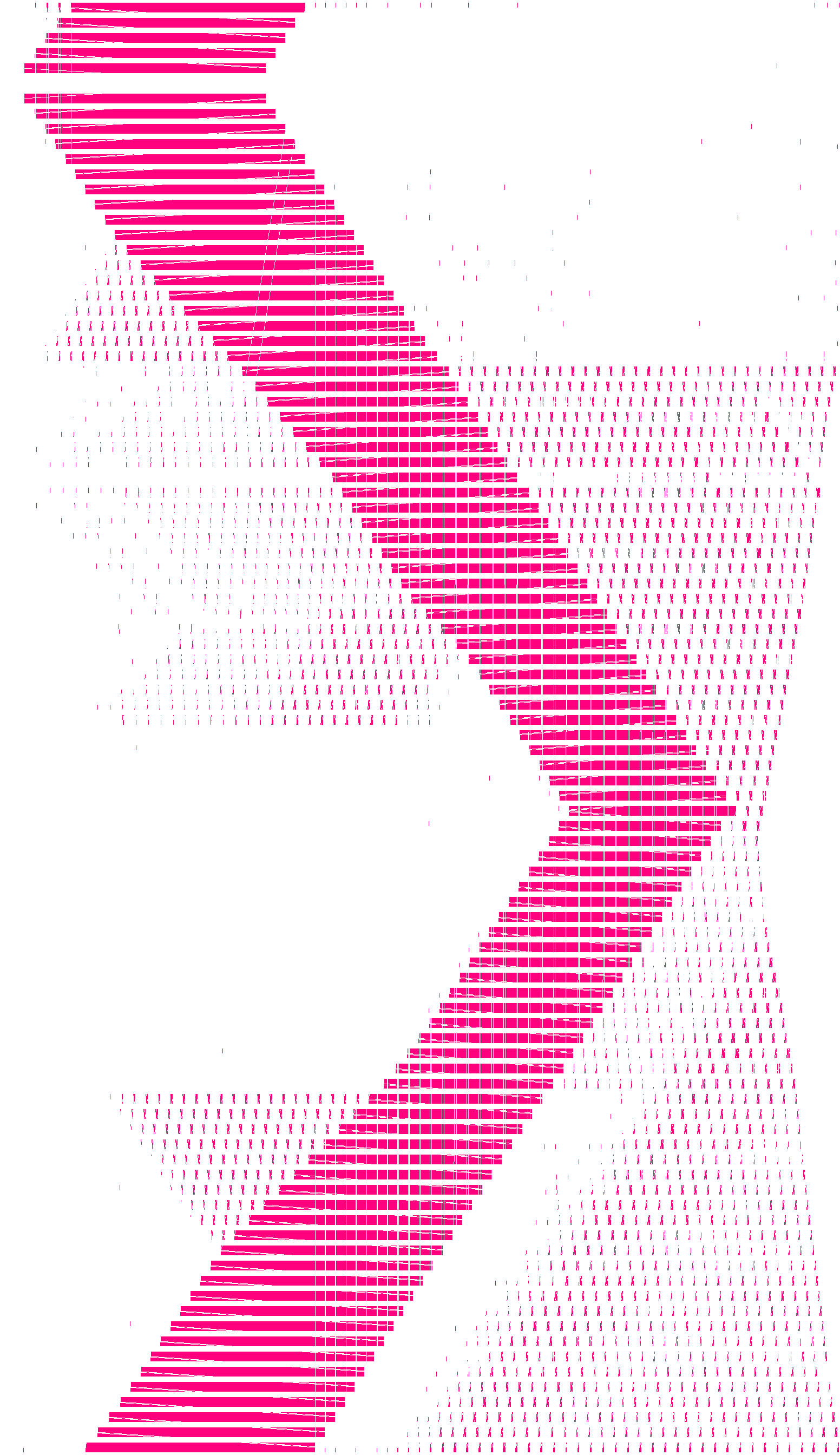}}
				\put(0,-3) {\begin{axis}[
						width=1.4\textwidth,height=0.26\textheight,
						y label style={at={(-0.1,0.5)}},
						xlabel = {Time step},
						x label style={at={(0.5,0.08)}},
						x tick label style={font=\scriptsize},
						y tick label style={font=\scriptsize}, 
						xmin=1, xmax=35,
						ymin=0, ymax=95,
						xtick={35},
						ytick={12,24,36,48,60,72,84,96}, 
						xticklabels={\textbf{50}},
						yticklabels={\textbf{83},,\textbf{59},,\textbf{34},,\textbf{11}},
						]
				\end{axis}}
				\node [font=\small] at (1.2,-1.1){(f) \protect\UseVerb{SuperMUCNG}@\protect\UseVerb{DIV}}; 
				\draw [semithick, dotted] (0,2.53 ) -- (2.5,2.53 );
				\draw [semithick, dotted] (0,1.65) -- (2.5,1.65);
				\draw [semithick, dotted] (0,0.7825) -- (2.5,0.7825); 
				\node [font=\small] at (0.42,3.16){\tikz \fill [blue] (0,0) rectangle (0.7,0.02);};
				\draw [thick,densely dashed,blue] (2.45,2.5) -- (2.45,3.3);
				\draw [thick,densely dashed,blue] (2.2,1.39) -- (2.45,2.5);
				\draw [thick,densely dashed,blue] (2.2,1.39) -- (2.45,-0.1);
				\end{tikzpicture}
			\end{subfigure}
		\end{minipage}
	\end{adjustbox}
	\caption{Saturation characteristics of benchmark platforms
          with different code and frequency settings and their
          influence on computational wave formation.  (a) Bandwidth
          saturation of microbenchmarks on a contention domain (MPI
          strong scaling) on four systems: STREAM triad on
          \protect\UseVerb{Emmy} with vs.\ without NT stores and on
          \protect\UseVerb{Meggie} using Turbo Mode vs.\ lowest core
          frequency.  On \protect\UseVerb{SuperMUCNG}, using STREAM
          triad and a ``slow'' Sch\"onauer triad, and standard STREAM
          triad on \protect\UseVerb{HazelHen}\@.  (b)-(f) Timeline
          visualization of idle wave-induced computational wave
          emergence under different saturation conditions.
          On \protect\UseVerb{HazelHen}, ITAC was not available
          so the trace was taken via explicit timing measurements.
          }
	\label{fig:SaturationCurves}
\end{figure}
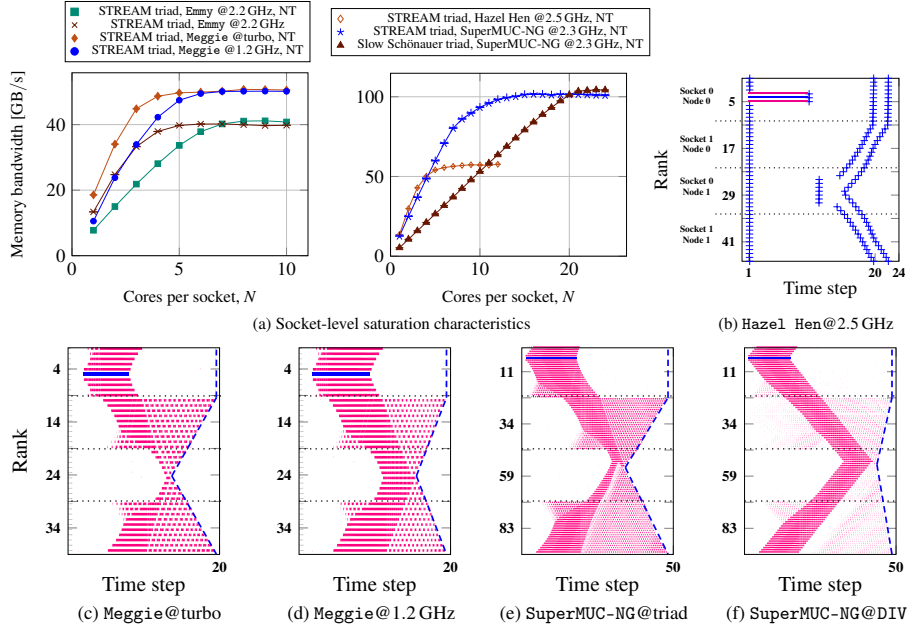

%% file: figures/CompWaveActiveProcesses.tex
\begin{figure}[tb]
	\centering
	\begin{adjustbox}{width=0.67\textwidth}
	\hspace{-6em}
	\begin{subfigure}[t]{0.705\textwidth}
		\begin{tikzpicture}
		\begin{axis}[trim axis left, trim axis right, scale only axis,
		width=0.73\textwidth,height=0.178\textheight,
		ylabel = {Average active processes},
		ymin=0,
		ymax=25,
		xmin=0,
		xmax=2,
		y label style={at={(0.05,0.5)},font=\footnotesize},
		x label style={font=\footnotesize},
		x tick label style={font=\footnotesize},
		ytick={0,5,10,15,20,25},
		xtick={0.4,1,1.6},
		xticklabels={Fig.~\ref{fig:ComputationalWave}(f),Fig.~\ref{fig:SaturationCurves}(e),Fig.~\ref{fig:SaturationCurves}(f)},
		ymajorgrids,
		legend columns = 1, 
		legend style = {
			nodes={inner sep=0.04em},
			draw=none,
			font=\tiny,
			cells={align=left},
			anchor=east,
			at={(0.71,0.8)},
			/tikz/column 1/.style={column sep=5pt,},
		},
		]
		
		\addplot [ybar,ybar legend, style={blue}, pattern=horizontal lines,mark=star, mark size=0.5pt,error bars/.cd, y dir=both, y explicit] 
		coordinates{
			(0.4,7) +- (1,1) 
		};
			
		\addplot [ybar,ybar legend, style={PineGreen}, pattern=dots,mark=diamond*, mark size=0.5pt,error bars/.cd, y dir=both, y explicit]
		coordinates{
			(1,13) +- (1,1) 
		};
		
		\addplot [ybar,ybar legend, style={Bittersweet}, pattern=north east lines, mark=square*,  mark size=0.5pt,error bars/.cd, y dir=both, y explicit] 
		coordinates{
			(1.6,20.5) +- (0.5,0.5) 
		};
		
		\addlegendentry{~\protect\UseVerb{STREAM} triad, \protect\UseVerb{Emmy} @\SI{2.2}{\giga \Hz}} 
		\addlegendentry{~\protect\UseVerb{STREAM} triad, SuperMUC-NG @\SI{2.3}{\giga \Hz}} 
		\addlegendentry{~slow Sch\"onauer triad, SuperMUC-NG @\SI{2.3}{\giga \Hz}} 
		\end{axis}
		\node [font=\small] at (3,-1.3){(a) Concurrently active cores};
		\end{tikzpicture}
		\end{subfigure}
	\hspace{-2.5em}
		\begin{subfigure}[t]{.265\textwidth}
			\begin{tikzpicture}
				\put(0,-2.2) {\includegraphics[width=0.9\textwidth,height=0.178 \textheight]{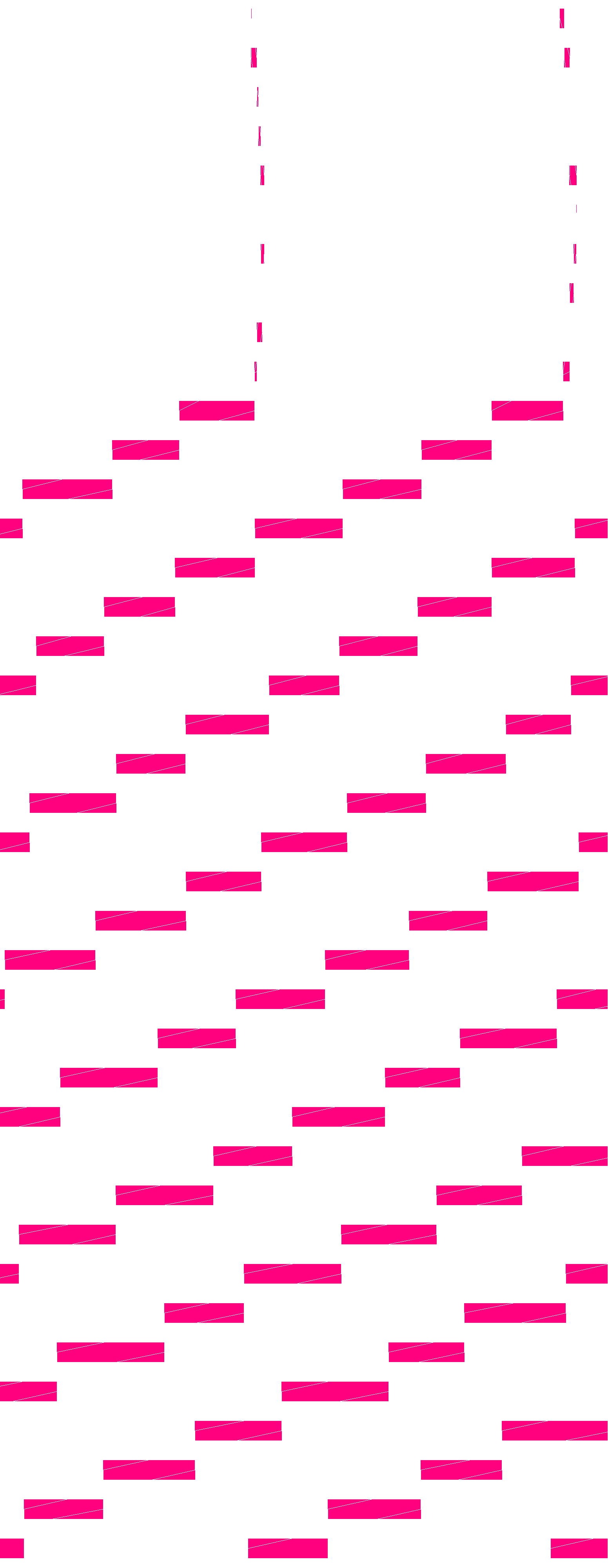}}
				\put(0,-3) {\begin{axis}[
					width=1.4\textwidth,height=0.26\textheight,
					ylabel = {Rank},
					y label style={at={(-0.1,0.5)}},
					xlabel = {Wall-clock time [s]},
					x label style={at={(0.5,0.08)}},
					x tick label style={font=\scriptsize},
					y tick label style={font=\scriptsize}, 
					xmin=1, xmax=35,
					ymin=0, ymax=39,
					xmajorticks=false,
					ytick={5,10,15,20,25,30,35,40}, 
					yticklabels={\textbf{34},,\textbf{24},,\textbf{14},,\textbf{4}},
					]
					\end{axis}}
				\node [font=\small] at (1,-1.3){(b) Exemplary timeline for Fig.~\ref{fig:ComputationalWave}(f)};
				\draw [semithick, dotted] (-0.5,2.53 ) -- (2.95,2.53 );
				\draw [semithick, dotted] (-0.5,1.65) -- (2.95,1.65);
				\draw [semithick, dotted] (-0.5,0.7825) -- (2.95,0.7825); 
				\node at (-0.85,3.1) {\tiny \textbf{Socket 0}}; 
				\node at (-0.85,2.9) {\tiny \textbf{Node 0}}; 
				\node at (-0.85,2.2) {\tiny \textbf{Socket 1}}; 
				\node at (-0.85,2) {\tiny \textbf{Node 0}}; 		
				\node at (-0.85,1.4) {\tiny \textbf{Socket 0}}; 
				\node at (-0.85,1.2) {\tiny \textbf{Node 1}}; 		
				\node at (-0.85,0.5) {\tiny \textbf{Socket 1}}; 
				\node at (-0.85,0.3) {\tiny \textbf{Node 1}}; 
    			\draw [thick,densely dashed,black] (2.33,-0.1) -- (2.33,3.3);
        		\draw [thick,densely dashed,black] (1.37,-0.1) -- (1.37,3.3);
        		\draw [thick,densely dashed,black] (0.37,-0.1) -- (0.37,3.3);	
        		\node at (0.25,0.5) {\tiny \textbf{\circledfilled{7}}};
        		\node at (0.25,1.3) {\tiny \textbf{\circledfilled{7}}};
        		\node at (0.25,2.15) {\tiny \textbf{\circledfilled{8}}};
        		\node at (1.25,0.3) {\tiny \textbf{\circledfilled{7}}};
				\node at (1.25,1.1) {\tiny \textbf{\circledfilled{7}}};
				\node at (1.25,1.99) {\tiny \textbf{\circledfilled{8}}};
				\node at (2.5,0.43) {\tiny \textbf{\circledfilled{7}}};
				\node at (2.5,1.2) {\tiny \textbf{\circledfilled{8}}};
				\node at (2.5,2.05) {\tiny \textbf{\circledfilled{7}}};	
			\end{tikzpicture}
		\end{subfigure}
		\end{adjustbox}
		\caption{(a) Average number of MPI processes executing user code
                  concurrently for the fully developed
                  steady state computational waves (wavelength
                  \protect\UseVerb{MPIcommsize}) in Figures~\ref{fig:ComputationalWave}(f),
                  \ref{fig:SaturationCurves}(e), and \ref{fig:SaturationCurves}(f)\@.
                  Minimum and maximum values among 60 samples along the timeline
                  are indicated as whiskers.
                  Data points were taken from the timeline data as shown in (b).
                  Numbers in circles denote number of active processes
                  at this point in time on this contention domain.
                }
		  \label{fig:CompWaveActiveProcesses}
\end{figure}
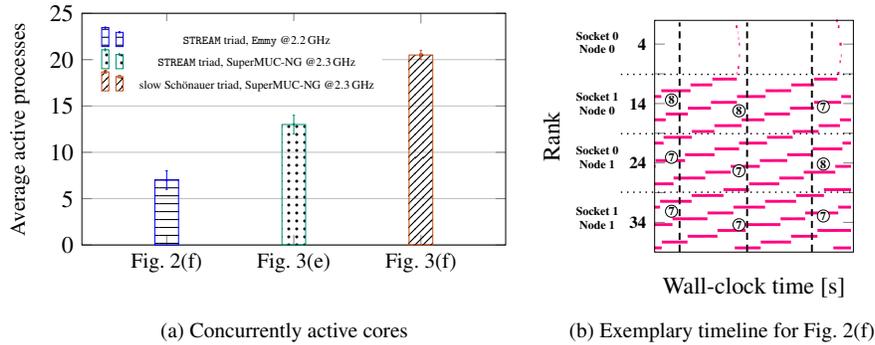

%% file: figures/CompWaveShapeAndSlope.tex
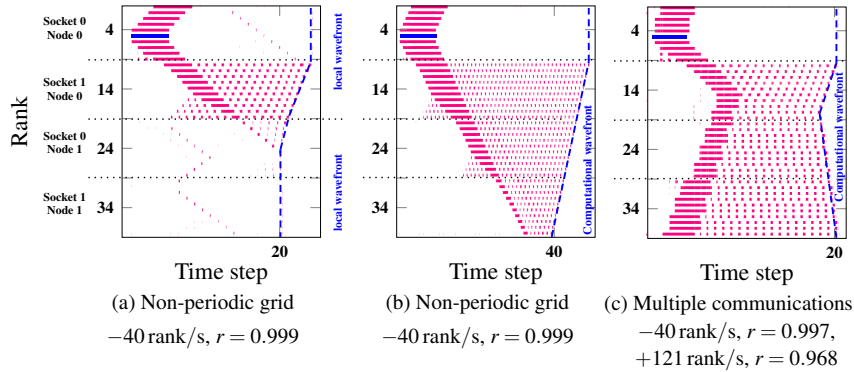
\begin{figure}[tb]
	\centering
		\begin{adjustbox}{width=0.8\textwidth}
		\begin{subfigure}[t]{.265\textwidth}
			\begin{tikzpicture}
		\put(-30,0){	
			\put(-2.9,-2.2) {\includegraphics[width=0.92\textwidth,height=0.178 \textheight]{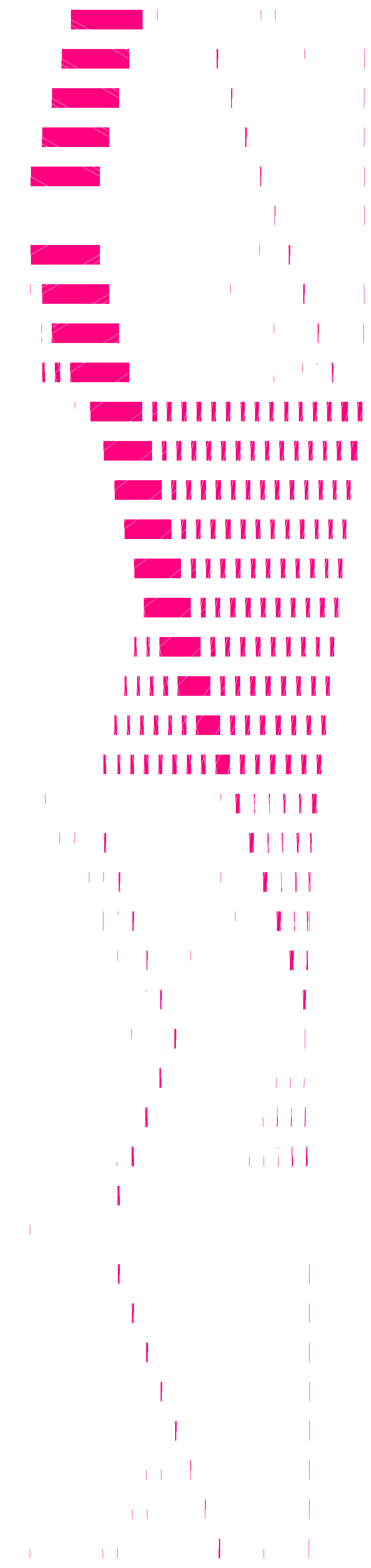}}
			\put(0,-3) {\begin{axis}[
				width=1.4\textwidth,height=0.26\textheight,
				ylabel = {Rank},
				y label style={at={(-0.1,0.5)}},
				xlabel = {Time step},
				x label style={at={(0.5,0.08)}},
				x tick label style={font=\scriptsize},
				y tick label style={font=\scriptsize}, 
				xmin=1, xmax=35,
				ymin=0, ymax=39,
				xtick={28},
				ytick={5,10,15,20,25,30,35,40}, 
				xticklabels={\textbf{20}},
				yticklabels={\textbf{34},,\textbf{24},,\textbf{14},,\textbf{4}},
				]
			\end{axis}}
			\node [font=\small] at (1.2,-1.1){(a) Non-periodic grid};
			\node [font=\small] at (1.2,-1.6){\SI{-40}{rank \per \second}, $r=0.999$}; 
			\draw [semithick, dotted] (-0.5,2.53 ) -- (2.95,2.53 );
			\draw [semithick, dotted] (-0.5,1.65) -- (3.3,1.65);
			\draw [semithick, dotted] (-0.5,0.7825) -- (2.95,0.7825); 
			\node [font=\small] at (0.42,2.89){\tikz \fill [blue] (1,0.11) rectangle (1.55,0.16);};
			\node at (-0.85,3.1) {\tiny \textbf{Socket 0}}; 
			\node at (-0.85,2.9) {\tiny \textbf{Node 0}}; 
			\node at (-0.85,2.2) {\tiny \textbf{Socket 1}}; 
			\node at (-0.85,2) {\tiny \textbf{Node 0}}; 		
			\node at (-0.85,1.4) {\tiny \textbf{Socket 0}}; 
			\node at (-0.85,1.2) {\tiny \textbf{Node 1}}; 		
			\node at (-0.85,0.5) {\tiny \textbf{Socket 1}}; 
			\node at (-0.85,0.3) {\tiny \textbf{Node 1}}; 
			\draw [thick,densely dashed,blue] (2.8,2.5) -- (2.8,3.3);
			\draw [thick,densely dashed,blue] (2.48,1.7) -- (2.8,2.5);
			\draw [thick,densely dashed,blue] (2.35,1.2) -- (2.48,1.7);
			\draw [thick,densely dashed,blue] (2.35,1.2) -- (2.35,-0.1);
			\node at (3.2,2.65) {\rotatebox{90}{\tiny \textbf{\textcolor{blue}{local wavefront}}}}; 
			\node at (3.2,0.56) {\rotatebox{90}{\tiny \textbf{\textcolor{blue}{local wavefront}}}}; }		
			\end{tikzpicture}
		\end{subfigure}
		\hspace{1.5em}
		\begin{subfigure}[t]{.265\textwidth}
			\begin{tikzpicture}
			\put(-2.9,-2.2) {\includegraphics[width=0.92\textwidth,height=0.178 \textheight]{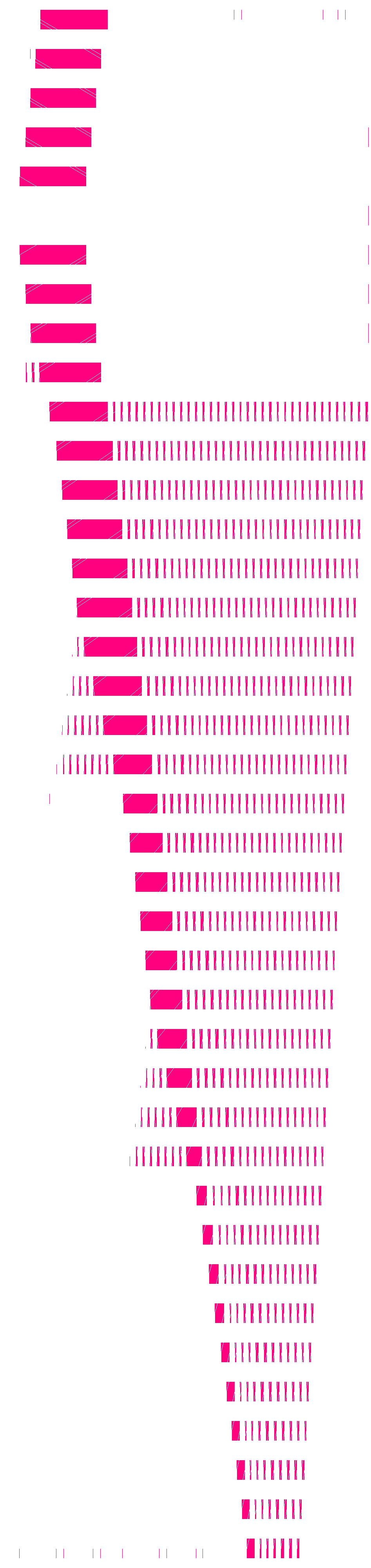}}
			\put(0,-3) {\begin{axis}[
				width=1.4\textwidth,height=0.26\textheight,
				y label style={at={(0.17,0.5)}},
				xlabel = {Time step},
				x label style={at={(0.5,0.08)}},
				x tick label style={font=\scriptsize},
				y tick label style={font=\scriptsize}, 
				xmin=1, xmax=35,
				ymin=0, ymax=39,
				xtick={28},
				ytick={5,10,15,20,25,30,35,40}, 
				xticklabels={\textbf{40}},
				yticklabels={\textbf{34},,\textbf{24},,\textbf{14},,\textbf{4}},
				]
			\end{axis}}
			\node [font=\small] at (1.2,-1.1){(b) Non-periodic grid};
			\node [font=\small] at (1.2,-1.6){\SI{-40}{rank \per \second}, $r=0.999$};
			\draw [semithick, dotted] (-0.5,2.53 ) -- (2.95,2.53 );
			\draw [semithick, dotted] (-0.5,1.65) -- (2.5,1.65);
			\draw [semithick, dotted] (-0.5,0.7825) -- (2.5,0.7825); 
			\node [font=\small] at (0.32,2.89){\tikz \fill [blue] (1,0.11) rectangle (1.54,0.16);};
			\draw [thick,densely dashed,blue] (2.85,2.5) -- (2.85,3.3);
			\draw [thick,densely dashed,blue] (2.3,-0.1) -- (2.85,2.5);
			\node at (2.85,0.9) {\rotatebox{90}{\tiny \textbf{\textcolor{blue}{Computational wavefront}}}}; 	
			\end{tikzpicture}
		\end{subfigure}
		\hspace{0.9em}
		\begin{subfigure}[t]{.265\textwidth}
			\begin{tikzpicture}
			\put(0,-9){
			\put(-2.9,-2.2) {\includegraphics[width=0.92\textwidth,height=0.178 \textheight]{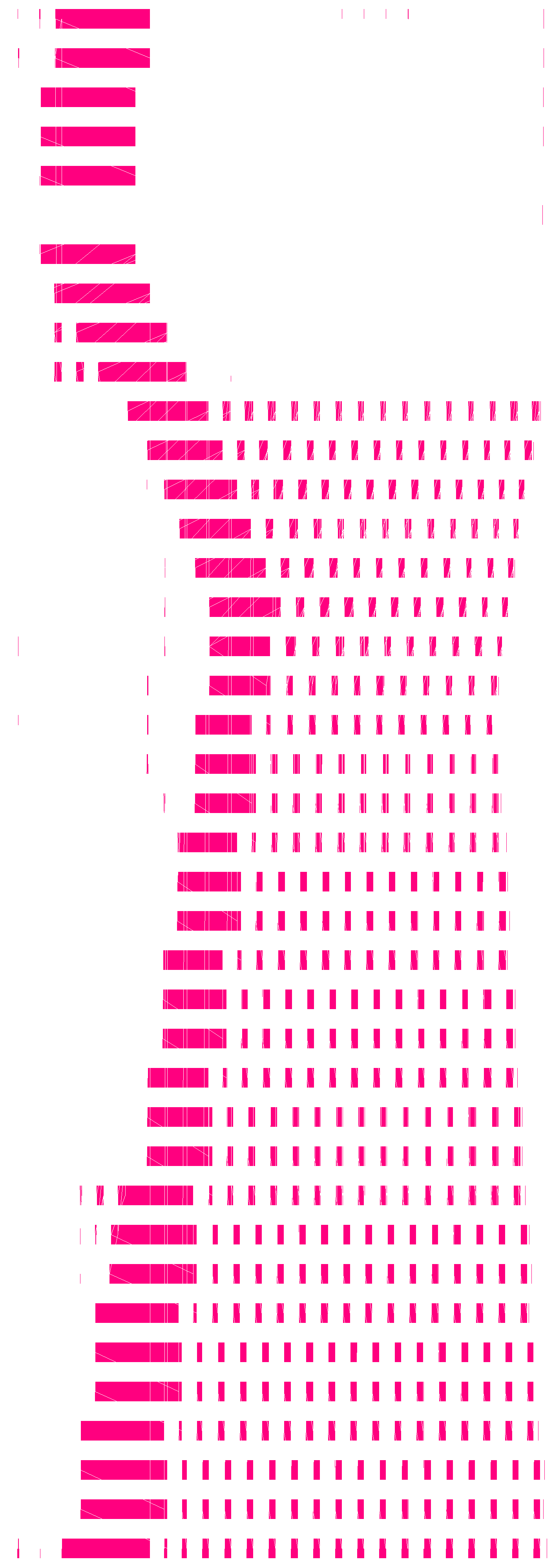}}
			\put(0,-3) {\begin{axis}[
				width=1.4\textwidth,height=0.26\textheight,
				y label style={at={(0.17,0.5)}},
				xlabel = {Time step},
				x label style={at={(0.5,0.08)}},
				x tick label style={font=\scriptsize},
				y tick label style={font=\scriptsize}, 
				xmin=1, xmax=35,
				ymin=0, ymax=39,
				xtick={33},
				ytick={5,10,15,20,25,30,35,40}, 
				xticklabels={\textbf{20}},
				yticklabels={\textbf{34},,\textbf{24},,\textbf{14},,\textbf{4}},
				]
				\end{axis}}
			\draw [semithick, dotted] (-0.5,2.53 ) -- (2.95,2.53 );
			\draw [semithick, dotted] (-0.5,1.65) -- (2.6,1.65);
			\draw [semithick, dotted] (-0.5,0.7825) -- (2.6,0.7825); 
			\node [font=\small] at (0.32,2.89){\tikz \fill [blue] (1,0.11) rectangle (1.51,0.16);};
			\draw [thick,densely dashed,blue] (2.8,2.5) -- (2.8,3.3);
			\draw [thick,densely dashed,blue] (2.55,1.75) -- (2.8,2.5);
			\draw [thick,densely dashed,blue] (2.55,1.75) -- (2.8,-0.1);
			\node at (2.85,1.4) {\rotatebox{90}{\tiny \textbf{\textcolor{blue}{Computational wavefront}}}}; 
			\node [font=\small] at (1.2,-1.1){(c) Multiple communications}; 
			\node [font=\small] at (1.2,-1.5){~~~\SI{-40}{rank \per \second}, $r=0.997$, }; 
			\node [font=\small] at (1.2,-1.9){~~~$+$\SI{121}{rank \per \second}, $r=0.968$};}
			\end{tikzpicture}
		\end{subfigure}
	\end{adjustbox}
	\caption{Shape and slope of
          memory-bound computational wavefront with different communication topologies and
          patterns on the \protect\UseVerb{Emmy} system.
          The measured slope(s) of the computational wave(s) in ranks per
          second is/are indicated together with correlation coefficients of linear fits.
          Code properties are the same as in \Cref{fig:ComputationalWave}\@.
          The $x$ axis shows walltime but the time step at which the
          computation was terminated is indicated. 
          (a) Open boundary conditions, next-neighbor communication,
          short one-off idle injection,
          (b) open boundary conditions, next-neighbor communication,
          long one-off idle injection, (c) periodic boundary conditions,
          next-neighbor communication along rising ranks, next- and next-to-next neighbor
          communication along falling ranks, short one-off idle injection.
        }
	\label{fig:CompWaveShapeAndSlope}
\end{figure}

%% file: figures/CompWaveStability.tex


%% file: figures/PureMPI.tex
\begin{figure}[tb]
	\centering
	\begin{adjustbox}{width=0.99\textwidth}
			\begin{tikzpicture}
			\put(270,-1) {\includegraphics[width=0.22\textwidth,height=0.178 \textheight]{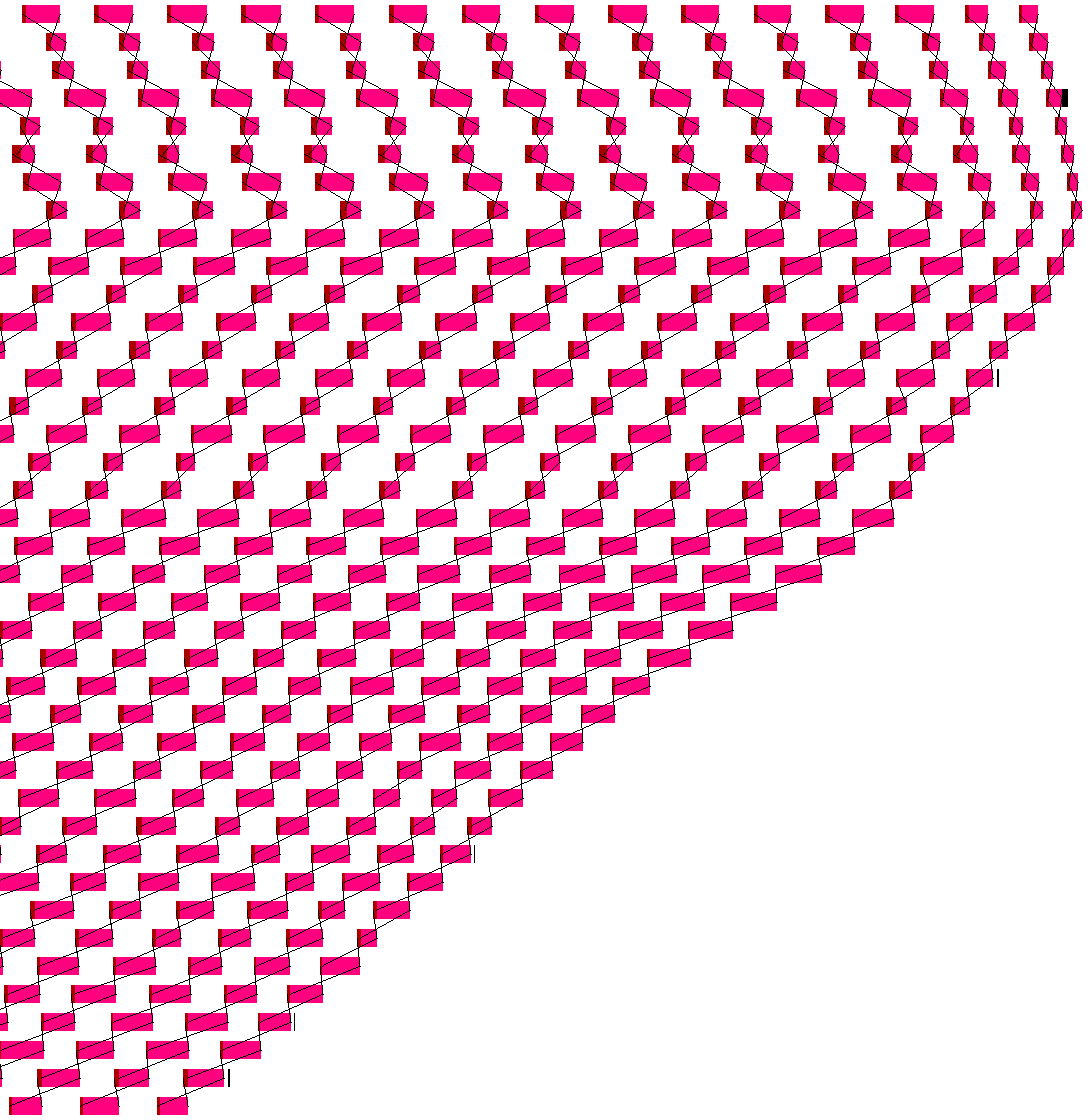}}
			\put(90,-1) {\includegraphics[width=0.22\textwidth,height=0.178 \textheight]{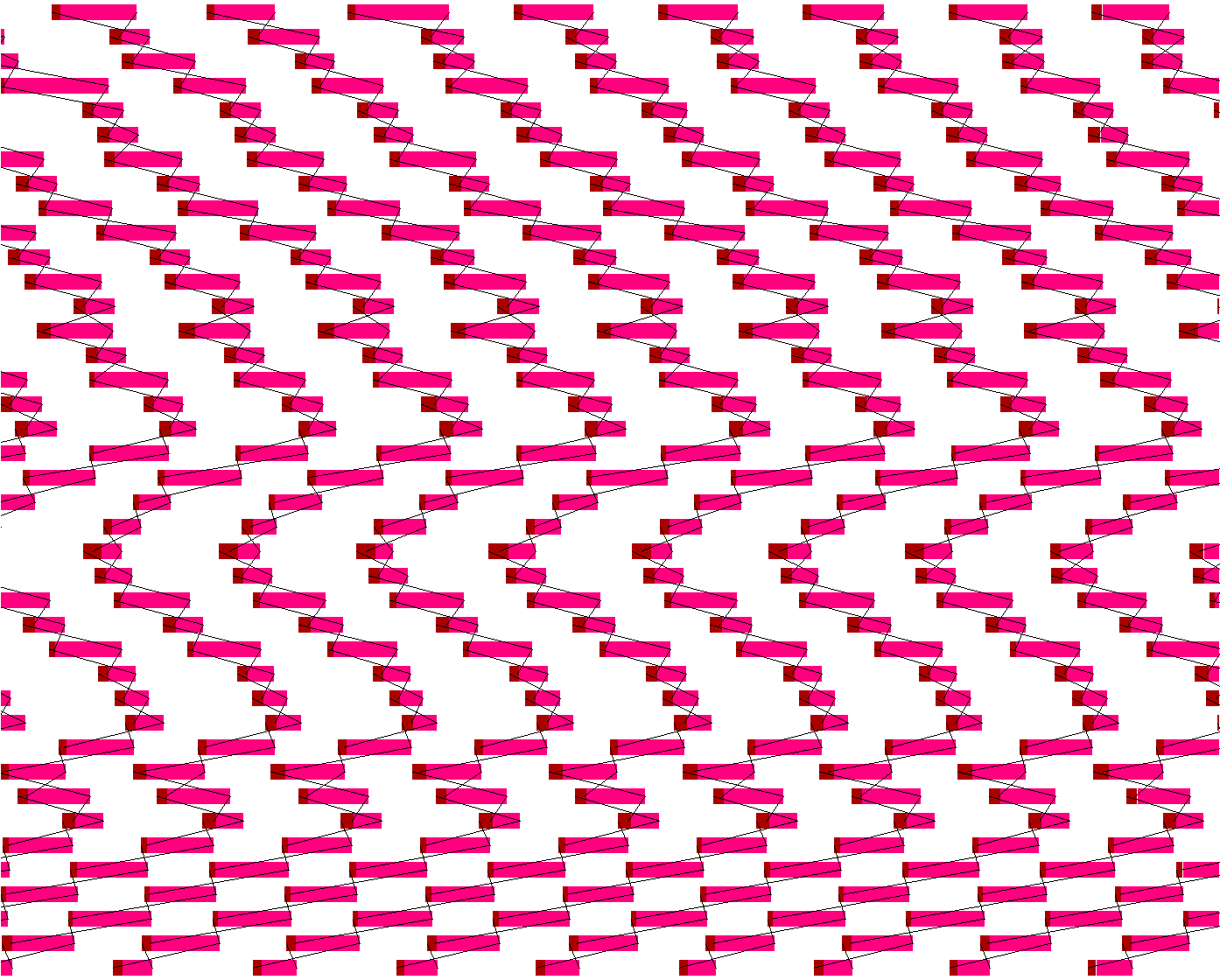}}
			\put(180,-1) {\includegraphics[width=0.22\textwidth,height=0.178 \textheight]{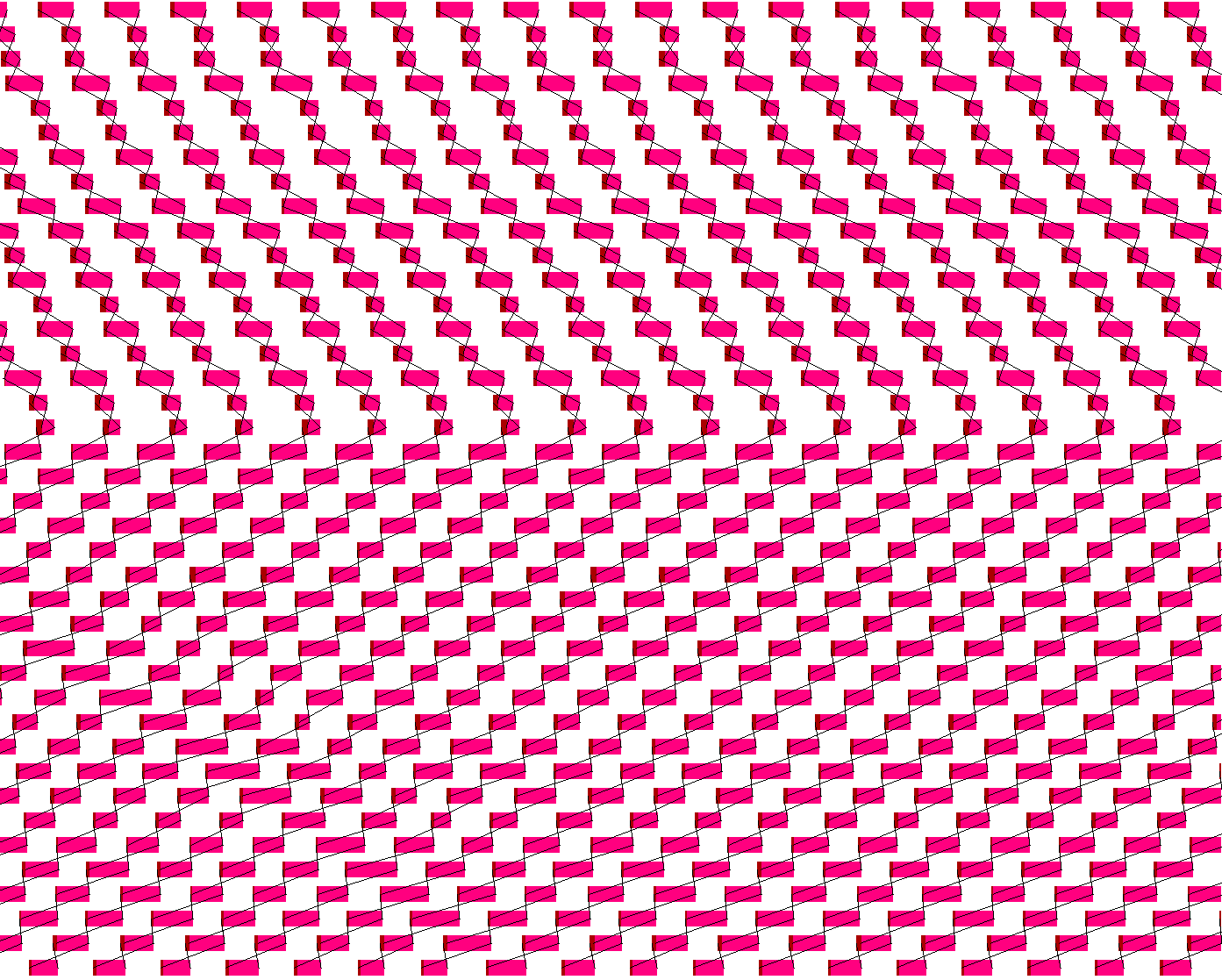}}
			\put(5.05,-1) {\includegraphics[width=0.22\textwidth,height=0.178 \textheight]{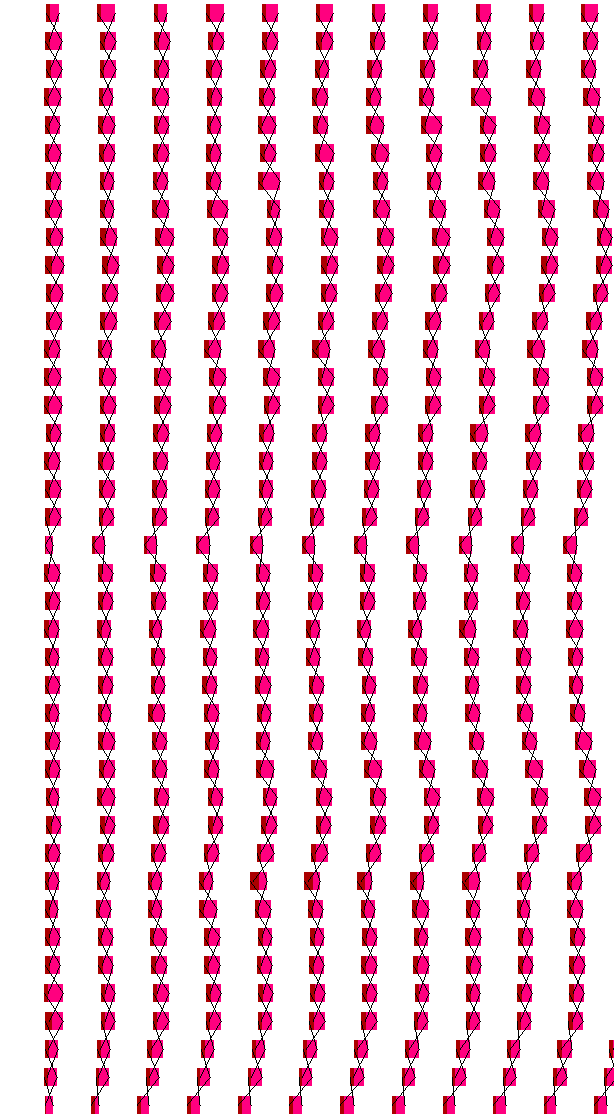}}
			\put(0,0) {\begin{axis}[
				trim axis left, trim axis right, scale only axis,
				width=\textwidth,height=0.178\textheight,
				title style={at={(0.5,-0.5)}},
				ylabel = {Rank},
				y label style={font=\footnotesize,at={(-0.04,0.5)}},
				xlabel = {Walltime for disconnected time step [s]},
				x label style={font=\footnotesize}, 			
				x tick label style={font=\scriptsize},
				y tick label style={font=\scriptsize}, 
				xmin=0, xmax=55,
				ymin=0, ymax=39,
				xtick={1,4,33.5,37.5,45,54.2},
				xticklabels={3.6,3.75,26.52, 26.74,1927.6,1928},
				ytick={5,10,15,20,25,30,35,40}, 
				yticklabels={\textbf{34},,\textbf{24},,\textbf{14},,\textbf{4}},
				]
				\end{axis}
				
				\draw [semithick, dotted] (-0.5,2.53 ) -- (12.15,2.53 );
				\draw [semithick, dotted] (-0.5,1.65) -- (12.15,1.65);
				\draw [semithick, dotted] (-0.5,0.7825) -- (12.15,0.7825); 
				\node at (-0.85,3.1) {\tiny \textbf{Socket 0}}; 
				\node at (-0.85,2.9) {\tiny \textbf{Node 0}}; 
				\node at (-0.85,2.2) {\tiny \textbf{Socket 1}}; 
				\node at (-0.85,2) {\tiny \textbf{Node 0}}; 		
				\node at (-0.85,1.4) {\tiny \textbf{Socket 0}}; 
				\node at (-0.85,1.2) {\tiny \textbf{Node 1}}; 		
				\node at (-0.85,0.5) {\tiny \textbf{Socket 1}}; 
				\node at (-0.85,0.3) {\tiny \textbf{Node 1}};
				\node at (1.5,3.8) {\small \textbf{global sync WF}};
				\node at (4.5,3.8) {\small \textbf{local de-sync WF}};
				\node at (7.5,3.8) {\small \textbf{global de-sync WF}};
				\node at (10.5,3.8) {\small \textbf{global de-sync WF}};
				\draw [thick,densely dashed,blue] (1.1,0) -- (1.1,3.4);
				\draw [thick,densely dashed,blue] (8.3,3.4) -- (8.9,2);
				\draw [thick,densely dashed,blue] (7.5,0) -- (8.9,1.9);
				\draw [thick,densely dashed,blue] (9.9,0) -- (11.65,1.8);
				\draw [thick,densely dashed,blue] (11.65,1.8) -- (12.1,2.6);
				\draw [thick,densely dashed,blue] (12.03,3.4) -- (12.15,2.65);
        		\draw [thick,densely dashed,black] (4.4,0) -- (4.4,3.4);	
				\node at (4.25,0.5) {\tiny \textbf{\circledfilled{6}}};
				\node at (4.25,1.3) {\tiny \textbf{\circledfilled{7}}};
				\node at (4.25,2.15) {\tiny \textbf{\circledfilled{8}}};
				\node at (4.25,3) {\tiny \textbf{\circledfilled{6}}};
				\draw [thick,densely dashed,black] (7.15,0) -- (7.15,3.4);	
				\node at (7,0.5) {\tiny \textbf{\circledfilled{5}}};
				\node at (7,1.3) {\tiny \textbf{\circledfilled{5}}};
				\node at (7,2.15) {\tiny \textbf{\circledfilled{6}}};
				\node at (7,3) {\tiny \textbf{\circledfilled{6}}};
				\draw [thick,densely dashed,black] (9.7,0) -- (9.7,3.4);	
				\node at (9.55,0.5) {\tiny \textbf{\circledfilled{5}}};
				\node at (9.55,1.3) {\tiny \textbf{\circledfilled{5}}};
				\node at (9.55,2.15) {\tiny \textbf{\circledfilled{6}}};
				\node at (9.55,3) {\tiny \textbf{\circledfilled{6}}};
			 }	
			\end{tikzpicture}
	\end{adjustbox}
	\caption{ $50\,000$ iterations run of MPI-parallel
          STREAM program (non-periodic grid, 4.8\,\GB\ overall data
          volume) on the \protect\UseVerb{Emmy} system
          (normal stores, saturation at 5--6 cores).  The four phases
          show different cutouts of the complete timeline near the indicated
          walltimes. Synchronized state (phase 1): 30{}+{}10\,ms
          average compute + communication intervals. Fully developed wavefront
          (phase 3,4): 20{}+{}17.5\,ms average compute + communication.
          Numbers of cuncurrent working processes per domain are indicated
          in circles.
        }\label{fig:PureMPI}
\end{figure}
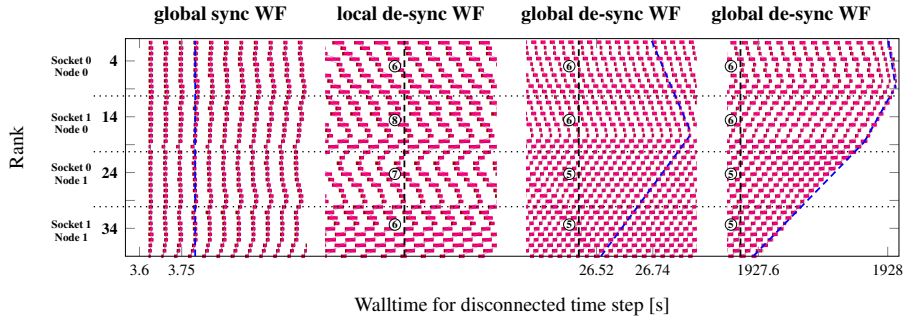

%% file: figures/Hybrid.tex
\begin{figure}[tb]
	\centering
	\begin{adjustbox}{width=0.33\textwidth}
	\begin{subfigure}[t]{.265\textwidth}
		\hspace{-8em}
		\begin{tikzpicture}
		\put(-0.4,-2.2) {\includegraphics[width=.92\textwidth,height=0.18 \textheight]{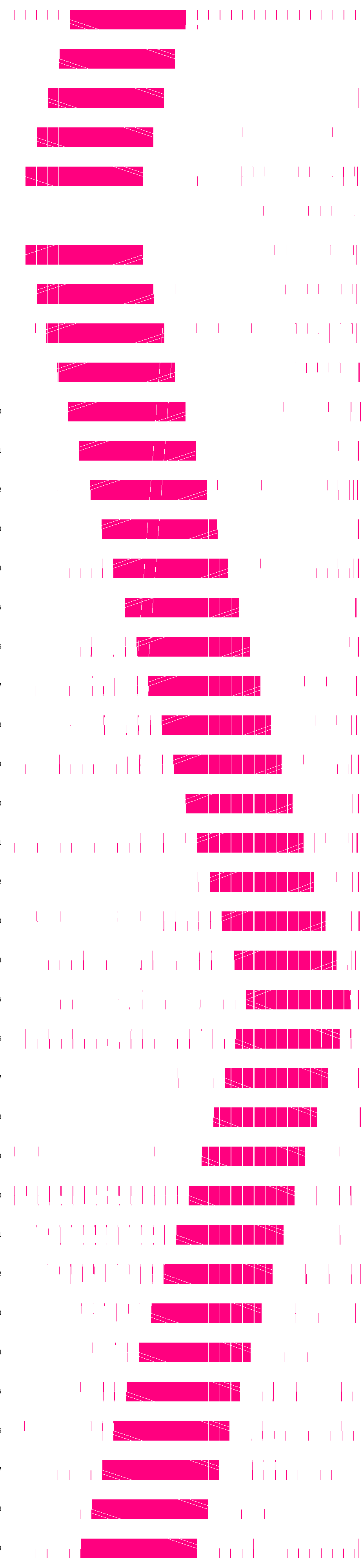}}
		\put(0,-3) {\begin{axis}[
		width=1.4\textwidth,height=0.265\textheight,
		ylabel = {Rank},
		y label style={at={(0.2,0.5)}},
		xlabel = {Walltime},
		x label style={at={(0.5,0.08)}},
		x tick label style={font=\scriptsize},
		y tick label style={font=\scriptsize}, 
		xmin=1, xmax=20,
		xtick={20}, 
		yticklabels={\textbf{20}},
		ymin=0, ymax=39,
		ytick={0,9,19,29,39}, 
		yticklabels={\textbf{39},\textbf{29},\textbf{19},\textbf{9},\textbf{0}},
		]
		\end{axis}}
		\node [font=\small] at (1.2,-1.1){(a) Induced wavefront}; 
		\node [font=\small] at (0.65,2.925){\tikz \fill [blue] (2,0.11) rectangle (1,0.16);};
		\node at (2.14,2.2) {\rotatebox{-41}{\tiny \textbf{leading edge}}};
		\node at (0.8,1.8) {\rotatebox{-41}{\tiny \textbf{trailing edge}}};
    		\draw [thick,densely dashed,blue] (2.9,-0.1) -- (2.9,3.4); 		
		\end{tikzpicture}
	\end{subfigure}
	\hspace{-4.4em}	
	\begin{subfigure}[t]{.265\textwidth}
		\begin{tikzpicture}
			\put(0,4) {\includegraphics[width=1.4\textwidth,height=0.085 \textheight]{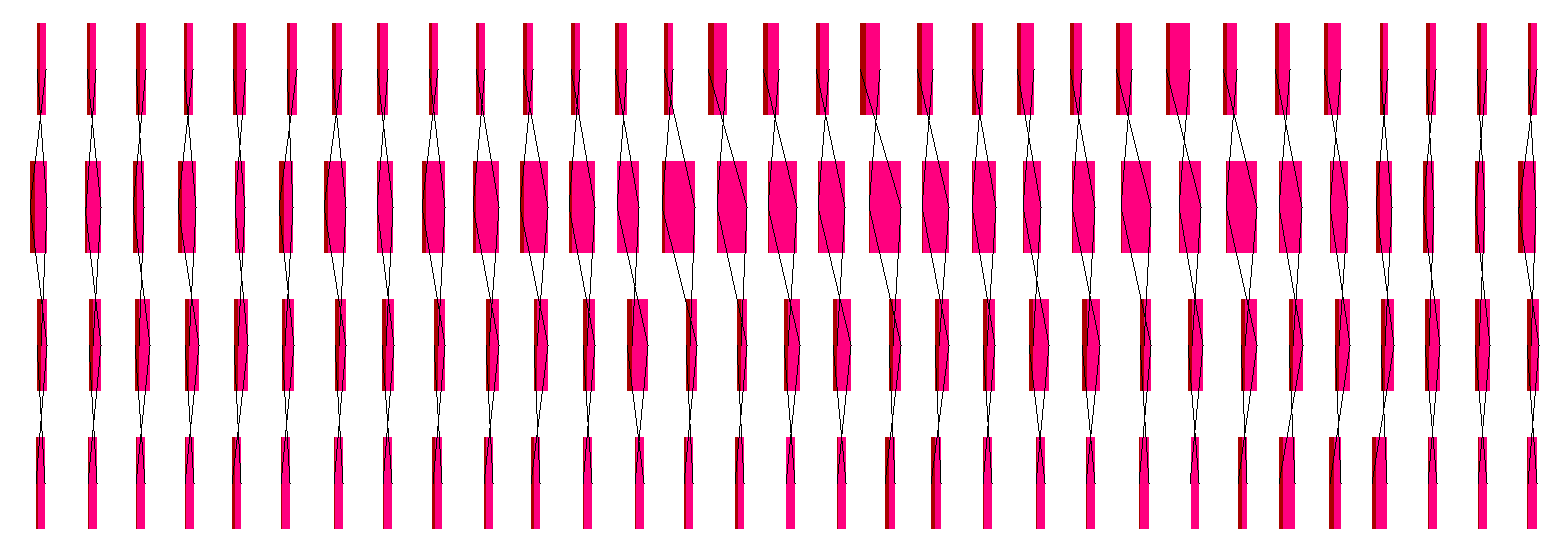}}
			\begin{axis}[
			trim axis left, trim axis right, scale only axis,
			width=1.4\textwidth,height=0.1\textheight,
			title style={at={(0.5,-1.22)}},
			ylabel = {Rank},
			y label style={font=\footnotesize,at={(-0.08,0.5)}},
			xlabel = {Intermediate time steps},
			x label style={font=\footnotesize,at={(0.5,-0.35)}}, 			
			x tick label style={font=\scriptsize},
			y tick label style={font=\scriptsize},
			xmin=0, xmax=55,
			ymin=0, ymax=5,
			xticklabel pos=right,
			xtick={0,55},
			xticklabels={48000,48031},
			ytick={1,2,3,4}, 
			yticklabels={\textbf{3},\textbf{2},\textbf{1},\textbf{0}},
			]
			\end{axis}
			\path[draw=black,solid,line width=0.5mm,fill=black,
			preaction={-triangle 90,thin,draw,shorten >=-1mm}
			] (0.1,-0.4) -- (0.1,-0.15);
			\path[draw=black,solid,line width=0.5mm,fill=black,
			preaction={-triangle 90,thin,draw,shorten >=-1mm}
			] (2.2,-0.4) -- (2.2,-0.15);
			\path[draw=black,solid,line width=0.5mm,fill=black,
			preaction={-triangle 90,thin,draw,shorten >=-1mm}
			] (4.35,-0.4) -- (4.35,-0.15);
			\node at (0,-0.7) {\scriptsize \textbf{Synchronized}}; 
			\node at (0,-0.95) {\scriptsize \textbf{WF}}; 
			\node at (2.2,-0.7) {\scriptsize \textbf{De-synchronization}}; 
			\node at (2.15,-0.95) {\scriptsize \textbf{with \enquote{natural noise}}}; 
			\node at (4.35,-0.7) {\scriptsize \textbf{Back to}};
			\node at (4.25,-0.95) {\scriptsize \textbf{synchronized WF}}; 
			
			\draw [semithick, dotted] (-0.5,1.38 ) -- (4.45,1.38 );
			\draw [semithick, dotted] (-0.5,0.95) -- (4.45,0.95);
			\draw [semithick, dotted] (-0.5,0.55) -- (4.45,0.55); 
			
			\node at (-0.85,1.8) {\tiny \textbf{Socket 0}}; 
			\node at (-0.85,1.6) {\tiny \textbf{Node 0}}; 
			\node at (-0.85,1.3) {\tiny \textbf{Socket 1}}; 
			\node at (-0.85,1.1) {\tiny \textbf{Node 0}}; 		
			\node at (-0.85,0.8) {\tiny \textbf{Socket 0}}; 
			\node at (-0.85,0.6) {\tiny \textbf{Node 1}}; 		
			\node at (-0.85,0.3) {\tiny \textbf{Socket 1}}; 
			\node at (-0.85,0.1) {\tiny \textbf{Node 1}}; 
			\node [font=\small] at (2,-2.1){(b) Spontaneous desynchronization}; 
		\end{tikzpicture}
	\end{subfigure}	
	\end{adjustbox}
	\caption{MPI+OpenMP hybrid execution of parallel STREAM triad
          on \protect\UseVerb{Emmy} with bidirectional next-neighbor communication,
          periodic boundary conditions, and the same overall data volume as in \Cref{fig:PureMPI}
          but with 10 threads per process and one process per
          contention domain.  (a) 40 processes on 20 nodes with negligible communication
          overhead and an idle
          injection on process 5 for first 20 iterations, (b) four
          processes on two nodes with 5\,\MB\ MPI message size for intermediate 31 iterations over a complete run of 50\,000 timesteps.}
	\label{fig:Hybrid}
\end{figure}
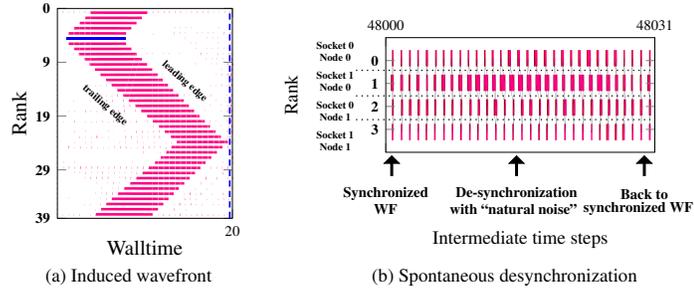

%% file: figures/ChebFD.tex
\begin{figure}[tb]
	\centering
	\begin{adjustbox}{width=0.92\textwidth}
		\hspace{-3em}
		\begin{subfigure}[t]{0.5\textwidth}
			\begin{tikzpicture}
			\pgfplotstableread{figures/Fig9_ChebFD_Emmy_SuperMUC_/Emmy_BW_topi128-64-64_ns128_nb2_Vector_chunkheight1_implAVX_aligned.txt}\Ablocksz;
			\pgfplotstableread{figures/Fig9_ChebFD_Emmy_SuperMUC_/Emmy_BW_topi128-64-64_ns128_nb32_Vector_chunkheight1_implAVX_aligned.txt}\Bblocksz;
			\begin{axis}[trim axis left, trim axis right, scale only axis,
			width=0.5\textwidth,height=0.178\textheight,
			xlabel = {Cores per socket, $N$},
			ybar, 
			ybar legend, 
			bar width=0.8mm,
			xmin=0,
			ymin=0,
			ymax=50,
			xmax=11,
			y label style={at={(0.05,0.5)},font=\footnotesize},
			x label style={font=\footnotesize},
			x tick label style={font=\footnotesize},
			ylabel = {Performance [\si{\giga \flop / \second}]}, 
			xtick={0,2,4,6,8,10},
			xticklabels={0,2,4,6,8,10},
			ytick={0,10,20,30,40,50},
			ymajorgrids,
			legend columns = 1, 
			legend style = {
				nodes={inner sep=0.04em},
				draw=none,
				font=\scriptsize,
				cells={align=left},
				anchor=east,
				at={(0.55,0.89)},
				/tikz/column 1/.style={column sep=5pt,},
			},
			]
			
%
%

			\addplot[ postaction={
				pattern=horizontal lines
			}, fill={myred!50!white}]
			table
			[
			x expr=\thisrow{Cores}, 
			y expr=\thisrow{GHOST_perf[GF/s]},
			]{\Bblocksz};
			\addlegendentry{~$n_\mathrm{b}=32$}	
					
					
			\end{axis}
			
			\begin{axis}[trim axis left, trim axis right, scale only axis,
			xshift=2.9pt,
			width=0.5\textwidth,height=0.178\textheight,
			ybar, 
			ybar legend, 
			bar width=0.8mm,
			xmin=0,
			ymin=0,
			ymax=50,
			xmax=11,
			ytick={},
			xmajorgrids,
			ymajorgrids,
			axis lines=none, 
			legend columns = 1, 
			legend style = {
				nodes={inner sep=0.04em},
				draw=none,
				font=\scriptsize,
				cells={align=left},
				anchor=east,
				at={(0.48,0.75)},
				/tikz/column 1/.style={column sep=5pt,},
			},
			]
%
%
%
%
			\addplot[ postaction={
				pattern=north east lines
			}, fill={myblue!50!white}]
			table
			[
			x expr=\thisrow{Cores}, 
			y expr=\thisrow{GHOST_perf[GF/s]},
			]{\Ablocksz};
			\addlegendentry{~$n_\mathrm{b}=2$}	
			\end{axis}
			
			\node [font=\small] at (1.5,-1.6){~(a) Single socket performance};
			\end{tikzpicture}
		\end{subfigure}
		\hspace{-5em}
		\begin{subfigure}[t]{.39\textwidth}
			\begin{tikzpicture}
			\pgfplotstableread{figures/Fig9_ChebFD_Emmy_SuperMUC_/Emmy_NUM_OMP_THREADS_topi128-64-64_ns128_nb2_np500_Vector_1.txt}\NbA;
			\pgfplotstableread{figures/Fig9_ChebFD_Emmy_SuperMUC_/Emmy_NUM_OMP_THREADS_topi128-64-64_ns128_nb2_np500_Vector_2.txt}\NbB;
			\pgfplotstableread{figures/Fig9_ChebFD_Emmy_SuperMUC_/Emmy_NUM_OMP_THREADS_topi128-64-64_ns128_nb2_np500_Vector_5.txt}\NbC;
			\pgfplotstableread{figures/Fig9_ChebFD_Emmy_SuperMUC_/Emmy_NUM_OMP_THREADS_topi128-64-64_ns128_nb2_np500_Vector_10.txt}\NbD;
			\begin{axis}[
			trim axis left, trim axis right, scale only axis,
			width=0.73\textwidth,height=0.178\textheight,
			xlabel = {Number of nodes},
			xmin=0,
			ymin=0,
			ymax=350,
			xtick={0,2,4,6,8,10},
			ylabel = {Performance [\si{\giga \flop / \second}]}, 
			y label style={font=\footnotesize,at={(0.08,0.5)}},
			x label style={font=\footnotesize}, 			
			x tick label style={font=\footnotesize},
			y tick label style={font=\scriptsize}, 
			xmajorgrids,
			ymajorgrids,
			legend columns = 1, 
			legend style = {
				nodes={inner sep=0.04em},
				draw=none,
				font=\scriptsize,
				cells={align=left},
				anchor=east,
				at={(0.52,0.83)},
				/tikz/column 1/.style={column sep=5pt,},
			},
			]
			
			\addlegendimage{only marks, mark=square*, PineGreen}
			\addlegendimage{only marks, mark=star, Sepia}
			\addlegendimage{only marks, mark=diamond*, Bittersweet}
			\addlegendimage{only marks, mark=*,blue}
			
			\addplot[ mark=square*,mark size =1.5 pt, PineGreen]
			table
			[
			x expr=\thisrow{nodes}, 
			y expr=\thisrow{Perf[GF/s]},
			]{\NbA};
			\addlegendentry{~1T, $n_\mathrm{b}=2$}
			
			\addplot[ mark=star,mark size =2.3 pt, Sepia]
			table
			[
			x expr=\thisrow{nodes}, 
			y expr=\thisrow{Perf[GF/s]},
			]{\NbB};
			\addlegendentry{~2T, $n_\mathrm{b}=2$}
			
			\addplot[mark=diamond*, mark size =2 pt, Bittersweet]
			table
			[
			x expr=\thisrow{nodes}, 
			y expr=\thisrow{Perf[GF/s]},
			]{\NbC};
			\addlegendentry{~5T, $n_\mathrm{b}=2$}
			
			\addplot[mark=*,mark size =1.5 pt, blue]
			table
			[
			x expr=\thisrow{nodes}, 
			y expr=\thisrow{Perf[GF/s]},
			]{\NbD};
			\addlegendentry{~10T, $n_\mathrm{b}=2$}
			
			%
			%
			%
			\end{axis}
			\node [font=\small] at (3,-1.6){~(b) MPI only vs.\ hybrid communication,  $n_\mathrm{b}=2$,};
			\node at (2.5,2.2) {\small \textbf{(c)}};
			\end{tikzpicture}
		\end{subfigure}
	\hspace{-1.5em}
		\begin{subfigure}[t]{.39\textwidth}
			\begin{tikzpicture}
			\pgfplotstableread{figures/Fig9_ChebFD_Emmy_SuperMUC_/Emmy_NUM_OMP_THREADS_topi128-64-64_ns128_nb32_np500_Vector_1.txt}\NbE;
			\pgfplotstableread{figures/Fig9_ChebFD_Emmy_SuperMUC_/Emmy_NUM_OMP_THREADS_topi128-64-64_ns128_nb32_np500_Vector_2.txt}\NbF;
			\pgfplotstableread{figures/Fig9_ChebFD_Emmy_SuperMUC_/Emmy_NUM_OMP_THREADS_topi128-64-64_ns128_nb32_np500_Vector_5.txt}\NbG;
			\pgfplotstableread{figures/Fig9_ChebFD_Emmy_SuperMUC_/Emmy_NUM_OMP_THREADS_topi128-64-64_ns128_nb32_np500_Vector_10.txt}\NbH;
			\begin{axis}[
			trim axis left, trim axis right, scale only axis,
			width=0.73\textwidth,height=0.178\textheight,
			xlabel = {Number of nodes},
			xmin=0,
			ymin=0,
			xmajorgrids,
			ymajorgrids,
			ymax=450,
			xtick={0,2,4,6,8,10},
			y label style={font=\footnotesize,at={(0.08,0.5)}},
			x label style={font=\footnotesize}, 			
			x tick label style={font=\footnotesize},
			y tick label style={font=\scriptsize}, 
			legend columns = 1, 
			legend style = {
				nodes={inner sep=0.04em},
				draw=none,
				font=\scriptsize,
				cells={align=left},
				anchor=east,
				at={(0.59,0.82)},
				/tikz/column 1/.style={column sep=5pt,},
			},
			]
			
			\addlegendimage{only marks, mark=square*, PineGreen}
			\addlegendimage{only marks, mark=star, Sepia}
			\addlegendimage{only marks, mark=diamond*, Bittersweet}
			\addlegendimage{only marks, mark=*,blue}
			
			\addplot[ mark=square*,mark size =1.5 pt, PineGreen]
			table
			[
			x expr=\thisrow{nodes}, 
			y expr=\thisrow{Perf[GF/s]},
			]{\NbE};
			\addlegendentry{~1T, $n_\mathrm{b}=32$}
			
			\addplot[ mark=star,mark size =2.3 pt, Sepia]
			table
			[
			x expr=\thisrow{nodes}, 
			y expr=\thisrow{Perf[GF/s]},
			]{\NbF};
			\addlegendentry{~2T, $n_\mathrm{b}=32$}
			
			\addplot[mark=diamond*, mark size =2 pt, Bittersweet]
			table
			[
			x expr=\thisrow{nodes}, 
			y expr=\thisrow{Perf[GF/s]},
			]{\NbG};
			\addlegendentry{~5T, $n_\mathrm{b}=32$}
			
			\addplot[mark=*,mark size =1.5 pt, blue]
			table
			[
			x expr=\thisrow{nodes}, 
			y expr=\thisrow{Perf[GF/s]},
			]{\NbH};
			\addlegendentry{~10T, $n_\mathrm{b}=32$}
			
			\end{axis}
			\node [font=\small] at (2.4,-1.6){~$n_\mathrm{b}=32$};
			\end{tikzpicture}
		\end{subfigure}
	\hspace{-0.5em}
			\begin{subfigure}[t]{.25\textwidth}
			\begin{tikzpicture}
						\put(0.1,-29) {\includegraphics[width=0.88\textwidth,height=0.08 \textheight]{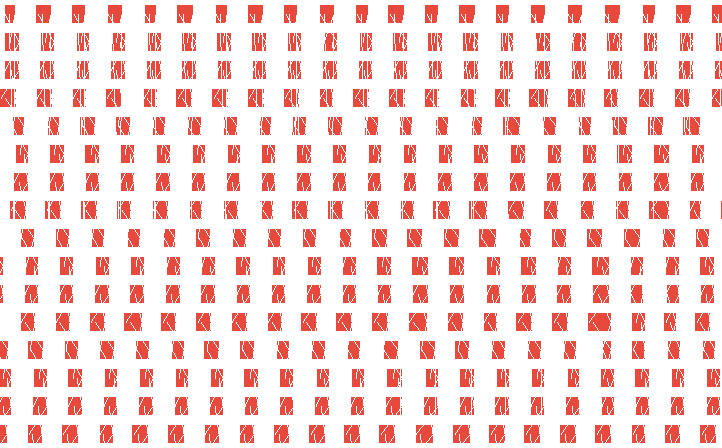}}
					\put(0,-29) {\begin{axis}[
						width=1.4\textwidth,height=0.16\textheight,
						ylabel = {Rank},
						y label style={at={(0.17,0.5)}},
						x label style={at={(0.5,0.08)}},
						x tick label style={font=\scriptsize},
						y tick label style={font=\scriptsize}, 
						xmin=1, xmax=35,
						ymin=0, ymax=15,
						xmajorticks=false, 
						ytick={0,5,10,15}, 
						yticklabels={\textbf{15},\textbf{10},\textbf{5},},
						]
				\end{axis}}	
			\put(0,19) {\includegraphics[width=0.88\textwidth,height=0.09 \textheight]{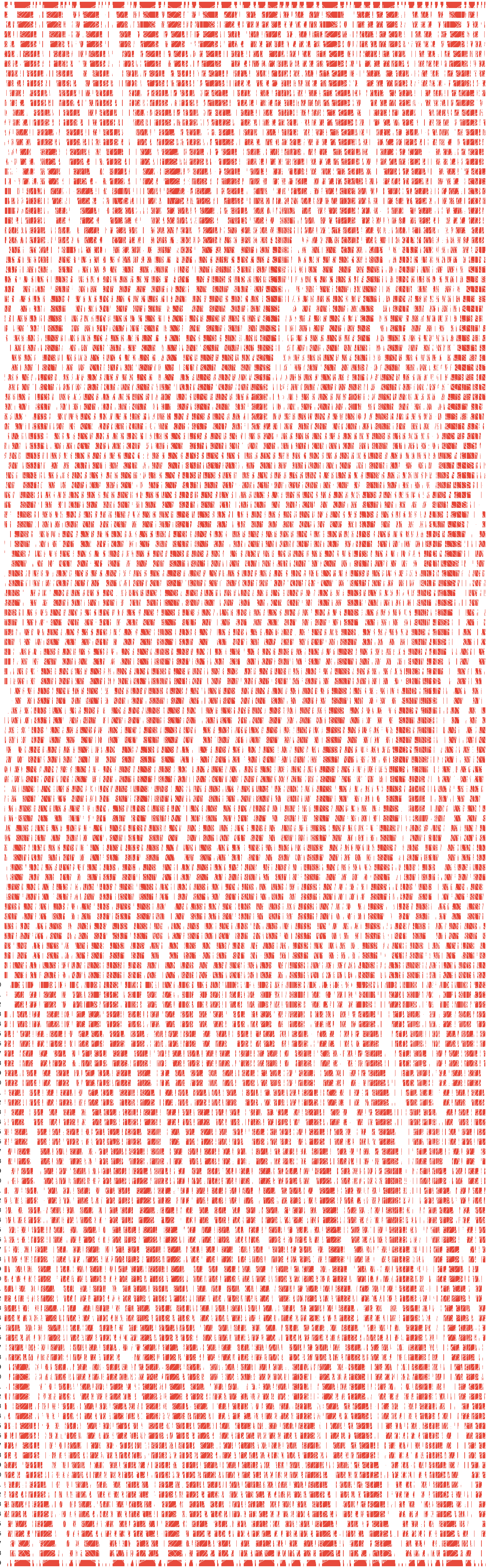}}
				\put(0,19) {\begin{axis}[
					width=1.4\textwidth,height=0.175\textheight,
					ylabel = {Rank},
					y label style={at={(0.17,0.5)}},
					xlabel = {Time step},
					x label style={at={(0.5,-0.8)}},
					x tick label style={font=\scriptsize},
					y tick label style={font=\scriptsize},
					xmajorticks=false, 
					xmin=1, xmax=35,
					ymin=0, ymax=159,
					ytick={0,40,80,120,159},   
					yticklabels={\textbf{159},\textbf{120},\textbf{80},\textbf{40},0},
					]
				\end{axis}}	
				\node [font=\small] at (1,-2.53){~(c) Timeline visulization};
				\node at (2.38,-0.83) {\small \textbf{~10T}};
				\node at (2.395,0.82) {\small \textbf{~1T}};
				\end{tikzpicture}
		\end{subfigure}			
	\end{adjustbox}
	\caption{ChebFD application for the 
	  topological insulator matrix \protect\UseVerb{2Topi128} (\protect\UseVerb{static}
          \protect\UseVerb{OpenMP} scheduling,
          AVX vectorized and aligned execution, $n_\mathrm{iter}=5$)
		running on (single leaf switch connected) homogeneous \protect\UseVerb{Emmy}\@ nodes.
		(a) Performance scaling with OpenMP on a contention domain for $n_b=2$ and $n_b=32$,
		(b) scaling up to 10 nodes for $n_b=2$ and $n_b=32$, and different numbers of threads per process,
                (c) timeline for a specific number of iterations of pure MPI vs.\ full hybrid execution for $n_b=2$ and 8 \protect\UseVerb{Emmy} nodes.
	}
	\label{fig:ChebFD}
\end{figure}
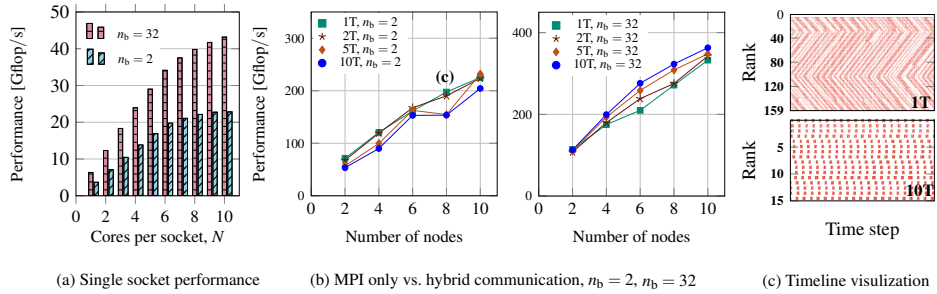